\documentclass[conference]{IEEEtran}
\IEEEoverridecommandlockouts
\usepackage{graphicx}
\usepackage{amsmath}
\usepackage{xcolor}
\usepackage{pdflscape}

\usepackage{cite}
\usepackage{amsmath,amssymb,amsfonts}
\usepackage{algorithmic}
\usepackage{graphicx}
\usepackage{textcomp}
\usepackage{xcolor}
\usepackage{graphicx}% Include figure files
\usepackage{dcolumn}% Align table columns on decimal point
\usepackage{bm}% bold math
\usepackage{cuted}
\usepackage{xcolor}
\usepackage[utf8]{inputenc}
\usepackage[T1]{fontenc}
\usepackage{mathptmx}
\usepackage{amsmath}

\newcommand{\ket}[1]{\left|#1\right\rangle}
\newcommand{\bra}[1]{\left\langle#1\right|}

\def\BibTeX{{\rm B\kern-.05em{\sc i\kern-.025em b}\kern-.08em
    T\kern-.1667em\lower.7ex\hbox{E}\kern-.125emX}}

\begin{document}

\title{ Analytic view on N body interaction in electrostatic quantum gates and decoherence effects in tight-binding model}
%Tight-binding approximation in description of electrostatic position-dependent Qbit, Quantum Swap and Quantum CNOT gate}
\author{Krzysztof Pomorski}

\author{\IEEEauthorblockN{ %%%%1\textsuperscript{st}
 Krzysztof Pomorski$^{a,b,c}$}
\IEEEauthorblockA{\textit{University College Dublin, Dublin, Ireland} \\ %\\
\textit{a: School of Computer Science}\\
\textit{b: School of Electrical and Electronic Engineering}\\
\textit{Dublin, Ireland}  \\
\textit{c: Quantum Hardware Systems} \\
Webpage: \textit{www.quantumhardwaresystems.com} \\
E-mail: \textit{kdvpomorski@gmail.com}}}

\maketitle

\begin{abstract}
Analytical solutions describing quantum swap and Hadamard gate are given with the use of tight-binding approximation. Decoherence effects are described analytically for 2 interacting electrons confined by local potentials with use of tight-binding simplistic model and in Schroedinger formalism with omission of spin degree of freedom. The obtained results can be generalized for the case of N electrostatically interacting quantum bodies confined by local potentials  (N-qubit) system representing any electrostatic quantum gate with $N_1$/N-$N_1$ inputs/outputs. The mathematical structure of system evolution with time is specified. 
\end{abstract}


\section{Technological motivation}
\begin{figure}
    \centering
    \includegraphics[width=0.6\columnwidth]{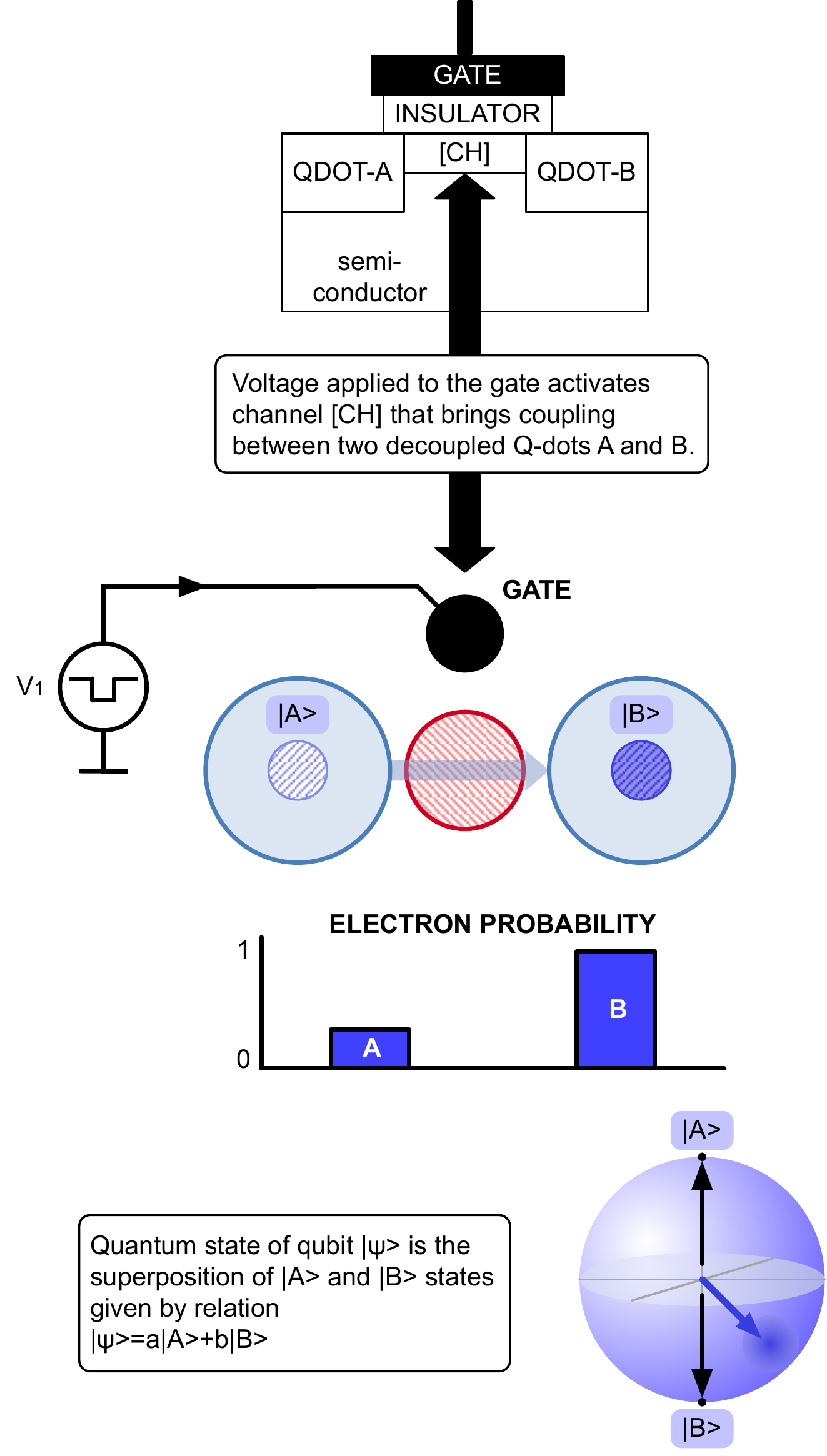} %{QUBIT_SUMMARY11QM.png}
    \caption{Basic concept of position based qubit \cite{Pomorski_spie}, \cite{Xu} and its correspondence to Bloch sphere \cite{Nbody}.}
 \label{central1}
\end{figure}
Systematic progress in implementation of superconducting quantum computer by IBM, Google and D-wave companies
is obtained but it faces the key challenges due to the fact that superconducting qubits are controlled by magnetic and RF fields.
The usage of magnetic field is main obstacle in scalability of such structures so very high integration quantum circuits are not
expected to take place. What is even more important the usage of low temperature superconductors that has superconducting coherence lenght of range of 300nm (it thus proportional to the size of Cooper pair) brings the limitation in further miniaturization of those structures \cite{Pomorski_compel}, \cite{PSSB2012}. Suprisingly semiconductor technologies have no such limitations and most advanced CMOS transistors has the channgel lenght between source and drain of 3nm. Therefore we have no longer temperature activation of electric carriers as it is the case of standard CMOS technologies.
What is more important in case of such small strucures the big gradient of electric field takes place and it serves as electric field activation of electric carriers even at mK temperatures. It motivates us to study the
physics of quantum information processing in such devices. First study of such structures was conducted by Fujisawa \cite{Fujisawa}, Petta \cite{Petta} and continued by many others as Pomorski \cite{Pomorski_spie}, \cite{Nbody}, Imran \cite{Bashir19}, Panagiotis, Leipold \cite{Panos}, \cite{Dirk}.

\section{Introduction into electrostatic position dependent qubit}

%\begin{figure}
%    \centering
%    \includegraphics[width=0.95\columnwidth]{QUBIT_SUMMARY11Q.pdf} %{QUBIT_SUMMARY11QM.png}
 %   \caption{Basic concept of position based qubit \cite{Pomorski_spie} and its correspondence to Bloch sphere \cite{Nbody}.}
% \label{central1}
%\end{figure}

%\newpage
The tight-binding Hamiltonian of this system
is given as %%%$\hat{H}(t)_{[x=(x_1,x_2)]}=$
%\begin{strip}
\begin{eqnarray}
\label{simplematrix}
%=\hat{H}(t)=
%\hat{H}(t)_{[x=(x_1,x_2)]}=
%\begin{split}
\hat{H}(t)_{[x=(x_1,x_2)]}=
\begin{pmatrix}
E_{p1}(t) & t_{s12}(t)=|t_{s12}|e^{+i\alpha(t)} \\
t_{s12}^{\dag}(t)=|t_{s12}|e^{-i\alpha(t)} & E_{p2}(t)
\end{pmatrix}=\nonumber \\
E_{p1}(t)\ket{x_1}\bra{x_1}+E_{p2}(t)\ket{x_2}\bra{x_2}+t_{s12}(t)\ket{x_1}\bra{x_2}+t_{s21}(t)\ket{x_2}\bra{x_1}
\nonumber \\
=(E_1(t)\ket{E_1}_t \bra{E_1}_t+E_2(t)\ket{E_2}\bra{E_2})_{[E=(E_1,E_2)]}.
\end{eqnarray}
%\end{strip}
%\end{split}
The Hamiltonian's eigenenergies $(E_1(t), E_2(t))$ with $(E_1<E_2)$ and eigenstates $(\ket{E_1(t)}, \ket{E_2(t)})$ are specified as \\
with $t_{s12}(t)=t_{sr}(t)+it_{si}(t)$ for any $(t_{sr}(t),t_{si}(t)) \in R$ and for any $(E_{p1}(t), E_{p2}(t)) \in R$. 
\begin{strip}
\begin{eqnarray}
E_1(t)= \left(-\sqrt{\frac{(E_{p1}(t)-E_{p2}(t))^2}{4}+|t_{s12}(t)|^2}+ \frac{E_{p1}(t)+E_{p2}(t)}{2}\right), %\nonumber \\
E_2(t)= \left(+\sqrt{\frac{(E_{p1}(t)-E_{p2}(t))^2}{4}+|t_{s12}(t)|^2}+\frac{E_{p1}(t)+E_{p2}(t)}{2}\right), \nonumber \\
%\end{eqnarray*}
%\end{strip}
%and energy eigenstates $\ket{E_1(t)}, \ket{E_2(t)}$ are given as  % expressed in terms of maximum localized state on the left and right node and have the following form
%\begin{strip}
%\begin{eqnarray}
\ket{E_g,t}_n=
%%\begin{pmatrix}
%%\frac{(E_{p2}(t)-E_{p1}(t))+\sqrt{\frac{(E_{p2}(t)-E_{p1}(t))^2}{2}+|t_{s12}(t)|^2}}{-i t_{sr}(t)+t_{si}(t)} \\
%%-1
%%\end{pmatrix}= \nonumber \\
\frac{(\frac{(E_{p2}(t)-E_{p1}(t))}{2}+\sqrt{\frac{(E_{p2}(t)-E_{p1}(t))^2}{2}+|t_{s12}(t)|^2})e^{-i \alpha(t)}\ket{x_1}-|t_{s12}(t)|\ket{x_2})}{\sqrt{|t_{s12}(t)|^2+(\frac{(E_{p2}(t)-E_{p1}(t))}{2}+\sqrt{\frac{(E_{p2}(t)-E_{p1}(t))^2}{2}+|t_{s12}(t)|^2})^2}}
=a(t)\ket{x_1}_n+b(t)\ket{x_2}_n,
\nonumber \\
\ket{E_e,t}_n=
%%\begin{pmatrix}
%%\frac{-(E_{p2}(t)-E_{p1}(t))+\sqrt{\frac{(E_{p2}(t)-E_{p1}(t))^2}{2}+|t_{s12}(t)|^2}}{t_{sr}(t) - i t_{si}(t)} \\
%%1
%%\end{pmatrix}= \nonumber \\
\frac{(-\frac{(E_{p2}(t)-E_{p1}(t))}{2}+\sqrt{\frac{(E_{p2}(t)-E_{p1}(t))^2}{2}+|t_{s12}(t)|^2})e^{+i\alpha(t)}\ket{x_1}+|t_{s12}(t)|\ket{x_2})}{\sqrt{|t_{s12}(t)|^2+(-\frac{(E_{p2}(t)-E_{p1}(t))}{2}+\sqrt{\frac{(E_{p2}(t)-E_{p1}(t))^2}{2}+|t_{s12}(t)|^2})^2}}=c(t)\ket{x_1}_n+d(t)\ket{x_2}_n, %\end{eqnarray}
%\begin{eqnarray}
\nonumber \\
a(t)=\frac{((\frac{E_{p2}(t)-E_{p1}(t)}{2})+\sqrt{(\frac{E_{p2}(t)-E_{p1}(t)}{2})^2+|t_{s12}(t)|^2})e^{-i \alpha(t)}}{\sqrt{|t_{s12}(t)|^2+(\frac{(E_{p2}(t)-E_{p1}(t))}{2}+\sqrt{\frac{(E_{p2}(t)-E_{p1}(t))^2}{2}+|t_{s12}(t)|^2})^2}}, \nonumber \\
b(t)=\frac{-|t_s(t)|}{\sqrt{|t_s(t)|^2+(\frac{(E_{p2}(t)-E_{p1}(t))}{2}+\sqrt{\frac{(E_{p2}(t)-E_{p1}(t))^2}{2}+|t_{s12}(t)|^2})^2}}, \nonumber \\
c(t)=\frac{-(\frac{E_{p2}(t)-E_{p1}(t)}{2})+\sqrt{(\frac{E_{p2}(t)-E_{p1}(t)}{2})^2+|t_{s}(t)|^2})e^{+i\alpha(t)}}{\sqrt{|t_s|^2+(-(E_{p2}(t)-E_{p1}(t))+\sqrt{\frac{(E_{p2}(t)-E_{p1}(t))^2}{2}+|t_{s12}(t)|^2})^2}}, \nonumber \\
d(t)=\frac{+|t_s(t)|}{\sqrt{|t_s|^2+(-\frac{(E_{p2}(t)-E_{p1}(t))}{2}+\sqrt{\frac{(E_{p2}(t)-E_{p1}(t))^2}{2}+|t_{s12}(t)|^2})^2}},\nonumber \\
N_1(t)=\frac{1}{\sqrt{|t_s(t)|^2+(\frac{(E_{p2}(t)-E_{p1}(t))}{2}+\sqrt{\frac{(E_{p2}(t)-E_{p1}(t))^2}{2}+|t_{s12}(t)|^2})^2}},\nonumber \\
N_2(t)=\frac{1}{\sqrt{|t_s|^2+(-\frac{(E_{p2}(t)-E_{p1}(t))}{2}+\sqrt{\frac{(E_{p2}(t)-E_{p1}(t))^2}{2}+|t_{s12}(t)|^2})^2}},
\nonumber \\
|a(t)|^2+|b(t)|^2=1, |c(t)|^2+|d(t)|^2=1, b(t) \in R, d(t) \in R,\nonumber \\
\rho(t)_{qubit}=\ket{\psi(t)}\bra{\psi(t)}=(c_g(t)\ket{E_g}+c_e(t)\ket{E_e})(c_g(t)^{*}\bra{E_g}+c_e(t)^{*}\bra{E_e}) %%=\nonumber \\
= \begin{pmatrix}
|c_g(t)|^2 & c_g(t)c_e(t)^{*} \\
 c_e(t)c_g(t)^{*} & |c_e(t)|^2
\end{pmatrix}, \nonumber \\
\rho(t)_{non-interacting,qubitA-qubit B}=
\begin{pmatrix}
|c_{g,A}(t)|^2 & c_{g,A}(t)c_{e,A}(t)^{*} \\
 c_{e,A}(t)c_{g,A}(t)^{*} & |c_{e,A}(t)|^2
\end{pmatrix} \times 
\begin{pmatrix}
|c_{g,B}(t)|^2 & c_{g,B}(t)c_{e,B}(t)^{*} \\
 c_{e,B}(t)c_{g,B}(t)^{*} & |c_{e,B}(t)|^2
\end{pmatrix}=\nonumber \\
\begin{pmatrix}
|c_{g,A}(t)|^2\begin{pmatrix}
|c_{g,B}(t)|^2 & c_{g,B}(t)c_{e,B}(t)^{*} \\
 c_{e,B}(t)c_{g,B}(t)^{*} & |c_{e,B}(t)|^2
\end{pmatrix} & c_{g,A}(t)c_{e,A}(t)^{*}\begin{pmatrix}
|c_{g,B}(t)|^2 & c_{g,B}(t)c_{e,B}(t)^{*} \\
c_{e,B}(t)c_{g,B}(t)^{*} & |c_{e,B}(t)|^2
\end{pmatrix} \\
c_{e,A}(t)c_{g,A}(t)^{*}
\begin{pmatrix}
|c_{g,B}(t)|^2 & c_{g,B}(t)c_{e,B}(t)^{*} \\
c_{e,B}(t)c_{g,B}(t)^{*} & |c_{e,B}(t)|^2
\end{pmatrix} &  |c_{e,B}(t)|^2\begin{pmatrix}
|c_{g,B}(t)|^2 & c_{g,B}(t)c_{e,B}(t)^{*} \\
 c_{e,B}(t)c_{g,B}(t)^{*} & |c_{e,A}(t)|^2
\end{pmatrix}
%\end{pmatrix} \\
% c_{e,A}(t)c_{g,A}(t)^{*} & |c_{e,A}(t)|^2
\end{pmatrix}=\nonumber \\ =
\begin{pmatrix}
|c_{g,A}(t)|^2|c_{g,B}(t)|^2 & |c_{g,A}(t)|^2 & c_{e,A}(t)^{*}c_{g,A}(t)|c_{g,B}(t)|^2 & c_{e,A}(t)^{*}c_{g,A}(t)c_{e,B}(t)^{*}c_{g,B}(t) \\
|c_{g,A}(t)|^2 & |c_{g,A}(t)|^2|c_{e,B}(t)|^2 & c_{e,A}(t)^{*}c_{g,A}(t)c_{e,B}(t)c_{g,B}(t)^{*} & c_{e,A}(t)^{*}c_{g,A}(t)|c_{e,B}(t)|^2 \\
c_{e,A}(t)c_{g,A}(t)^{*}|c_{g,B}(t)|^2 & c_{e,A}(t)c_{g,A}(t)^{*}c_{g,B}(t)c_{e,B}(t)^{*} & |c_{e,A}(t)|^2|c_{g,B}(t)|^2 & |c_{e,A}(t)|^2c_{g,B}(t)c_{e,B}(t)^{*} \\
c_{e,A}(t)c_{g,A}(t)^{*} c_{e,B}(t)c_{g,B}(t)^{*} & c_{e,A}(t)c_{g,A}(t)^{*}|c_{e,B}(t)|^2 & |c_{e,A}(t)|^2c_{g,B}(t)^{*}c_{e,B}(t) & |c_{e,A}(t)|^2|c_{e,B}(t)|^2 \\
\end{pmatrix}=\nonumber \\
\hat{U}(t,t_0)\rho(t_0)\hat{U}(t,t_0)^{\dag}=
\begin{pmatrix}
e^{\frac{1}{\hbar i}\int_{t0}^{t}dt'(E_{g,A}(t')+E_{g,B}(t'))dt'} & 0 & 0 & 0 \\
0 & e^{\frac{1}{\hbar i}\int_{t0}^{t}dt'(E_{g,A}(t')+E_{e,B}(t'))dt'} & 0 & 0 \\
0 & 0 & e^{\frac{1}{\hbar i}\int_{t0}^{t}dt'(E_{e,A}(t')+E_{g,B}(t'))dt'} & 0 \\
0 & 0 & 0 & e^{\frac{1}{\hbar i}\int_{t0}^{t}dt'(E_{e,A}(t')+E_{e,B}(t'))dt'} \\
\end{pmatrix} \times \nonumber \\ 
\times \rho(t_0) \times
\begin{pmatrix}
e^{-\frac{1}{\hbar i}\int_{t0}^{t}dt'(E_{g,A}(t')+E_{g,B}(t'))dt'} & 0 & 0 & 0 \\
0 & e^{-\frac{1}{\hbar i}\int_{t0}^{t}dt'(E_{g,A}(t')+E_{e,B}(t'))dt'} & 0 & 0 \\
0 & 0 & e^{-\frac{1}{\hbar i}\int_{t0}^{t}dt'(E_{e,A}(t')+E_{g,B}(t'))dt'} & 0 \\
0 & 0 & 0 & e^{-\frac{1}{\hbar i}\int_{t0}^{t}dt'(E_{e,A}(t')+E_{e,B}(t'))dt'}\\
\end{pmatrix}.
\end{eqnarray}
\end{strip}

Energy eigenstates $\ket{E_1(t)}, \ket{E_2(t)}$ are expressed in terms of maximum localized state on the left ($1 \rightarrow \ket{x_1}$) and right node ($1 \rightarrow \ket{x_2}$) as depicted in Fig.\ref{central1}.

%
%%\onecolumn
The last expressions can be written in a compact form
\begin{strip}
\begin{eqnarray}
\begin{pmatrix}
\ket{E_1,t} \\
\ket{E_2,t} \\
\end{pmatrix}=
\hat{S}_{2 \times 2}
\begin{pmatrix}
\ket{x_1} \\
\ket{x_2} \\
\end{pmatrix}
=%\nonumber \\
\begin{pmatrix}
\frac{(E_{p2}(t)-E_{p1}(t))+\sqrt{\frac{(E_{p2}(t)-E_{p1}(t))^2}{2}+|t_{s12}(t)|^2}}{-i t_{sr}(t)+t_{si}(t)} \frac{1}{N_1} & \frac{-1}{N_1} \\
\frac{-(E_{p2}(t)-E_{p1}(t))+\sqrt{\frac{(E_{p2}(t)-E_{p1}(t))^2}{2}+|t_{s12}(t)|^2}}{t_{sr}(t) - i t_{si}(t)}\frac{1}{N_2} & \frac{+1}{N_2} \\
\end{pmatrix}
\begin{pmatrix}
\ket{x_1} \\
\ket{x_2} \\
\end{pmatrix}.
\end{eqnarray}
\end{strip}
Setting $t_{si}(t)=1, t_{sr}(t)=0$ and $E_{p1}(t)=E_{p2}(t)=E_p$ we obtain
%\begin{eqnarray}
$\begin{pmatrix}
\ket{E_2}_n \\
\ket{E_1}_n \\
\end{pmatrix}=\frac{1}{\sqrt{2}}
\begin{pmatrix}
1 &  +1 \\
1 &  -1 \\
\end{pmatrix}
\begin{pmatrix}
\ket{x_2} \\
\ket{x_1} \\
\end{pmatrix}$
%\end{eqnarray}
which brings Hadamard matrix as relating q-state in the position base and in the energy base, where $\ket{E_{1(2)}}_n=\frac{1}{\sqrt{2}}\ket{E_{1(2)}}$.
If we associate logic state 0 with occupancy of node 1 (spanned by $\ket{x_1}$) and logic state 1 with occupancy of node 2 spanned by $\ket{x_2}$, then Hadamard operation on logic state $0$ brings occupancy of $E_2$ (so it is spanned by $\ket{E_2}$) and Hadamard operation on
logic state $1$ brings the entire occupancy of energy level $E_1$ (that is spanned by $\ket{E_1}$).

It shall be underlined that in the most simple case of position-based qubit $E_{p1}=E_{p2}=E_p=\rm const_1$ and $t_{s12}=|t|=\rm const_2$ and we obtain
%
%%\begin{eqnarray}
$\ket{\psi(t)}=\frac{1}{\sqrt{2}}(c_{E_1}e^{\frac{E_1}{\hbar}t}+c_{E2}e^{\frac{E_2}{\hbar}t})\ket{x_1}
+\frac{1}{\sqrt{2}}(-c_{E1}e^{\frac{E_1}{\hbar}t}+c_{E_2}e^{\frac{E_2}{\hbar}t})\ket{x_2}$.
%%\end{eqnarray}
%
It implies an oscillation of probabilities for the electron presence at node 1 (quantum logical 0) and 2 (quantum logical 1) with frequency $2|t|=E_2-E_1$, where $|c_{E_1}|^2 (|c_{E_2}|^2)$ is the probability for the quantum state to be in the ground (excited) state. % denoted by $E_1 (E_2)$. %%% $$.
It is possible to determine the qubit state under any evolution of two eigenergies $E_1(t)$ and $E_2(t)$ that are dependent on $E_{p1}(t), E_{p2}(t), t_{s12}(t)=t_{sr}(t)+t_{si}(t)i$. Simply, we have the state at any time instant given by \begin{eqnarray}
\ket{\psi_t}=%%\nonumber \\
e^{\int_{t_0}^{t}\frac{1}{\hbar i}\hat{H}(t')dt'}\ket{\psi_{t_0}}\nonumber \\
=\hat{U}(t,t_0)\ket{\psi_{t_0}}=
\begin{pmatrix}
e^{\frac{1}{\hbar i}\int_{t_0}^{t}E_1(t')dt'}, 0 \\
0 & e^{\frac{1}{\hbar i}\int_{t_0}^{t}E_2(t')dt'} \\
\end{pmatrix}\ket{\psi_{t_0}},
\end{eqnarray}
We notice that in case of qubit the evolution operator is given as
\begin{strip}
\begin{eqnarray}
\hat{U}(t,t_0)= %\nonumber \\
\begin{pmatrix}
e^{\frac{1}{\hbar i}\int_{t_0}^{t}\left(-\sqrt{\frac{(E_{p1}(t')-E_{p2}(t'))^2}{4}+|t_{s12}(t')|^2}+\frac{E_{p1}(t')+E_{p2}(t')}{2}\right)dt'} & 0 \\
0 & e^{\frac{1}{\hbar i}\int_{t_0}^{t}\left(+\sqrt{\frac{(E_{p1}(t')-E_{p2}(t'))^2}{4}+|t_{s12}(t')|^2}+\frac{E_{p1}(t')+E_{p2}(t')}{2}\right)dt'} \\
\end{pmatrix},
\end{eqnarray}
%%\end{strip}
\begin{eqnarray}
%%%\nonumber \\
\ket{\psi_t}=c_{e1}(t_0)e^{\frac{1}{\hbar i}\int_{t_0}^{t}\left(-\sqrt{\frac{(E_{p1}(t')-E_{p2}(t'))^2}{4}+|t_{s12}(t')|^2}+\frac{E_{p1}(t')+E_{p2}(t')}{2}\right)dt'}\ket{E_1(t)}+ \nonumber \\ +c_{e2}(t_0)e^{\frac{1}{\hbar i}\int_{t_0}^{t}\left(+\sqrt{\frac{(E_{p1}(t')-E_{p2}(t'))^2}{4}+|t_{s12}(t')|^2}+\frac{E_{p1}(t')+E_{p2}(t')}{2}\right)dt'}\ket{E_2(t)}=, \nonumber \\
=c_{e1}(t_0)e^{\frac{1}{\hbar i}\int_{t_0}^{t}\left(-\sqrt{\frac{(E_{p1}(t')-E_{p2}(t'))^2}{4}+|t_{s12}(t')|^2}+\frac{E_{p1}(t')+E_{p2}(t')}{2}\right)dt'}
\begin{pmatrix}
\frac{((E_{p2}(t)-E_{p1}(t))+\sqrt{\frac{(E_{p2}(t)-E_{p1}(t))^2}{2}+|t_{s12}(t)|^2})e^{i phase(t_{s12}(t))}i)}{\sqrt{|t_s(t)|^2+((E_{p2}(t)-E_{p1}(t))+\sqrt{\frac{(E_{p2}(t)-E_{p1}(t))^2}{2}+|t_{s12}(t)|^2})^2}} \\
\frac{-|t_s(t)|)}{\sqrt{|t_s(t)|^2+((E_{p2}(t)-E_{p1}(t))+\sqrt{\frac{(E_{p2}(t)-E_{p1}(t))^2}{2}+|t_{s12}(t)|^2})^2}}\\
\end{pmatrix}_{x}+\nonumber \\
+c_{e2}(t_0)e^{\frac{1}{\hbar i}\int_{t_0}^{t}\left(+\sqrt{\frac{(E_{p1}(t')-E_{p2}(t'))^2}{4}+|t_{s12}(t')|^2}+\frac{E_{p1}(t')+E_{p2}(t')}{2}\right)dt'}
\begin{pmatrix}
\frac{(-(E_{p2}(t)-E_{p1}(t))+\sqrt{\frac{(E_{p2}(t)-E_{p1}(t))^2}{2}+|t_{s}(t)|^2})e^{-iphase(t_{s12}(t))}}{\sqrt{|t_s|^2+(-(E_{p2}(t)-E_{p1}(t))+\sqrt{\frac{(E_{p2}(t)-E_{p1}(t))^2}{2}+|t_{s12}(t)|^2})^2}} \\
\frac{+|t_s(t)|}{\sqrt{|t_s|^2+(-(E_{p2}(t)-E_{p1}(t))+\sqrt{\frac{(E_{p2}(t)-E_{p1}(t))^2}{2}+|t_{s12}(t)|^2})^2}} \\
\end{pmatrix}_{x}.\nonumber \\
\end{eqnarray}
\end{strip}
Here, $c_{e1}(t_0)$ and $c_{e2}(t_0)$ describe the qubit in the energy representation at the initial time $t_0$, so $|c_{e1}(t_0)|^2+|c_{e2}(t_0)|^2=1$. Such presented evolution of position-based qubit is under the circumstances of small adiabatic changes in $t_s(t)$ and in
$E_{p1}(t)$, $E_{p2}(t)$. It is not the case of a qubit subjected to the rapid AC field that will support the existence of resonant states \cite{Pomorski_spie}.
The presented tight-binding approach can be seen as simplified version of Hubbard model when omission of spin was done \cite{Spalek}. 
%This sample document demonstrates proper use of REV\TeX~4.1 (and
%\LaTeXe) in manuscripts prepared for submission to AIP
%conference proceedings. Further information can be found in the documentation included in the distribution or available at
%\url{http://authors.aip.org} and in the documentation for
%REV\TeX~4.1 itself.

%When commands are referred to in this example file, they are always
%shown with their required arguments, using normal \TeX{} format. In
%this format, \verb+#1+, \verb+#2+, etc. stand for required
%author-supplied arguments to commands. For example, in
%\verb+\section{#1}+ the \verb+#1+ stands for the title text of the
%author's section heading, and in %\verb+\title{#1}+ the \verb+#1+
%stands for the title text of the paper.

%Line breaks in section headings at all levels can be introduced using
%\textbackslash\textbackslash. A blank input line tells \TeX\ that the
%paragraph has ended.

\vspace{+10mm}

\subsection{Qubit in external AC field}

%%In case of 2 well representing coupled quantum dot one can write the Hamiltonian in the tight-binding approximation in the following way
%%\begin{equation}
 %%   \label{simple_equation}
%%    H=(t_{1,2}|1><2| + t_{2,1}|2><1|)+ Ep_{1}|1><1| + Ep_{2}|2><2|.
%%\end{equation}
The quantum state is given as $|\psi>=c_0|1>+c_1|2>$, where $|c_0|^2$ is occupancy probability in q-dot 1 and $|c_1|^2$ is occupancy probability in q-dot 2 so $|c_0|^2+|c_1|^2=1$.
We have assumed that quantum dots can be asymmetric but in case of symmetric quantum dots we have $E_p(1)=E_p(2)=E_p$ and $t_{1,2}=t_{2,1}=t$.
%The Hamiltonian has the matrix representation given as
%\begin{equation}
  %  \label{simple_equation1}
  %  H=
%\begin{pmatrix}
%Ep_{1} & t_{2,1} \\
%t_{1,2} & Ep_{2}
%\end{pmatrix}
%\end{equation}
%We have 2 different eigenenergy values $(E_1,E_2)$ given as
%\begin{equation}
%E_{1(2)}=1/2 (Ep_1 + Ep_2 \pm \sqrt{4 t_{1,2} t_{2,1} + Ep_1^2 - 2 Ep_1 Ep_2 + Ep_2^2})
%\end{equation}
%with eigenvalues
%\begin{equation}
  %  \label{simple_equation2}
   % |\psi_1>=
%\begin{pmatrix}
%-(-Ep_1 + Ep_2 - \sqrt{4 t_{1,2} t_{2,1} + Ep_1^2 - 2 Ep_1 Ep_2 + Ep_2^2})/
%  2 t_{1,2} \\
%1
%\end{pmatrix}
%\end{equation}
%and
%\begin{equation}
  %  \label{simple_equation3}
   % |\psi_2>=
%\begin{pmatrix}
%-((-Ep_1 + Ep_2 + \sqrt{4 t_{1,2} t_{2,1} + Ep_1^2 - 2 Ep_1 Ep_2 + Ep_2^2})/(
 %2 t_{1,2})) \\ 1
%\end{pmatrix}
%\end{equation}
In case of symmetric wells we have eigenergies $E_{1(2)}=E_p \pm t$.
and eigenstates are $\psi_1=(-1,1)$ and $\psi_2=(1,1)$.
%\begin{equation}
%E_2=1/2 (Ep_1 + Ep_2 - \sqrt{4 t_{1,2} t_{2,1} + Ep_1^2 - 2 Ep_1 Ep_2 + Ep_2^2}).
%\end{equation}
%Two eigenstates correspond to simultaneous occupancy of 1 and 2 region so we van wrote
%\begin{eqnarray}
  %  \label{simple_equation2Q}
  %  |\psi_1(t)>=
%\begin{pmatrix}
%-\frac{-Ep_1 + Ep_2 - \sqrt{4 t_{1,2}t_{2,1} + (Ep_1 - Ep_2)^2}}{
%  2 t_{1,2}} \\
%1
%\end{pmatrix}=  \nonumber \\
%\frac{-(-Ep_1 + Ep_2 - \sqrt{4 t_{1,2} t_{2,1} + (Ep_1 - Ep_2)^2})}{2 t_{1,2}}
%\begin{pmatrix}
%1 \\
%0
%\end{pmatrix}+
%\begin{pmatrix}
%0 \\
%1
%\end{pmatrix}=  \nonumber \\
%\frac{-(-Ep_1 + Ep_2 - \sqrt{4 t_{1,2} t_{2,1} + (Ep_1 - Ep_2)^2})}{2 t_{1,2}}|1,0>+|0,1>
%\end{eqnarray}
%and
%\begin{eqnarray}
 %   \label{simple_equation3Q}
 %   |\psi_2(t)>=
%\begin{pmatrix}
%-(\frac{(-Ep_1 + Ep_2 + \sqrt{4 t_{1,2} t_{2,1} + (Ep_1^2 - Ep_2)^2})}{(
% 2 t_{1,2})}) \\ 1
%\end{pmatrix}
%= \nonumber \\
%=-(\frac{(-Ep_1 + Ep_2 + \sqrt{4 t_{1,2} t_{2,1} + (Ep_1^2 - Ep_2)^2})}{(
% 2 t_{1,2})})
% \begin{pmatrix}
%1 \\
%0
%\end{pmatrix}+
%\begin{pmatrix}
%0 \\
%1
%\end{pmatrix}
%= \nonumber \\
%=
%-(\frac{(-Ep_1 + Ep_2 + \sqrt{4 t_{1,2} t_{2,1} + (Ep_1^2 - Ep_2)^2})}{(
% 2 t_{1,2})})
%|1,0>+|0,1>.
%\end{eqnarray}
\subsubsection{Double Q-Dot in external non-uniform time-dependent potential}
\begin{figure}
\centering
\includegraphics[scale=0.7]{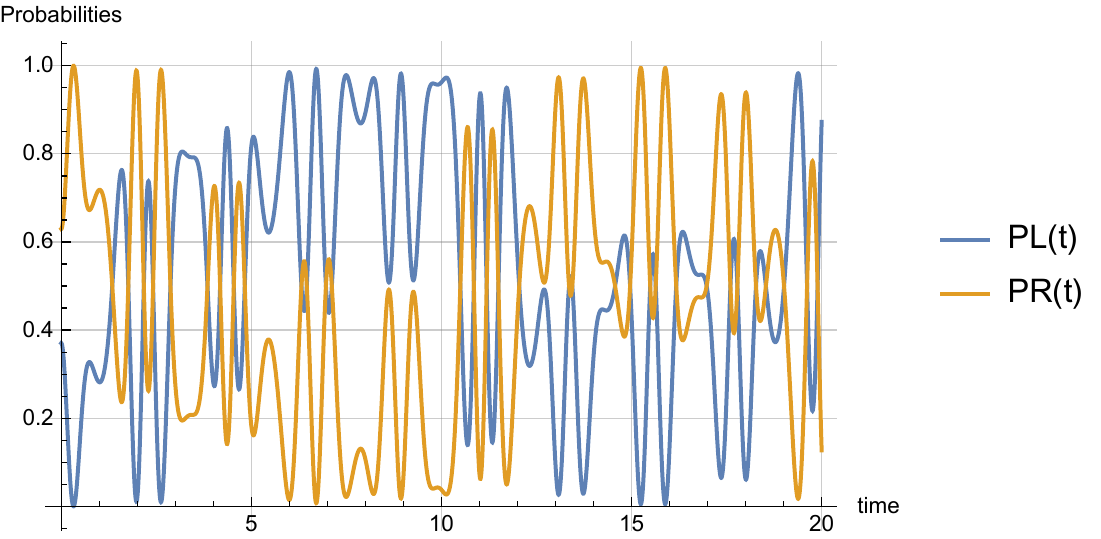}
\includegraphics[scale=0.7]{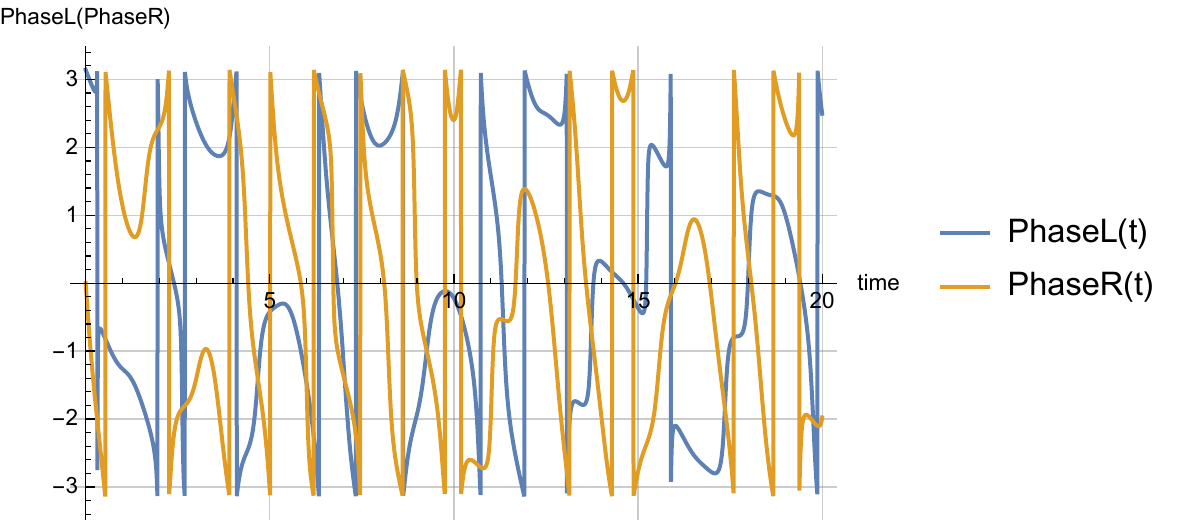}
\includegraphics[scale=0.7]{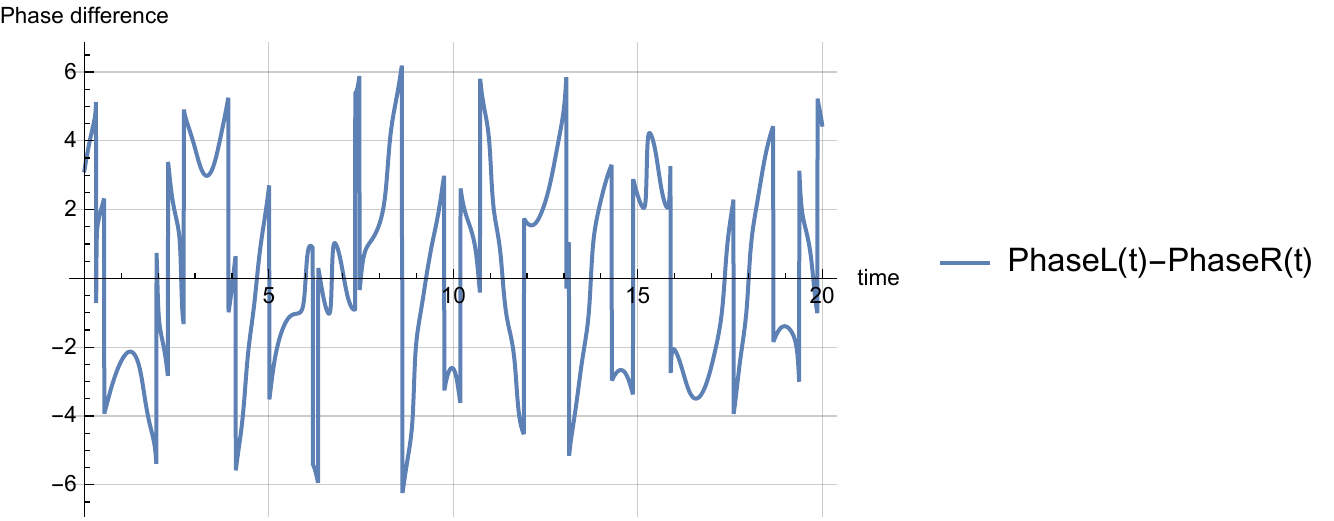}
\includegraphics[scale=0.7]{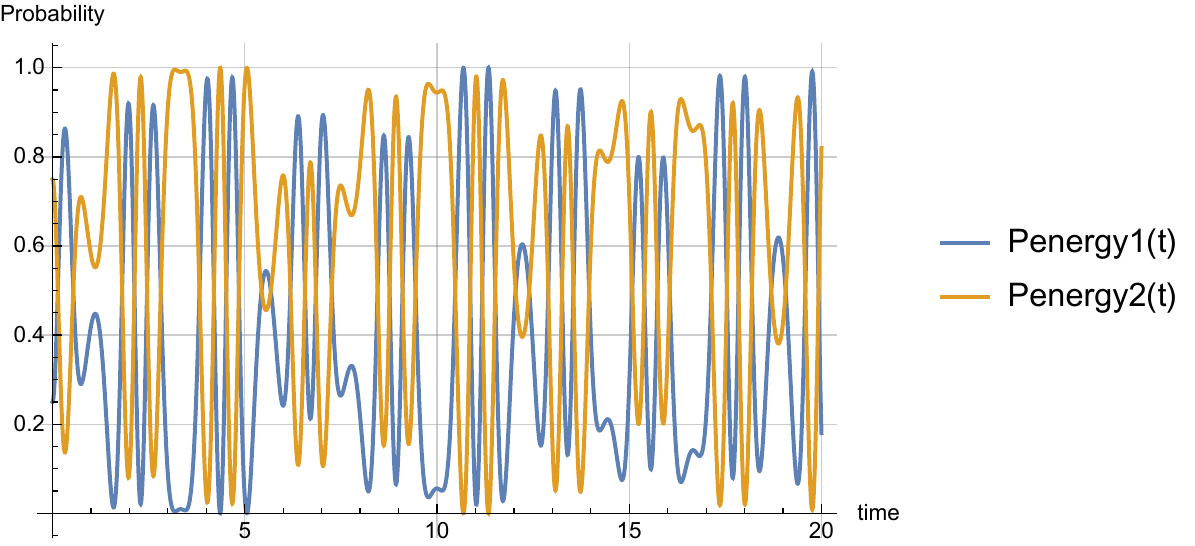}
\includegraphics[scale=0.7]{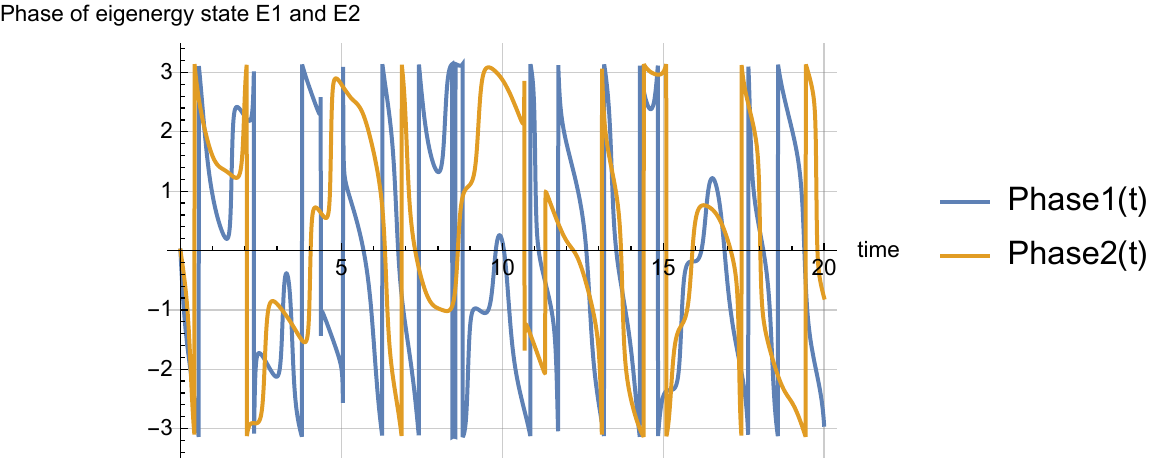}
\includegraphics[scale=0.7]{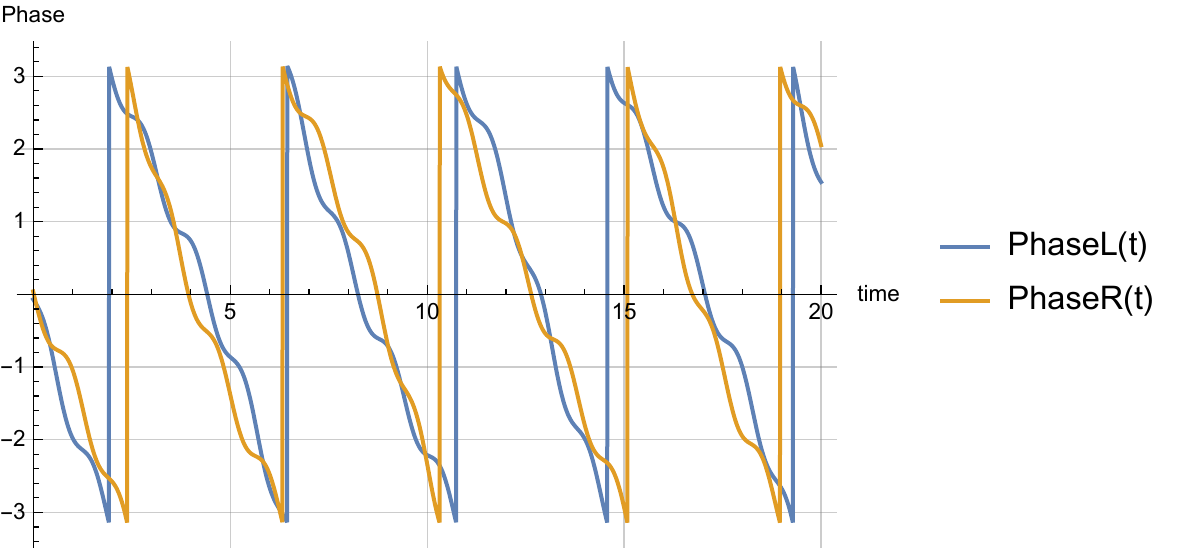}
\caption{Qubit in resonant state and Rabi oscillations shown by the probability of occupancy E1 and E2 state and by phase difference evolution with time for two different eigenenergies.Different $E_{p1}$ and $E_{p2}$ were used. First plot describes probability of particles at left and right node (upper), phase of particle wavefunction at left and right node, difference in phases, probability of occupancy of E1 and E2 energy levels and wavefunction phases in energy representation. Last plot describes the phase difference between left and right node in evolution of quantum system in microwavefield with same $E_{p1}=E_{p2}=E_{p}$. }
\end{figure}
We can introduce external time-dependent potential acting on points 1 and 2 such as $V(1,t)$ and $V(2,t)$.
The Hamiltonian of the system is then given as
\begin{equation}
    \label{simple_equation1QQ}
    H=
\begin{pmatrix}
E_{p1}+V(1,t) & t_{2,1}(t) \\
t_{1,2}(t) & E_{p2}+V(2,t)
\end{pmatrix}.
\end{equation}
If external time-dependent potential is weak than $t_{1,2}(t)\approx t_{1,2}$ and $t_{2,1}(t)\approx t_{2,1}$.
\begin{equation}
    \label{simple_equation1QQ1}
    H=
\begin{pmatrix}
E_{p1}+V(1,t) & t_{2,1} \\
t_{1,2} & E_{p2}+V(2,t)
\end{pmatrix}.
\end{equation}
We have 2 different time dependent eigenenergy values $(E_1(t),E_2(t))$ given as
\begin{eqnarray}
E_{1(2)}(t)=1/2 ( (E_{p1} + E_{p2}+\frac{1}{2}(V(1,t)+V(2,t)) \nonumber \\
 \pm \sqrt{4 t_{1,2}(t) t_{2,1}(t) + ((E_{p1} - E_{p2})+\frac{1}{2}(V(1,t)+V(2,t)))^2})).
\end{eqnarray}

with eigenvalues
\begin{eqnarray}
    \label{simple_equation2QQ}
    \ket{\psi_1(t)}=\nonumber \\
\begin{pmatrix}
-\frac{(-E_{p1}-V(1,t) + E_{p2}+V(2,t)
- \sqrt{4 t_{1,2}(t) t_{2,1}(t) + (E_{p1} - E_{p2}+(V(1,t)-V(2,t)))^2})}{
  2 t_{1,2}} \\
1
\end{pmatrix}
\end{eqnarray}
and
\begin{eqnarray}
    \label{simple_equation3QQ}
    \ket{\psi_2(t)}=\nonumber \\
\begin{pmatrix}
-(\frac{(-E_{p1}-V(1,t) + E_{p2}+V(2,t) + \sqrt{4 |t_{2,1}(t)|^2+ (E_{p1} - Ep_2+(V(1,t)-V(2,t)))^2})}{(
 2 t_{1,2})} \\ 1
\end{pmatrix}
\end{eqnarray}
At any time instant the quantum state is given as
\begin{equation}
|\psi(t)>=c1(t)|\psi_{1n}(t)>+c2(t)|\psi_{2n}(t)>,
\end{equation}
where $|c1(t)|^2+|c2(t)|^2=1$ and normalized eigenstates are denoted as $|\psi_{1n}(t)>$ and $|\psi_{2n}(t)>$.
Equivalently we obtain
\begin{equation}
<\psi_{1n}(t)|\psi(t)>=c1(t) ,<\psi_{2n}(t)|\psi(t)>=c2(t).
\end{equation}
The equation of motion can be written as $|\psi(t+dt)>=|\psi(t)>+\frac{dt}{i \hbar}(H(t) |\psi(t)>)$.
Equivalently we obtain 2 coupled recurrent relations for coefficients $c1(t+dt)$ and $c2(t+dt)$ depending on coefficients $c1(t)$ and $c2(t)$. We have
\begin{eqnarray}
c_1(t+dt)=<\psi_{1n}(t+dt)||\psi(t+dt)>= \nonumber \\ <\psi_{1n}(t+dt)||\psi(t)> + \frac{dt}{i \hbar}( <\psi_{1n}(t+dt)| H(t) |\psi(t)>)= \nonumber \\,
=<\psi_{1n}(t+dt)|(c_1(t)|\psi_{1n}(t)>+c2(t)|\psi_{2n}(t)>)+ \nonumber \\
+\frac{dt}{i \hbar}( <\psi_{1n}(t+dt)| \begin{pmatrix}
E_{p1}+V(1,t) & t_{2,1}(t) \\
t_{1,2}(t) & E_{p2}+V(2,t)
\end{pmatrix}  \nonumber \\ (c1(t)|\psi_{1n}(t)>+c2(t)|\psi_{2n}(t)>)).
\end{eqnarray} and
\begin{eqnarray}
c_2(t+dt)=<\psi_{2n}(t+dt)||\psi(t+dt)>= \nonumber \\ <\psi_{2n}(t+dt)||\psi(t)> + \frac{dt}{i \hbar}( <\psi_{2n}(t+dt)| H(t) |\psi(t)>)= \nonumber \\,
=<\psi_{2n}(t+dt)|(c_1(t)|\psi_{1n}(t)>+c_2(t)|\psi_{2n}(t)>)+ \nonumber \\
+\frac{dt}{i \hbar}( <\psi_{1n}(t+dt)| \begin{pmatrix}
E_{p1}+V(1,t) & t_{2,1}(t) \\
t_{1,2}(t) & Ep_{p2}+V(2,t)
\end{pmatrix} \nonumber \\ (c1(t)|\psi_{1n}(t)>+c2(t)|\psi_{2n}(t)>)).
\end{eqnarray}
In case of symmetric wells $E_{p1}=E_{p2}=E_p, t_{1,2}=|t|=t_{2,1}$ and time-independent Hamiltonian we have
\begin{eqnarray}
\frac{d}{dt}c_1(t)=\frac{1}{i\hbar}( E_p c_1(t) + |t| c_2(t) ) \nonumber \\
\frac{d}{dt}c_2(t)=\frac{1}{i\hbar}( |t| c_1(t) + E_p c_2(t) ).
\end{eqnarray}
Adding and substracting those two equations give us relations $|t|$-hopping (kinetic) term
\begin{eqnarray}
\frac{d}{dt}(c_1(t)+c_2(t))=-\frac{i}{\hbar}[+(E_p+|t|)(c_1(t) + c_2(t))] \nonumber \\
\frac{d}{dt}(c_1(t)-c_2(t))=-\frac{i}{\hbar}[+(E_p-|t|)(c_1(t) - c_2(t))].
\end{eqnarray}
In analogy to the evolution operator in Schroedinger equation we recognize two analytic solutions
\begin{eqnarray}
[ c_1(t) + c_2(t) ] = e^{ -\frac{i}{\hbar}(E_p + |t|) (t-t_0)   } [ c_1(t_0) + c_2(t_0) ]
\end{eqnarray}
\begin{eqnarray}
[ c_1(t) - c_2(t) ] =  e^{ -\frac{i}{\hbar}(E_p - |t|) (t-t_0)   } [ c_1(t_0) - c_2(t_0) ] .
\end{eqnarray}

By adding or substracting 2 equations and multiply by $1/2$ we obtain $c_1(t)$ or $c_2(t)$ given as
\begin{eqnarray}
 c_1(t)  = (e^{ -\frac{i}{\hbar}(E_p + |t|) (t-t_0)   } + e^{ -\frac{i}{\hbar}(E_p - |t|) (t-t_0)   } )\frac{c_1(t_0)}{2} + \nonumber \\
(e^{ -\frac{i}{\hbar}(E_p + |t|) (t-t_0)   } - e^{ -\frac{i}{\hbar}(E_p - |t|) (t-t_0)   } ) \frac{c_2(t_0)}{2}
\end{eqnarray}
\begin{eqnarray}
c_2(t)  =  (e^{ -\frac{i}{\hbar}(E_p + |t|) (t-t_0)   } - e^{ -\frac{i}{\hbar}(E_p - |t|) (t-t_0)   } )\frac{c_1(t_0)}{2} +  \nonumber \\
(e^{ -\frac{i}{\hbar}(E_p + |t|) (t-t_0)   } + e^{ -\frac{i}{\hbar}(E_p - |t|) (t-t_0)   } ) \frac{c_2(t_0)}{2} .
\end{eqnarray}
Presence of weak-time dependent potential present in 1 and 2 will change the analytic solutions into
\begin{eqnarray}
 c_1(t)  = e^{-\frac{i}{\hbar}\int_{t_0}^{t} V(1,t)dt}e^{ -\frac{i}{\hbar}E_p(t-t_0)}[(e^{ -\frac{i}{\hbar}|t|(t-t_0)} + e^{ +\frac{i}{\hbar}|t|(t-t_0)   } )\frac{c_1(t_0)}{2}+\nonumber \\
(e^{-\frac{i}{\hbar}|t| (t-t_0)   } - e^{\frac{i}{\hbar}|t|(t-t_0)   } ) \frac{c_2(t_0)}{2}]
\end{eqnarray}
\begin{eqnarray}
 c_2(t)  = e^{-\frac{i}{\hbar}\int_{t_0}^{t} V(2,t)dt}e^{ -\frac{i}{\hbar}E_p(t-t_0)}[(e^{ -\frac{i}{\hbar}|t|(t-t_0)} - e^{ +\frac{i}{\hbar}|t|(t-t_0)   } )\frac{c_1(t_0)}{2} \nonumber \\
+(e^{-\frac{i}{\hbar}|t| (t-t_0)   } + e^{\frac{i}{\hbar}|t|(t-t_0)   } ) \frac{c_2(t_0)}{2}]
\end{eqnarray}
The last two equations are equivalent to
\begin{strip}
\begin{eqnarray}
 c_1(t)  = e^{-\frac{i}{\hbar}\int_{t_0}^{t} V(1,t)dt}e^{ -\frac{i}{\hbar}E_p(t-t_0)}[ cos(\frac{|t|}{\hbar}(t-t_0)) c_1(t_0)- sin(\frac{|t|}{\hbar}(t-t_0)) i c_2(t_0)] = \nonumber \\
 e^{-\frac{i}{\hbar}\int_{t_0}^{t} V(1,t)dt}(cos(\frac{E_p(t-t_0)}{\hbar})-sin(\frac{E_p|t-t_0|}{\hbar})i )[ cos(\frac{|t|}{\hbar}(t-t_0)) c_1(t_0)- sin(\frac{|t|}{\hbar}(t-t_0)) i c_2(t_0)]= \nonumber \\
 e^{-\frac{i}{\hbar}\int_{t_0}^{t} V(1,t)dt}(c_1(t_0)cos(\frac{E_p(t-t_0)}{\hbar})^2 - c_2(t_0) sin(\frac{E_p(t-t_0)}{\hbar})^2) + \nonumber \\
- i e^{-\frac{i}{\hbar}\int_{t_0}^{t} V(1,t)dt}(c_1(t_0)sin(\frac{E_p(t-t_0)}{\hbar})cos(\frac{|t|}{\hbar}(t-t_0))+c_2(t_0)cos(\frac{E_p(t-t_0)}{\hbar}t)sin(\frac{|t|}{\hbar}(t-t_0))) = \nonumber \\
=e^{-\frac{i}{\hbar}\int_{t_0}^{t} V(1,t)dt}(c_1(t_0) - (c_2(t_0)+c_1(t_0)) sin(\frac{E_p(t-t_0)}{\hbar})^2) + \nonumber \\
- i e^{-\frac{i}{\hbar}\int_{t_0}^{t} V(1,t)dt}c_1(t_0)(sin(\frac{E_p(t-t_0)}{\hbar})cos(\frac{|t|}{\hbar}(t-t_0))+cos(\frac{E_p(t-t_0)}{\hbar}t)sin(\frac{|t|}{\hbar}(t-t_0))) + \nonumber \\
+(-i)e^{-\frac{i}{\hbar}\int_{t_0}^{t} V(1,t)dt}(c_2(t_0)-c_1(t_0)) cos(\frac{E_p(t-t_0)}{\hbar}t)sin(\frac{|t|}{\hbar}(t-t_0)) = \nonumber \\
=e^{-\frac{i}{\hbar}\int_{t_0}^{t} V(1,t)dt}(c_1(t_0) - (c_2(t_0)+c_1(t_0)) sin(\frac{E_p(t-t_0)}{\hbar})^2)
- i e^{-\frac{i}{\hbar}\int_{t_0}^{t} V(1,t)dt}c_1(t_0)sin(\frac{(|t|+E_p)}{\hbar}(t-t_0)) + \nonumber \\
+\frac{-i}{2}e^{-\frac{i}{\hbar}\int_{t_0}^{t} V(1,t)dt}(c_2(t_0)-c_1(t_0))[sin(\frac{(E_p+|t|)(t-t_0)}{\hbar}t)+sin(\frac{(|t|-E_p)}{\hbar}(t-t_0))]. \nonumber
\end{eqnarray}
\end{strip}
In quite analogical way we deal with coefficient $c_2(t)$ whose analytic solution can be written as
\begin{eqnarray}
 c_2(t)  = %\nonumber \\
e^{-\frac{i}{\hbar}\int_{t_0}^{t} V(2,t)dt}e^{ -\frac{i}{\hbar}E_p(t-t_0)}(sin(\frac{|t|}{\hbar}(t-t_0))(-i c_1(t_0))\nonumber \\
+cos(\frac{|t|(t-t_0)}{\hbar}) c_2(t_0).
\end{eqnarray}

\subsection{Position qubit in resonant state and Rabi oscillations}

We can define the general form of energy evolution of 2 level qubit  preserving its two level states. It will have the form 
\begin{eqnarray}
\hat{H}=
\begin{pmatrix}
E_1(t) & E_{12}(t) \\
E_{12}(t)^{*} & E_2(t) 
\end{pmatrix}
\end{eqnarray}
and one obtains the following energy eigenstates $E_{1m}(t)=E_1(t)$ that are in first order approximation can be approximated by $E_{2m}(t)=\frac{E_{1}(t) + E_{2}(t)}{2} - \sqrt{(\frac{E_{1}(t)-E_{2}(t)}{2})^2 + |E_{12}(t)|^2}$ and $E_{1m}(t)=\frac{E_{1}(t) + E_{2}(t)}{2} + \sqrt{(\frac{E_{1}-E_{2}}{2})^2 + |E_{12}(t)|^2}$ and for small values of resonant field $|E_{12}|<<E_1(t), E_2(t)$ we can assume $E_1(t) \approx E_{1m}(t)$ and $E_2(t) \approx E_{2m}(t)$.
In general case one shall assume the following evolution matrix of qubit system 
\begin{eqnarray}
\hat{U}_{2,1}(t,t_0)= \nonumber \\
-\frac{ (\int_{t0}^{t}dt'E_{12}(t')^{*}dt') e^{-\frac{i \int_{t0}^{t}(E_1(t')+E_2(t'))dt'}{2 \hbar}}}{\sqrt{\left(|\int_{t0}^{t}E_{12}(t')dt'|^2\right)^2+( \int_{t0}^{t}dt'\frac{E_{1}(t')-E_{2}(t')}{2})^2} } \times \nonumber \\
\times   \sinh (\frac{i \sqrt{( \int_{t0}^{t}dt'\frac{E_{1}(t')-E_{2}(t')}{2})^2+ \left(|\int_{t0}^{t}E_{12}(t')dt'|^2\right)}}{\hbar})
   %%%{2\hbar}\right)}{\sqrt{-(E_{1}-E_{2})^2-4 \left(|E_{12}(t)|^2\right)}}
\end{eqnarray}
\begin{strip}
\begin{eqnarray}
\hat{U}_{2,2}(t,t_0)= %\nonumber \\
\frac{e^{\left(-\frac{i\sqrt{(\int_{t0}^{t}dt'\frac{E_1(t')+E_2(t')}{2})^2+\left(|\int_{t0}^{t}dt'E_{12}(t')dt'|^2\right)}+i \frac{1}{2}\int_{t0}^{t}(E_1(t')+E_2(t'))dt'
   }{\hbar}\right)}}{4\sqrt{(\int_{t0}^{t}dt'\frac{E_1(t')-E_2(t')}{2})^2+|\int_{t0}^{t}dt'E_{12}(t')dt'|^2}} \times \nonumber \\
\times \Bigg[ i\frac{\int_{t0}^{t}dt'(E_{p2}(t')-E_{p1}(t'))(1-e^{\left(-\frac{2i\sqrt{(\int_{t0}^{t}dt'\frac{E_1(t')-E_2(t')}{2})^2+\left(|\int_{t0}^{t}dt'E_{12}(t')dt'|^2\right)}+i \frac{1}{2}\int_{t0}^{t}(E_1(t')-E_2(t'))dt'
   }{\hbar}\right)})}{2}+ \nonumber \\+(1+e^{i\frac{2\sqrt{(\int_{t0}^{t}dt'\frac{E_1(t')-E_2(t')}{2})^2+\left(|\int_{t0}^{t}dt'E_{12}(t')dt'|^2\right)}}{\hbar}}) i\sqrt{(\int_{t0}^{t}dt'\frac{E_1(t')-E_2(t')}{2})^2+\left(|\int_{t0}^{t}dt'E_{12}(t')dt'|^2\right)} \Bigg], \nonumber \\
\end{eqnarray}
\begin{eqnarray}   
   \hat{U}_{1,1}(t,t_0)= %\nonumber \\
\frac{e^{\left(-\frac{i\sqrt{(\int_{t0}^{t}dt'\frac{E_1(t')-E_2(t')}{2})^2+\left(|\int_{t0}^{t}dt'E_{12}(t')dt'|^2\right)}+i \frac{1}{2}\int_{t0}^{t}(E_1(t')+E_2(t'))dt'
   }{\hbar}\right)}}{4\sqrt{(\int_{t0}^{t}dt'\frac{E_1(t')-E_2(t')}{2})^2+|\int_{t0}^{t}dt'E_{12}(t')dt'|^2}} \times \nonumber \\
\times \Bigg[i \Big[ (\int_{t0}^{t}dt'(E_{p1}(t')-E_{p2}(t'))) +2  \sqrt{(\int_{t0}^{t}dt'\frac{E_1(t')-E_2(t')}{2})^2+\left(|\int_{t0}^{t}dt'E_{12}(t')dt'|^2\right)}\Big]+\nonumber \\ 
+i\Big[(\int_{t0}^{t}dt'(E_{p1}(t')-E_{p2}(t'))) + \nonumber \\ 2\sqrt{(\int_{t0}^{t}dt'\frac{E_1(t')-E_2(t')}{2})^2+\left(|\int_{t0}^{t}dt'E_{12}(t')dt'|^2\right)}  \Big]e^{\left(-\frac{2i\sqrt{(\int_{t0}^{t}dt'\frac{E_1(t')-E_2(t')}{2})^2+\left(|\int_{t0}^{t}dt'E_{12}(t')dt'|^2\right)} }{\hbar}\right)}\Bigg]
%-e^{\left(-\frac{2i\sqrt{(\int_{t0}^{t}dt'\frac{E_1(t')-E_2(t')}{2})^2+\left(|\int_{t0}^{t}dt'E_{12}(t')dt'|^2\right)}+i }{\hbar}\right)})}{2}+ \nonumber \\
%+(1+e^{i\sqrt{(\int_{t0}^{t}dt'\frac{E_1(t')-E_2(t')}{2})^2+\left(|\int_{t0}^{t}dt'E_{12}(t')dt'|^2\right)}}) i\sqrt{(\int_{t0}^{t}dt'\frac{E_1(t')-E_2(t')}{2})^2+\left(|\int_{t0}^{t}dt'E_{12}(t')dt'|^2\right)} \Bigg]
\end{eqnarray} 
\end{strip}

%\newpage
%\onecolumn
%\onecolumn
\begin{strip}
Now we analyze the heating effects in time-dependent 2 level qubit Hamiltonian. We have
\newpage 
\begin{eqnarray}
\label{simplematrix}
%=\hat{H}(t)=
%\hat{H}(t)_{[x=(x_1,x_2)]}=
%\begin{split}
%%\hat{H}(t)_{[x=(x_1,x_2)]}=
%%\begin{pmatrix}
%%E_{p1}(t) & t_{s12}(t)=|t_{s12}|e^{+i\alpha(t)} \\
%%t_{s12}^{\dag}(t)=|t_{s12}|e^{-i\alpha(t)} & E_{p2}(t)
%%\end{pmatrix}=\nonumber \\
%%E_{p1}(t)\ket{x_1}\bra{x_1}+E_{p2}(t)\ket{x_2}\bra{x_2}+t_{s12}(t)\ket{x_1}\bra{x_2}+t_{s21}(t)\ket{x_2}\bra{x_1}
%%\nonumber \\
\hat{H}_{qubit}=(E_1(t)\ket{E_1}_t \bra{E_1}_t+E_2(t)\ket{E_2}_t\bra{E_2}_t+E_{12}(t)\ket{E_1}_t\bra{E_2}_t+E_{21}(t)\ket{E_2}_t\bra{E_1}_t)_{[E=(E_1,E_2)]}=\nonumber \\
(E_1(t)\ket{E_1}_t \bra{E_1}_t+E_2(t)\ket{E_2}_t\bra{E_2}_t+e^{\frac{\int_{t_0}^{t}(E_2(t')-E_1(t'))dt'}{-i\hbar}}E_{12}(t)\ket{E_1}_{t0}\bra{E_2}_{t0}+E_{21}(t)\ket{E_2}_{t0}\bra{E_1}_{t0})_{[E=(E_1,E_2)]}=\nonumber \\
=E_1(t)(\ket{x_1}a+\ket{x_2}b)(\bra{x_1}a^{*}+\bra{x_2}b^{*})+E_2(t)(\ket{x_1}c+\ket{x_2}d)(\bra{x_1}c^{*}+\bra{x_2}d^{*})+ \nonumber \\.
+[E_{12}(t)e^{i\frac{\int_{t_0}^{t}(E_2(t')-E_1(t'))dt'}{\hbar}}(a(t_0)\ket{x_1}+b(t_0)\ket{x_2})(c(t_0)^{*}\bra{x_1}+d(t_0)^{*}\bra{x_2})+\nonumber \\
+E_{21}(t)e^{-i\frac{\int_{t_0}^{t}(E_2(t')-E_1(t'))dt'}{\hbar}}(c(t_0)\ket{x_1}+d(t_0)\ket{x_2})(a(t_0)^{*}\bra{x_1}+b(t_0)^{*}\bra{x_2})]=\nonumber \\
=\ket{x_1}\bra{x_1}(|a|^2 E_1(t)+|c|^2 E_2(t)+(a(t_0)c(t_0)^{*}E_{12}(t)e^{i\frac{\int_{t_0}^{t}(E_2(t')-E_1(t'))dt'}{\hbar}}+a(t_0)^{*}c(t_0)E_{12}^{*}(t)e^{-i\frac{\int_{t_0}^{t}(E_2(t')-E_1(t'))dt'}{\hbar}}))+\nonumber \\
+\ket{x_2}\bra{x_2}(|b|^2 E_1(t)+|d|^2 E_2(t)+(a(t_0)c(t_0)^{*}E_{12}(t)e^{i\frac{\int_{t_0}^{t}(E_2(t')-E_1(t'))dt'}{\hbar}}+a(t_0)^{*}c(t_0)E_{12}^{*}(t)e^{i\frac{\int_{t_0}^{t}(E_2(t')-E_1(t'))dt'}{-\hbar}}))+\nonumber \\
+\ket{x_1}\bra{x_2}(ab^{*}E_1(t)+cd^{*}E_2(t)+E_{12}(t)e^{+i\frac{(E_2-E_1)(t-t_0)}{\hbar}}a(t_0)d(t_0)^{*})+\nonumber \\
+\ket{x_2}\bra{x_1}(a^{*}bE_1(t)+c^{*}dE_2(t)+E_{12}^{*}(t)e^{-i(E_2-E_1)(t-t_0)}a(t_0)^{*}d(t_0))=\nonumber \\
\ket{x_1}\bra{x_1}E_{p1_{eff}}(t)+\ket{x_2}\bra{x_2}E_{p2_{eff}}(t)+\ket{x_1}\bra{x_2}t_{s21_{eff}}(t)+\ket{x_2}\bra{x_1}t_{s12_{eff}}(t), \nonumber \\
E_{p1_{eff}}(t)=|a|^2 E_1(t)+|c|^2 E_2(t)+(a(t_0)c(t_0)^{*}E_{12}(t)e^{i(E_2-E_1)(t-t_0)}+a(t_0)^{*}c(t_0)E_{12}^{*}(t)e^{-i(E_2-E_1)(t-t_0)})=\nonumber \\
=|a|^2 E_1(t)+|c|^2 E_2(t)+\cos((E_2-E_1)(t-t_0))[(Re(a(t_0)c(t_0)^{*}E_{12}(t))+(Re(a(t_0)^{*}c(t_0)E_{12}(t)^{*}))]+\nonumber \\
+(-1)\sin((E_2-E_1)(t-t_0))[(Im(a(t_0)c(t_0)^{*}E_{12}(t))-(Im(a(t_0)^{*}c(t_0)E_{12}^{*}(t)))],\nonumber \\
E_{p2_{eff}}(t)=|b_t|^2 E_1(t)+|d_t|^2 E_2(t)+(b(t_0)d(t_0)^{*}E_{12}(t)e^{i(E_2-E_1)(t-t_0)}+b(t_0)^{*}d(t_0)E_{12}^{*}(t)e^{-i(E_2-E_1)(t-t_0)})=\nonumber \\
=|b_t|^2 E_1(t)+|d_t|^2 E_2(t)+\cos((E_2-E_1)(t-t_0))[(Re(b(t_0)d(t_0)^{*}E_{12}(t))+(Re(b(t_0)^{*}d(t_0)E_{12}(t)^{*}))]+\nonumber \\
+(-1)\sin((E_2-E_1)(t-t_0))[(Im(b(t_0)d(t_0)^{*}E_{12}(t))-(Im(b(t_0)^{*}d(t_0)E_{12}^{*}(t)))], \nonumber \\
t_{s21}(t)=(a(t)b^{*}(t)E_1(t)+c(t)d^{*}(t)E_2(t)+E_{12}(t)a(t)d(t)^{*})=(a(t)b(t)E_1(t)+c(t)d(t)E_2(t)+E_{12}(t)a(t)d(t)), \nonumber \\
t_{s12}(t)=(a(t)^{*}b(t)E_1(t)+c(t)^{*}d(t)E_2(t)+E_{12}(t)^{*}a(t)^{*}d(t)), (b,d, E_1(t),E_2(t))\in R.
\end{eqnarray}
\end{strip}
%\newpage

We observe that from knowledge of $t_{s21}(t)$ and $t_{s21}(t)$ we can extract the knowledge on $E_{12}(t)$ function. 
We have 
\begin{strip}
\begin{eqnarray}
(t_{s21}(t)+t_{s21}(t))\in R =(a(t)+a(t)^{*})b(t)E_1(t)+(c(t)+c(t)^{*})d(t)E_2(t)+(E_{12}(t)a(t)+E_{12}(t)^{*}a(t)^{*})d(t)= \nonumber \\
=2Re(a(t))b(t)E_1(t)+2Re(c(t))d(t)E_2(t)+2d(t)([Re(E_{12}(t))Re(a(t))-Im(E_{12}(t))Im(a(t))]),\nonumber \\
(t_{s21}(t)-t_{s21}(t))=2Im(a(t))b(t)E_1(t)+2Im(c(t))d(t)E_2(t)+2d(t)([Re(E_{12}(t))Im(a(t))+Im(E_{12}(t))Re(a(t))]).
 %+[Re(E_{12}(t))Re(a(t))-Im(E_{12}(t))Im(a(t))]).\nonumber \\
\end{eqnarray}
\end{strip}

Last equation implies 
%Let us assume that there exist only real value component of $E_{12}(t)$ so $E_{12}(t) \in R$. We have
\begin{strip}
\begin{eqnarray}
\frac{(t_{s21}(t)+t_{s21}(t))-2Re(a(t))b(t)E_1(t)-2Re(c(t))d(t)E_2(t)}{2d(t)}=([Re(E_{12}(t))Re(a(t))-Im(E_{12}(t))Im(a(t))]), \nonumber \\
\frac{(t_{s21}(t)-t_{s21}(t))-2Im(a(t))b(t)E_1(t)-2Im(c(t))d(t)E_2(t)}{2d(t)}=([Re(E_{12}(t))Im(a(t))+Im(E_{12}(t))Re(a(t))]). 
\end{eqnarray}
\end{strip}
and we finally can obtain the unique formula for  $Re(E_{12}(t)$ and $Im(E_{12}(t)$ functions that control the Rabi oscillations of switching the occupancy between $E_g$ and $E_g$ level.
\begin{strip}
\begin{eqnarray}
\frac{1}{2 Re(a(t))Im(a(t)}[Im(a(t))\frac{(t_{s21}(t)+t_{s21}(t))-2Re(a(t))b(t)E_1(t)-2Re(c(t))d(t)E_2(t)}{2d(t)}+ \nonumber \\
Re(a(t))\frac{(t_{s21}(t)-t_{s12}(t))-2Re(a(t))b(t)E_1(t)-2Re(c(t))d(t)E_2(t)}{2d(t)}] %\nonumber \\
=Re(E_{12}(t)), \nonumber \\
\frac{1}{2 Re(a(t))Im(a(t)}[-Re(a(t))\frac{(t_{s21}(t)+t_{s21}(t))-2Re(a(t))b(t)E_1(t)-2Re(c(t))d(t)E_2(t)}{2d(t)}+ \nonumber \\
Im(a(t))\frac{(t_{s21}(t)-t_{s12}(t))-2Re(a(t))b(t)E_1(t)-2Re(c(t))d(t)E_2(t)}{2d(t)}] %\nonumber \\
=Im(E_{12}(t)), \nonumber \\
%+\frac{t_{s21}(t)-2Re(a(t))b(t)E_1(t)-2Re(c(t))d(t)E_2(t)}{2d(t)}=Re(E_{12}(t))Re(a(t))Im(a(t))
\end{eqnarray}
\end{strip}
In particular we can observe that changing phase imprint by means of $t_{s12}(t)=|t_s|e^{\alpha(t)}=|t_s|\cos(\alpha(t))+i|t_s|\sin(\alpha(t))$ with $t_s=constant$ and non-constant $\alpha(t)$ what implies $t_{s21}(t)+t_{s12}(t)=2|t_s|\cos(\alpha(t))$, $t_{s21}(t)-t_{s12}(t)=-i 2|t_s|\sin(\alpha(t))$.
Therefore it is possible to obtain the heating or cooling down the quantum state only by change of phase imprint expressed by $\alpha(t)$.  
In particular case we when $E_1(t)$ and $E_2(t)$ are time independent we have Rabi oscillations with constant frequency since $(E_2-E_1)(t-t_0)=\int_{t_0}^{t}(E_2(t')-E_1(t'))dt'$.

Last Hamiltonian preserves Hermicity if $E_{12}(t)=E_{12,r}(t)+iE_{12,im}(t) =E_{21}(t)^{*},(E_{12,r}(t),E_{12,im}(t))\in R$. The term $E_{12}(t)\ket{E_1}\bra{E_2}$  describes the flow of energy from excited energy level into ground level (or lower energy level) what practically means that qubits system is cooling down.
In quite analogical way term  $E_{12}(t)\ket{E_1}\bra{E_2}$  describes the flow of energy from ground  energy level (or lower energy level) into excited energy level (or higher energy level) what practically means that qubits system is heating up.
 Let us investigate the case when $|E_1(t)|, |E_2(t)| >> |E_{12}(t)|^2, |E_{21}(t)|^2$ so Hamitonian term responsible for energy flow between considered energy levels is small.
\section{Quantum electrostatic Swap gate}
\begin{figure}
    \centering
    \includegraphics[width=3.0in]{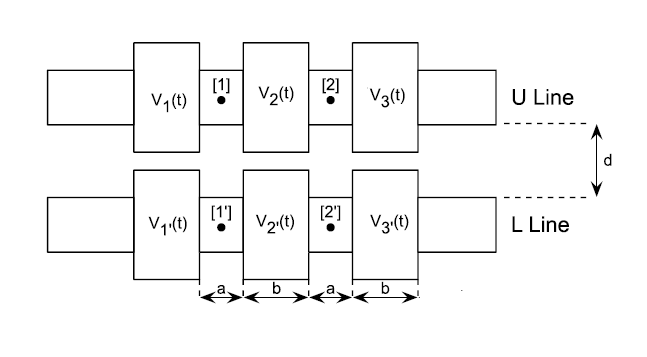}
    \caption{2 electrons confined in two separated in space double interconnected quantum dots interacting electrostatically.}
    \label{QSwap}
\end{figure}

\begin{figure}
\label{DqdotsOnLine}
\centering
\includegraphics[scale=0.4]{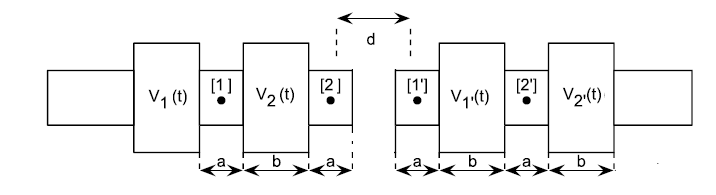}
\caption{Case of 2 double quantum dots on the line.}
\end{figure}
We consider the situation as depicted in Fig. \ref{QSwap}. We have 2 separated in space systems of double quantum dots U (Upper) and L (Lower). One electron is in L system and one electron is in U system.
If U system is far away from L system than the quantum states of U is independent from L quantum system and in
such case we can write $|\psi>_U=c_1|1,0>_U+c_2|0,1>_U$ and $|\psi>_L=c_3|1,0>_L+c_4|0,1>_L$ and normalization conditions $|c_1|^2+|c_2|^2=1$ and
$|c_3|^2+|c_4|^2=1$. In case of separated systems we can write the total Hilbert space by factorization so
$|\psi>=|\psi>_U|\psi>_L$. However it is not true when we bring L and U systems sufficiently close so Coulomb interaction has no longer perturbative character. In such case we have the most general form of quantum state given as
\begin{eqnarray}
\ket{\psi}=c_a|0,1>_U|0,1>_L+c_b|0,1>_U|1,0>_L+ \nonumber \\
c_c|1,0>_U|0,1>_L+c_d|1,0>_U|1,0>_U.
\end{eqnarray}
Since we have 2 electrons we can write $|c_a|^2+|c_b|^2+|c_c|^2+|c_d|^2=1$.
In rare and special case we can write
\begin{eqnarray}
|\psi>=(c_{1_U}|0,1>_U+c_{2_U}|1,0>_U) \times \nonumber \\
\times (c_{1_L}|0,1>_L+c_{2_L}|1,0>_L)=\nonumber \\
=c_{1_U}c_{1_L}|0,1>_U|0,1>_L+c_{1_U}c_{2_L}|0,1>_U|1,0>_L+ \nonumber \\
c_{2_U}c_{1_L}|1,0>_U|0,1>_L+c_{2_U}c_{2_L}|1,0>_U|1,0>_L. \nonumber \\
\end{eqnarray}
If $c_a=c_{1_U}c_{1_L}$,$c_b=c_{1_U}c_{2_L}$, $c_c=c_{2_U}c_{1_L}$ and $c_d=c_{2_U}c_{2_L}$ and we are dealing with non-entangled state since total quantum state can be factorized
as product of 2 quantum subsystems. Such case is rare and in most cases our system depicted in Fig.1 is entangled what especially takes place if U and L are not far away.

Now we need to write down the system Hamiltonian from Fig.1. One electron from U can be in points 1 and 2 and second electron can be in points 1' and 2'.
We make assumptions that U and L have the same physical structure and are symmetric.

We have total system Hamiltonian as the sum of all potential and kinetic energies given as
\begin{eqnarray}
H=(E_p(1)|1,0>_U<1,0|_U+E_p(2)|0,1>_U<0,1|_U)I_L + \nonumber \\
I_U(E_p(1')|1,0>_L<1,0|_L)+\nonumber \\ +E_p(2')|0,1>_L<0,1|_L) %\nonumber \\
+t_U(|1,0>_U<0,1|_U \nonumber \\ +|0,1>_U<1,0|_U)I_L+\nonumber \\
 I_U t_L(|1,0>_L<0,1|_L  +|0,1>_L<1,0|_L)  \nonumber \\
+E_c(1,1')(|1,0>_U|1,0>_L <1,0|_U<1,0|_L)+  \nonumber \\ +  E_c(2,2')|0,1>_U|0,1>_L<0,1|_U<0,1|_L + \nonumber \\ + E_c(2,1')(|0,1>_U|1,0>_L<0,1|_U<1,0|_L)  \nonumber \\
+E_c(1,2')|1,0>_U|0,1>_L <1,0|_U <0,1|_L.
\end{eqnarray}
The last 4 terms are Coulomb interaction terms between one electron confined in L system and one electron confined in U system.
The expressions $ E_c(1,1'), E_c(2,2'), E_c(2,1'), E_c(1,2')$ have values Coulomb classical energy between points (1,1'),(2,2'),(2,1'),(1,2').
Two double quantum dots are geometrically parametrized by constants $d1,d,a, b$ and thus we have distances $d_{1,1'}=d_1=d_{2,2'}$ and $d_{1,2'}=d_{1',2}=\sqrt{d_1^2+(b+a)^2}$. Therefore $E_c(1,1')=\frac{q^2}{d_1}=E_c(2,2')$ and $E_c(1,2')=\frac{q^2}{\sqrt{(d_1)^2+(b+a)^2}}=E_c(2,1')$.
The electron kinetic energy in U system is parametrized by $t_U$ and in L system by $t_L$. In simplified case we have $t_U=t_L=t$.
We also denoted
$I_L=(|0,1><0,1|_L+|1,0><1,0|_L)$ and $I_U=(|0,1><0,1|_U+|1,0><1,0|_U)$. $I_L$ is identity operator and is projection of L state on itself. The same is with $I_U$ operator that is identity operator and is projection of state U on itself. It is convenient to express total system Hamiltonian in matrix representation. We have 4 by 4 matrix given as
\begin{strip}
\begin{equation}
H=
\begin{pmatrix}
E_{p2}+E_{p2'}+E_c(2,2') & t_{L} & t_U & 0 \\
t_{L} & E_{p2}+E_{p1'}+E_c(2,1') & 0 & t_U \\
t_U & 0 & E_{p1}+E_{p2'}+E_c(1,2') & t_L \\
0 & t_U & t_L & E_{p1}+E_{p1'}+E_c(1,1') \\
\end{pmatrix}
\end{equation}
\end{strip}
\normalsize
Now we need to find system 4 eigenvalues and eigenstates(4 orthogonal 4-dimensional vectors)  so we are dealing with matrix eigenvalue problem) what is the subject of classical algebra. Let us assume that 2 double quantum dot systems are symmetric and biased by the same voltages generating potential bottoms $V_s$ so we have $Ep(1)=Ep(2)=Ep(1')=Ep(2')=V_s$ and that $t_L=t_U=t_s$. Denoting $E_c(1,1')=E_c(2,2')=Ec1s$ and $E_c(1,2')=E_c(2,1')=Ec2s$  we are obtaining 4 orthogonal eigenvectors
\begin{eqnarray}
V_1=
\begin{pmatrix}
-1 \\
0 \\
0 \\
1
\end{pmatrix}
, V_2=
\begin{pmatrix}
0 \\
-1 \\
1 \\
0
\end{pmatrix}
, \nonumber \\
V_{3(4)}=
\begin{pmatrix}
1 \\
\mp\frac{4 t_s}{\pm(-E_{c1s} + E_{c2s}) + \sqrt{(E_{c1s} - E_{c2s})^2 + 16 t_s^2}} \\
\mp\frac{4 t_s}{\pm(-E_{c1s} + E_{c2s}) + \sqrt{(E_{c1s} - E_{c2s})^2 + 16 t_s^2}} \\
1
\end{pmatrix}
\end{eqnarray}
corresponding to 4 eigenenergies
\begin{eqnarray}
E_1=E_{c1s} + 2 V_s, E_2=E_{c2s} + 2 V_s, E_1 > E_2 \nonumber \\
E_3= \frac{1}{2} ( (E_{c1s} + E_{c2s}) - \sqrt{(E_{c1s} -E_{c2s})^2 + 16 t_s^2} + 4 V_s)  \nonumber \\
E_4 = \frac{1}{2} ( (E_{c1s} + E_{c2s})+ \sqrt{(E_{c1s} -E_{c2s})^2 + 16 t_s^2} + 4 V_s), E_4 > E_3 .
\end{eqnarray}
We observe that setting quantum state to vector $V_1$ we obtain the state
\begin{equation}
|\psi>=-|0,1>_U|0,1>_L+|1,0>_U|1,0>_L
\end{equation}
and such state is indeed entangled.
Setting the quantum state to vector $V_2$ we also obtain the entangled state
\begin{equation}
|\psi>=-|0,1>_U|1,0>_L+|1,0>_U|0,1>_L.
\end{equation}
We also notice that the state $V_3$ and $V_4$ does not have its classical counterpart since upper electron exists at both positions 1 and 2 and lower electron exists at both positions at the same time. We observe that when distance between two systems of double quantum dots goes into infinity the energy difference between quantum state corresponding to $V_1$ and $V_2$ goes to zero. This makes those two entangled states to be degenerate.
We notice that vectors
$
\begin{pmatrix}
1 \\
0 \\
0 \\
0
\end{pmatrix},
\begin{pmatrix}
0 \\
1 \\
0 \\
0
\end{pmatrix},
\begin{pmatrix}
0 \\
0 \\
1 \\
0
\end{pmatrix},
\begin{pmatrix}
0 \\
0 \\
0 \\
1
\end{pmatrix}
$
correspond to states \\  $(|0,1>_U|0,1>_L),(|0,1>_U|1,0>_L),(|1,0>_U|0,1>_L),(|1,0>_U|1,0>_U)$.

We observe that $|v_1|^2=2$, $|v_2|^2=2$
and that

\begin{equation}
|v_3|^2= \frac{(E_{c2s}-E_{c1s}) + 8 t_s - \sqrt{ (E_{c1s} - E_{c2s})^2 + 16 t_s^2}}{4 t_s}
\end{equation}

with
\begin{equation}
|v_4|^2= \frac{(E_{c1s} - E_{c2s}) + 8 t_s - \sqrt{ (E_{c1s} - E_{c2s})^2 + 16 t_s^2}}{4 t_s}.
\end{equation}

Thus normalized 4 eigenvectors are of the following form

\begin{equation}
V_{1n}=\frac{1}{\sqrt{2}}
\begin{pmatrix}
-1 \\
0 \\
0 \\
1
\end{pmatrix}
, V_{2n}=\frac{1}{\sqrt{2}}
\begin{pmatrix}
0 \\
-1 \\
1 \\
0
\end{pmatrix}
\end{equation}
\begin{eqnarray}
, V_{3n}=\sqrt{\frac{4t_s}{(E_{c2s}-E_{c1s}) + 8 t_s - \sqrt{ (E_{c1s} - E_{c2s})^2 + 16 t_s^2}}} \times \nonumber \\
\times
\begin{pmatrix}
1 \\
-\frac{4 t_s}{(-E_{c1s} + E_{c2s}) + \sqrt{(E_{c1s} - E_{c2s})^2 + 16 t_s^2}} \\
-\frac{4 t_s}{(-E_{c1s} + E_{c2s}) + \sqrt{(E_{c1s} - E_{c2s})^2 + 16 t_s^2}} \\
1
\end{pmatrix},
\end{eqnarray}

\begin{eqnarray}
V_{4n}=\sqrt{\frac{4t_s}{(E_{c1s}-E_{c2s}) + 8 t_s - \sqrt{ (E_{c1s} - E_{c2s})^2 + 16 t_s^2}}} \times \nonumber \\
\begin{pmatrix}
1 \\
\frac{4 t_s}{(E_{c1s} - E_{c2s}) + \sqrt{(E_{c1s} - E_{c2s})^2 + 16 t_s^2}} \\
\frac{4 t_s}{(E_{c1s} - E_{c2s}) + \sqrt{(E_{c1s} - E_{c2s})^2 + 16 t_s^2}} \\
1
\end{pmatrix}
\end{eqnarray}
%\newpage 

\subsubsection{Case of 2 double quantum dots on the line}
We consider the situation as depicted in Fig.\ref{DqdotsOnLine}.

We have the following Coulomb interaction terms $E_C(2,1')=q^2/d, E_C(2,1')=q^2/(d+2b+2a), E_C(2,2')=q^2/(d+b+a)=E_C(1,1')$.
We assume $t_L=t_R=|t|$. All energies at the nodes 1,2, 1' and 2' are controlled with biasing voltage $V_s$. We obtain the following Hamiltonian of the system
\begin{eqnarray}
H=(E_p(1)|1,0>_L<1,0|_L+E_p(2)|0,1>_L<0,1|_L))I_R +  \nonumber \\ I_L(E_p(1')|1,0>_R<1,0|_R)+
t_{1 \rightarrow 2}(|0,1>_L<1,0|_L)I_R + \nonumber \\ + t_{2 \rightarrow 1}(|1,0>_L<0,1|_L)I_R + \nonumber \\ + E_c(1,1')(|1,0>_L|1,0>_R<1,0|_L<1,0|_R)\nonumber \\
+E_c(2,2')(|0,1>_L|0,1>_R <0,1|_L<0,1|_R) + \nonumber \\
+ E_c(1,2')(|1,0>_L|0,1>_R <1,0|_L<0,1|_R)\nonumber \\ +  E_c(2,1')(|0,1>_L|1,0>_R <0,1|_L<1,0|_R) .
%+\nonumber \\ +E_p(2')|0,1>_L<0,1|_L) %\nonumber \\
%+t_U(|1,0>_U<0,1|_U+|0,1>_U<1,0|_U)I_L+ I_U t_L(|1,0>_L<0,1|_L +  \nonumber \\ +|0,1>_L<1,0|_L)
%+E_c(1,1')(|1,0>_U|1,0>_L <1,0|_U<1,0|_L)+  \nonumber \\ +  E_c(2,2')|0,1>_U|0,1>_L<0,1|_U<0,1|_L + \nonumber \\ + E_c(2,1')(|0,1>_U|1,0>_L<0,1|_U<1,0|_L)  +E_c(1,2')|1,0>_U|0,1>_L <1,0|_U <0,1|_L.
\end{eqnarray}
The quantum state of the system can be written as
\begin{eqnarray}
|\psi>=\nonumber \\
=c_a |0,1>_L|0,1>_R+ c_ b |0,1>_L|1,0>_R+ \nonumber \\
 c_c |1,0>_L|0,1>_R + c_d |1,0>_L |1,0>_R.
\end{eqnarray}
The coefficients $|c_a|^2+|c_b|^2+|c_c|^2+|c_d|^2=1$ since \\ $<\psi|\psi>=1$.
We assume that

\begin{eqnarray}
\ket{0,1}_L\ket{0,1}_R =
\begin{pmatrix}
1 \\
0 \\
0 \\
0
\end{pmatrix}
, \ket{0,1}_L\ket{1,0}_R =
\begin{pmatrix}
0 \\
1 \\
0 \\
0
\end{pmatrix}, \nonumber \\
%\end{equation}
%\begin{equation}
 \ket{1,0}_L\ket{0,1}_R =
\begin{pmatrix}
0 \\
0 \\
1 \\
0
\end{pmatrix},
 \ket{1,0}_L\ket{1,0}_R =
\begin{pmatrix}
0 \\
0 \\
0 \\
1
\end{pmatrix} .
\end{eqnarray}

In such case the system Hamiltonian is given as
%\begin{strip}
%\begin{equation}
%H=
%\begin{pmatrix}
%E_p(2) + E_p(2') + \textcolor{red}{E_c(2,2')}  & t_R(2'\rightarrow 1') & t_L(2 \rightarrow 1) & 0 \\
%t_R(1'\rightarrow 2') & E_p(2) + E_p(1') + E_c(2,1') & 0 & t_L(2 \rightarrow 1) \\
%t_L(1 \rightarrow 2 ) & 0 & E_p(1) + E_p(2') + E_c(1,2') & t_R(2' \rightarrow 1') \\
%0 & t_L(1 \rightarrow 2) & t_R(1' \rightarrow 2') & E_p(1) + E_p(1')+ \textcolor{red}{E_c(1,1')} \\
%\end{pmatrix} .
%\end{equation}
%\end{strip}
%\normalsize
We notice that $E_c(2,2')=E_c(1,1')$ since two double symmetric qubits are on the same line.
Now we place the dependence of Coulomb energy on geometry. 
%\onecolumn
We obtain
%\tiny
\begin{strip}
\begin{eqnarray}
H=
\begin{pmatrix}
E_p(2) + E_p(2') + \textcolor{red}{E_c(2,2')}  & t_R(2'\rightarrow 1') & t_L(2 \rightarrow 1) & 0 \\
t_R(1'\rightarrow 2') & E_p(2) + E_p(1') + E_c(2,1') & 0 & t_L(2 \rightarrow 1) \\
t_L(1 \rightarrow 2 ) & 0 & E_p(1) + E_p(2') + E_c(1,2') & t_R(2' \rightarrow 1') \\
0 & t_L(1 \rightarrow 2) & t_R(1' \rightarrow 2') & E_p(1) + E_p(1')+ \textcolor{red}{E_c(1,1')} \\
\end{pmatrix} = \nonumber \\
\begin{pmatrix}
E_p(2) + E_p(2') + \frac{q^2}{d+b+a}  & t_R(2'\rightarrow 1') & t_L(2 \rightarrow 1) & 0 \\
t_R(1'\rightarrow 2') & E_p(2) + E_p(1') + \frac{q^2}{d} & 0 & t_L(2 \rightarrow 1) \\
t_L(1 \rightarrow 2 ) & 0 & E_p(1) + E_p(2') + \frac{q^2}{d+2(b+a)} & t_R(2' \rightarrow 1') \\
0 & t_L(1 \rightarrow 2) & t_R(1' \rightarrow 2') & E_p(1) + E_p(1')+ \frac{q^2}{d+b+a} \\
\end{pmatrix}=\nonumber 
\end{eqnarray}
\end{strip}
\begin{strip}
\begin{eqnarray}
\begin{pmatrix}
2 V_s + \frac{q^2}{d+b+a}  & t & t & 0 \\
t & 2 V_s + \frac{q^2}{d} & 0 & t \\
t & 0 & 2 V_s + \frac{q^2}{d+2(b+a)} & t \\
0 & t & t & 2 V_s + \frac{q^2}{d+b+a} \\
\end{pmatrix}  .
\end{eqnarray}
\end{strip}
\normalsize
where we set all hoping coefficients to |t| and we set $E_p(1)=E_p(2)=E_p(1')=E_p(2')=V_s$. 

\normalsize
%\newpage
\subsubsection{Case of 2 perpendicular double quantum dots}

It is important to consider the situation as depicted in Fig.1A.

\begin{figure}
\centering
\includegraphics[scale=0.4]{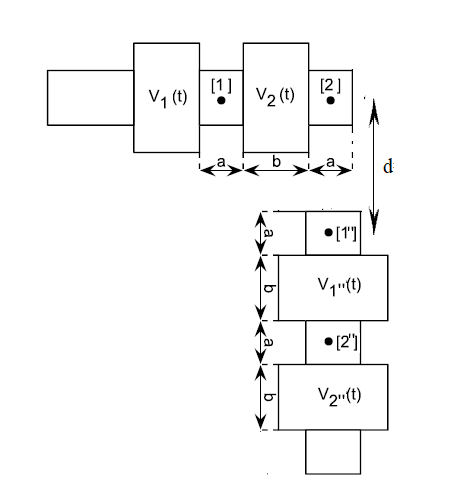} %%%{TwoDoubleQdots.png}
\caption{Case of 2 interacting double quantum dots}
\end{figure}
In highest simplified case we have
\begin{strip}
\begin{equation}
H=
\begin{pmatrix}
2V_s+\frac{q^2}{\sqrt{(d_1+a+\frac{3}{2}b)^2+(d_2)^2}} & t & t & 0 \\
t & 2V_s+\frac{q^2}{\sqrt{(d_1+\frac{1}{2}b)^2+(d_2)^2}}  & 0 & t \\
t & 0 & 2V_s+ \frac{q^2}{\sqrt{(d_1+\frac{3}{2}b+a)^2+(d_2+b+a)^2}} & t \\
0 & t & t & 2V_s+\frac{q^2}{\sqrt{(d_1+\frac{1}{2}b)^2+(d_2+b+a)^2}} \\
\end{pmatrix}
\end{equation}
\end{strip}
\normalsize

\subsection{Quantum dynamics with time for Q-SWAP gate}
\normalsize
\subsection{Extraction of 1-body quantum state from 2-body quantum state}
Now we need to be able to extract one body wavefunction from 2-body wavefunction. Let us extract the wavefunction for U system.
%We need to apply the following operators $|1,0>_U(<1,0|_U*I_L)$ , $|0,1>_U(<0,1|_U*I_L)$.
Thus we need to apply the following projection operators $P1_U$ and $P2_U$ on general quantum state $(U,L)$ to obtain only U wavefunction subcomponents:
\begin{equation}
P1_U=\frac{1}{\sqrt{2}}|1,0>_U((<1,0|_U<1,0|_L)+(<1,0|_U<0,1|_L))
\end{equation}
and
\begin{equation}
P2_U=\frac{1}{\sqrt{2}}|0,1>_U((<0,1|_U<1,0|_L)+(<0,1|_U<0,1|_L)).
\end{equation}
Consequently the whole U 1-body wavefunction is given as follows
\begin{eqnarray}
|\psi>_U=(P1_U+P2_U)|\psi>=\nonumber \\
(\frac{1}{\sqrt{2}}|1,0>_U((<1,0|_U<1,0|_L)+(<1,0|_U<0,1|_L))+ \nonumber \\
\frac{1}{\sqrt{2}}|0,1>_U((<0,1|_U<1,0|_L)+(<0,1|_U<0,1|_L)))|\psi>.
\end{eqnarray}
In quite analogical way we can introduce the following projection operators $P1_L$ and $P2_L$ on general quantum state $(U,L)$ to obtain only L wavefunction subcomponents:
\begin{equation}
P1_L=\frac{1}{\sqrt{2}}|1,0>_L((<1,0|_U<1,0|_L)+(<0,1|_U<1,0|_L))
\end{equation}
and
\begin{eqnarray}
P2_L=\frac{1}{\sqrt{2}}|0,1>_L((<1,0|_U<0,1|_L)+ \nonumber \\
+(<0,1|_U<0,1|_L)).
\end{eqnarray}
Consequently the whole L: 1-body wavefunction is given as follows
\begin{eqnarray}
|\psi>_L=(P1_L+P2_L)|\psi>= \nonumber \\
(\frac{1}{\sqrt{2}}|1,0>_L((<1,0|_U<1,0|_L)+(<0,1|_U<1,0|_L))+ \nonumber \\
\frac{1}{\sqrt{2}}|0,1>_L((<1,0|_U<0,1|_L)+\nonumber \\
(<0,1|_U<0,1|_L))|\psi>.
\end{eqnarray}
%\newpage
\subsection{The action of strong measurement on one of the subsystems L and U}
Making the measurement determining the position of particle from U system on the left side is represented by the projection
\begin{eqnarray}
PL_U=\frac{1}{\sqrt{2}}(|1,0>_U(|1,0>_L + |0,1>_L)) \nonumber \\
(<1,0|_{U}(<1,0|_{L}+<0,1|_{L})).
\end{eqnarray}
Thus after the determination of the state of electron in U to be on the left side we have the total quantum state after measurement $|\psi'>$ expressed by the state before measurement $|\psi>$ to be of the form
\begin{eqnarray}
|\psi'>=PL_U |\psi> = \nonumber \\
=\frac{1}{\sqrt{2}}(|1,0>_U(|1,0>_L + |0,1>_L)) \nonumber \\
(<1,0|_{U}(<1,0|_{L}+<0,1|_{L}))|\psi>.
\end{eqnarray}
Making the measurement determining the position of particle from U system on the right side is represented by the projection
\begin{eqnarray}
PR_U=\frac{1}{\sqrt{2}}(|0,1>_U(|1,0>_L + |0,1>_L)) \nonumber \\
(<0,1|_{U}(<1,0|_{L}+<0,1|_{L})).
\end{eqnarray}
Thus after the determination of the state of electron in U to be on the left side we have the total quantum state after measurement $|\psi'>$ expressed by the state before measurement $|\psi>$ to be of the form
\begin{eqnarray}
|\psi'>=PR_U |\psi> = \nonumber \\
=\frac{1}{\sqrt{2}}(|0,1>_U(|1,0>_L + |0,1>_L))\nonumber \\
(<0,1|_{U}(<1,0|_{L}+<0,1|_{L}))|\psi>.
\end{eqnarray}
Quite obviously instead of measurement of position of electron from U system we can make the determination of electron state from the L system to be on the left what we obtain with the projection operator $PL_U$ that is represented as following
\begin{equation}
PL_L=\frac{1}{\sqrt{2}}(|1,0>_U+ |0,1>_U)|1,0>_L(|1,0>_U+|0,1>_U)<1,0|_L.
\end{equation}
In similar way we can introduce the projection measurement determining the position of electron to be on the right side of system L in the way as
\begin{eqnarray}
PR_L=\frac{1}{\sqrt{2}}(|1,0>_U+ |0,1>_U)\times \nonumber \\
|0,1>_L(|1,0>_U+|0,1>_U)<0,1|_L.
\end{eqnarray}
In similar way as before the state of the quantum system after determination of particle position in L subsystem to be on the left side is $PL_L|\psi>$ while the state of the total quantum system after measurement determination of particle position in L subsystem to be on the right side is $PR_L|\psi>$.

\subsection{Combined approach of tight-binding + integro-differential equations for Q-CNOT gate}
The effect of 1-st quantum system from Fig.4 on the second quantum system (1-qubit:double Q-Dot system) can be accounted by the following Hamiltonian.
We assume that $d_3>d_2$.
The Hamiltonian of 2-nd quantum system that is CNOT output in its functional dependence from Hamiltonian of 1-st quantum system [Q-SWAP gate]. 2-nd quantum system Hamiltonian has the matrix
%\tiny
%\begin{eqnarray*}
%H_{2nd}[H_{1nd},|\psi>_{1nd}]=H_{2,non-interaction}+H_{1 \rightarrow 2}= \nonumber \\
%\begin{pmatrix}
%Ep_{1''}+E_c(1\rightarrow 1'')+E_c(2\rightarrow 1'')+E_c(1'\rightarrow 1'')+E_c(2'\rightarrow 1'') & t2 \\
%t2 & Ep_{2''}+E_c(1\rightarrow 2'')+E_c(2\rightarrow 2'')+E_c(1'\rightarrow 2'')+E_c(2'\rightarrow 2'')
%\end{pmatrix}=
%\begin{pmatrix}
%V_{s2} & t2 \\
%t2 & V_{s2}
%\end{pmatrix}
%\nonumber \\
%+ \nonumber
%\end{eqnarray*}
%\begin{eqnarray}
%\begin{pmatrix}
%\frac{q^2(2d_3+b+a)}{(d_3+b+a)d_3}+\frac{q^2}{\sqrt((d_{32})^2+(\frac{2d_1+b}{2})^2)}+\frac{q^2}{\sqrt((d_{32})^2+(d_1+\frac{2a+3b}{2})^2)} & 0 \\
%0 & \frac{q^2(2d_3+3(b+a))}{(d_3+2(b+a))(d_3+b+a)}+\frac{q^2}{\sqrt{|d_{32}|^2+|\frac{b+2d_1}{2}|^2}}+\frac{q^2}{\sqrt{|d_{32}+a+b|^2+(\frac{3b+2a}{2}+d_1)^2}}. %E_c(1'\rightarrow 2'')+  %+E_c(2'\rightarrow 2'')
%\end{pmatrix}
%\end{eqnarray}
%\begin{eqnarray}
%\end{eqnarray}
%=
\begin{strip}
\begin{eqnarray*}
H_{2,non-interaction}+H_{1 \rightarrow 2}[1] \nonumber \\
+H_{1 \rightarrow 2}[2]=
\begin{pmatrix}
V_{s2} & t_2 \\
t_2 & V_{s2}
\end{pmatrix}
\nonumber \\
+
\begin{pmatrix}
\frac{q^2|c_1(t)|^2}{d_3+b+a}+\frac{q^2|c_2(t)|^2}{d_3} & 0 \\
0 & \frac{q^2|c_1(t)|^2}{d_3+2(b+a)}+\frac{q^2|c_2(t)|^2}{d_3+b+a}
\end{pmatrix}
+ \nonumber \\
+
\begin{pmatrix}
+\frac{q^2|c_{1'}(t)|^2}{\sqrt((d_{32})^2+(\frac{2d_1+b}{2})^2)}+\frac{q^2|c_{2'}(t)|^2}{\sqrt((d_{32})^2+(d_1+\frac{2a+3b}{2})^2)} & 0 \\
0 & +\frac{q^2|c_{1'}(t)|^2}{\sqrt{|d_{32}|^2+|\frac{b+2d_1}{2}|^2}}+\frac{q^2|c_{2'}(t)|^2}{\sqrt{|d_{32}+a+b|^2+(\frac{3b+2a}{2}+d_1)^2}}.
\end{pmatrix}
= \nonumber \\
\begin{pmatrix}
Ep_{1''}+E_c(1\rightarrow 1'')+E_c(2\rightarrow 1'')+E_c(1'\rightarrow 1'')+E_c(2'\rightarrow 1'') & t2 \\
t2 & Ep_{2''}+E_c(1\rightarrow 2'')+E_c(2\rightarrow 2'')+E_c(1'\rightarrow 2'')+E_c(2'\rightarrow 2'')
\end{pmatrix}
\end{eqnarray*}
\end{strip}
\normalsize
representation, where $d_{32}=d_3-d_2$) and $|c_1|^2,|c_2|^2,|c_{1'}|^2,|c_{2'}|^2$ %|c_{1''}|^2,|c_{2''}|^2$
are the probabilities of occupancies of nodes 1,1',2,2' by electrons. Thus we need to know dynamics of system 1 to determine dynamics of system 2.
We need to use formulas $\ref{occupancy1}$ and $\ref{occupancy2}$.
The terms $H_{1 \rightarrow 2}[1], H_{1 \rightarrow 2}[2]$ can be treated as the perturbation to $H_{2,non-interaction}$. Analyzing more precisely we can introduce pertubation to the system 1 coming from system 2. This perturbation of system 1 from system 2 is later affecting the dynamics of system 2 as well. However in the first level of approximation we can recognize that only system 1 is affecting system 2 and that system 2 is no-having impact on the system 1.

\section*{Propagator in tight binding model }

Without presence of microwave field we have the Hamiltonian H given as
\begin{eqnarray}
H=E_p(1)|1,0><1,0|+ \nonumber \\
E_p(2)|0,1><0,1|+|t|_{2 \rightarrow 1}|1,0><0,1|+ \nonumber \\ +|t|_{2 \rightarrow 1}|0,1><1,0| \nonumber \\
+ f_1(t)|g><e|+ f_2(t)|e><g|.
\end{eqnarray}
Here $f_1(t)$ and $f_2(t)$ are time-dependent signals and $(|1,0>=w_L(x),|0,1>=w_R(x))$ are position based functions (Wannier functions), while $H|e>=E_2|e>$ and $H|g>=E_1|g>$. There is continous unitary transformation from $(|1,0>,|0,1>)$ to $(|g>,|e>)$ bases.
Spectral representation of system Hamiltonian with no external microwave field is given as
\begin{equation}
H=
\begin{pmatrix}
E_p(1) & |t|_{2 \rightarrow 1} \\
|t|_{1 \rightarrow 2}  & E_p(2)
\end{pmatrix}.
\end{equation}
and for the symmetric case we have
\begin{equation}
H=
\begin{pmatrix}
E_p & |t| \\
|t|  & E_p
\end{pmatrix}.
\end{equation}
The system eigenergies are given as $(E_p - \epsilon)^2-|t|^2=0$ what brings
$\epsilon_1=E_p-|t|$ and $\epsilon_2=E_p+|t|$. From the condition
\begin{equation}
\begin{pmatrix}
E_p & |t| \\
|t| & E_p
\end{pmatrix}
\begin{pmatrix}
a \\
b
\end{pmatrix}
=
\epsilon
\begin{pmatrix}
a \\
b
\end{pmatrix}
\end{equation}
We have
$\frac{b}{a}=\frac{\epsilon - E_p}{t}$ that gives either -1 or 1 for $\epsilon_1$ and $\epsilon_2$.

We have the following eigenvectors
\begin{equation}
|g>=\frac{1}{\sqrt{2}}(|1,0>-|0,1>)=\frac{1}{\sqrt{2}}(
\begin{pmatrix}
1 \\
0
\end{pmatrix}
+
\begin{pmatrix}
0 \\
-1
\end{pmatrix})
=\frac{1}{\sqrt{2}}
\begin{pmatrix}
1 \\
-1
\end{pmatrix}
.
\end{equation}
and
\begin{equation}
|e>=\frac{1}{\sqrt{2}}(|1,0>+|0,1>)= \frac{1}{\sqrt{2}}
(
\begin{pmatrix}
1 \\
1
\end{pmatrix})
\end{equation}
and such that $H\ket{g}=\epsilon_1\ket{g}$ and $H\ket{e}=\epsilon_2\ket{e}$, where $|g>$ denotes the quantum system energy ground eigenstate and $|e>$ denotes quantum system excited state.

Zero tunneling case between 1 and 2 or 2 and 1 is by presence of infinite barrier between 1 and 2.
In such case all kinetic and potential energy components are encoded in $E_{p}(1)$ and in $E_{p}(2)$ Hamiltonian components as components describing two insulated quantum systems.

Writing quantum state as
\begin{equation}
|\psi>=c_g|g> +c_e|e>.
\end{equation}
with eigenvector of energies

\begin{equation}
H_{(|g>,|e>)}=
\begin{pmatrix}
E_g & 0 \\
0 & E_e
\end{pmatrix}.
\end{equation}
We have
\begin{eqnarray}
H_{(|g>,|e>)}
\begin{pmatrix}
|g> \\
|e>
\end{pmatrix}
= \nonumber \\
\begin{pmatrix}
E_g & 0 \\
0 & E_e
\end{pmatrix}.
\begin{pmatrix}
|g> \\
|e>
\end{pmatrix}
=\nonumber \\
\begin{pmatrix}
E_g & 0 \\
0 & E_e
\end{pmatrix}
\frac{1}{\sqrt{2}}
\begin{pmatrix}
-1 & 1 \\
1 & 1
\end{pmatrix}
\begin{pmatrix}
|0,1>\\
|1,0>
\end{pmatrix}
\end{eqnarray}
Last expression can be written as
\begin{equation}
\begin{pmatrix}
|g> \\
|e>
\end{pmatrix}
=\frac{1}{\sqrt{2}}
\begin{pmatrix}
-1 & 1 \\
1 & 1
\end{pmatrix}
\begin{pmatrix}
|0,1>\\
|1,0>
\end{pmatrix}
\end{equation}
or equivalently
\begin{equation}
\begin{pmatrix}
|e> \\
|g>
\end{pmatrix}
=\frac{1}{\sqrt{2}}
\begin{pmatrix}
1 & +1 \\
1 & -1
\end{pmatrix}
\begin{pmatrix}
|1,0>\\
|0,1>
\end{pmatrix}
=U_{Hadamard}
\begin{pmatrix}
|1,0>\\
|0,1>
\end{pmatrix}
\end{equation}
Quantum state can be given as
\begin{eqnarray}
|\psi>=
\begin{pmatrix}
c_e & c_g \\
\end{pmatrix}
\begin{pmatrix}
|e> \\
|g>
\end{pmatrix} \nonumber \\
=c_e|e>+c_g|g>=
\begin{pmatrix}
c_e & c_g \\
\end{pmatrix}
\frac{1}{\sqrt{2}}
\begin{pmatrix}
1 & +1 \\
1 & -1
\end{pmatrix}
\begin{pmatrix}
|1,0>\\
|0,1>
\end{pmatrix}
= \nonumber \\
\frac{1}{\sqrt{2}}
\begin{pmatrix}
c_e & c_g \\
\end{pmatrix}
\begin{pmatrix}
|1,0> + |0,1> \\
|1,0> - |0,1>
\end{pmatrix} \nonumber \\
=\frac{1}{\sqrt{2}}(c_e+c_g)|1,0>+\frac{1}{\sqrt{2}}(c_e-c_g)|0,1>.
%=U_{Hadamard}
%\begin{pmatrix}
%|1,0>\\
%|0,1>
%\end{pmatrix}
\end{eqnarray}
Controlling the occupancy of $|g>$ and $|e>$ by external microwave signal that is encoded in $c_g$ and $c_e$ coefficients we can obtain the occupancy of the left and right well by the coefficients $\frac{1}{\sqrt{2}}(c_e+c_g)$ and $\frac{1}{\sqrt{2}}(c_e-c_g)$.
We have normalization condition that is
\begin{eqnarray}
[\frac{1}{\sqrt{2}}(c_e+c_g)][\frac{1}{\sqrt{2}}(c_e+c_g)]^{*}+[\frac{1}{\sqrt{2}}(c_e+c_g)][\frac{1}{\sqrt{2}}(c_e-c_g)]^{*}\nonumber \\
=|c_e|^2+|c_g|^2=1.
\end{eqnarray}
Now we need to evaluate
the time dependent Hamiltonian that is about evaluation of the terms
\begin{eqnarray}
f_1(t)|e><g|=f_1(t)\frac{1}{2}(|1,0>+|0,1>)(<1,0|-<0,1|)= \nonumber \\
=f_1(t)\frac{1}{2}(|1,0><1,0|-|0,1><0,1|\nonumber \\
-|1,0><0,1|+|0,1><1,0|)
\end{eqnarray}
and
\begin{eqnarray}
f_2(t)|g><e|=f_2(t)\frac{1}{2}(|1,0>-|0,1>)(<1,0|+<0,1|) = \nonumber \\
=f_2(t)\frac{1}{2}(|1,0><1,0|\nonumber \\
-|0,1><0,1|+|1,0><0,1|-|0,1><1,0|)
\end{eqnarray}
Final Hamiltonian of 2 symmetric qdots that is under microwave field becomes
\begin{eqnarray}
H=
\begin{pmatrix}
E_p+ \frac{1}{2}(f_1+f_2)[t] & |t|-\frac{1}{2}(f_1-f_2)[t] \\
|t|-\frac{1}{2}(f_1-f_2)[t]  & E_p - \frac{1}{2}(f_1+f_2)[t]
\end{pmatrix}
\end{eqnarray}
The most general Hamiltonian becomes
\begin{eqnarray}
H=
\begin{pmatrix}
E_p(1,t)+ \frac{1}{2}(f_1+f_2)[t] & |t_{2 \rightarrow 1}|-\frac{1}{2}(f_1-f_2)[t] \\
|t_{1 \rightarrow 2}|-\frac{1}{2}(f_1-f_2)[t]  & E_p(2,t) - \frac{1}{2}(f_1+f_2)[t]
\end{pmatrix}
\end{eqnarray}

The hopping coefficient becomes renormalized as well as $E_p$ coefficients.
 %  well of length L with walls of infinite height.
%%We describe the situation of particle under the influence of external magnetic field
%%\begin{equation}
%%H=-\frac{\hbar^2}{2 m_1}\frac{d^2}{dx_1^2}+V(x)=\frac{1}{2 m_1}p_x^2+V(x).
%%\end{equation}
Let us find the eigenvalues in the simplified case. We have
\begin{eqnarray}
(E_p+ \frac{1}{2}(f_1+f_2)[t]-\epsilon)(E_p- \frac{1}{2}(f_1+f_2)[t]-\epsilon) \nonumber \\
-(|t|-\frac{1}{2}(f_1-f_2)[t])(|t|-\frac{1}{2}(f_1-f_2)[t])=0.
\end{eqnarray}
It is equivalent to
\begin{equation}
((E_p-\epsilon)^2- (\frac{1}{2}(f_1+f_2)[t])^2)-(|t|-\frac{1}{2}(f_1-f_2)[t])^2=0.
\end{equation}
or to
\begin{equation}
((E_p-\epsilon)^2- (\frac{1}{2}(f_1+f_2)[t])^2)-(|t|^2+\frac{1}{4}(f_1-f_2)^2[t]-|t|(f_1-f_2))=0.
\end{equation}
that is equivalent to
\begin{equation}
(E_p-\epsilon)^2-(|t|^2-|t|(f_1-f_2))=0.
\end{equation}
The last equation has two solutions
\begin{eqnarray}
\epsilon=E_p \pm \sqrt{|t|^2-|t|(f_1(time)-f_2(time))}
\end{eqnarray}
Therefore we have the renormalized |t| function by means of $f_1$ and $f_2$ functions.
The system Hamiltonian with no presence of microwave field can be written as
\begin{equation}
H=(E_p+|t|)|e><e|+(E_p-|t|)|g><g|.
\end{equation}
Adding the microwave field we obtain
\begin{eqnarray}
H=(E_p+|t|)|e><e|+(E_p-|t|)|g><g|+\nonumber \\
f_1(t)|e><g|+f_2(t)|g><e|.
\end{eqnarray}
Let us assume that at given time instant the quantum system has the equations
$|\psi(t)>=c_e(t)|e>+c_g(t)|g>$

If $f_1(t)=f_2(t)=0$ we have $c_e(t)=c_e(0)e^{-\frac{i}{\hbar}t}, c_g(t)=c_g(0)e^{-\frac{i}{\hbar}t}$ with normalization condition $|c_e|^2+|c_g|^2=1$.
Adding time dependent functions $f_1$ and $f_2$ to Hamiltonian brings
\begin{eqnarray}
\frac{i \hbar}{dt}(c_e(t+dt)-c_e(t))|e>+(c_g(t+dt)-c_g(t))|g>= \nonumber \\  (E_p+|t|)c_e(t)|e>+(E_p-|t|)c_g(t)|g>+ \nonumber \\
f1(t)c_g(t)|e>+f2(t)c_e(t)|g>.
\end{eqnarray}

We have
\begin{equation}
\frac{i \hbar}{dt}(c_e(t+dt)-c_e(t))=(E_p+|t|)c_e(t)+f1(t)c_g(t)=0
\end{equation}
and
\begin{equation}
\frac{i \hbar}{dt}(c_g(t+dt)-c_g(t))=(E_p-|t|)c_g(t)+f2(t)c_e(t)=0.
\end{equation}
Having $f_1(t)=f_2(t)=0$ we have two analytic solutions
\begin{eqnarray}
c_e(t)=c_e(0)e^{-\frac{i}{\hbar}t (E_p+|t|)}, \nonumber \\
c_g(t)=c_g(0)e^{-\frac{i}{\hbar}t (E_p-|t|)}.
\end{eqnarray}
We notice that in such case we have $|c_e|^2=const1$ and $|c_g|^2=const2$.
Now we are dealing with time-dependent case.
We notice $\frac{d}{dt}c_e(t)=\frac{-i}{\hbar}(E_p+|t|)c_e(t)+f_1(t)c_g(t)$.
We use the fact $c_e^{*}(t)\frac{d}{dt}c_e(t)+c(t)\frac{d}{dt}c^{*}_e(t)=-c_g^{*}(t)\frac{d}{dt}c_g(t)-c_g(t)\frac{d}{dt}c_g(t)^{*}$.
In time dependent case we can solve
\begin{equation}
\frac{i \hbar }{dt}(c_e(t+dt)-c_e(t))=(E_p+|t|)c_e(t)+f1(t)c_g(t)=0
\end{equation}
and
\begin{equation}
\frac{i \hbar }{dt}(c_g(t+dt)-c_g(t))=(E_p-|t|)c_g(t)+f1(t)c_e(t)=0.
\end{equation}
Adding two equations we get
\begin{eqnarray}
i \hbar \frac{ d}{dt}(c_e(t)+c_e(t))=(E_p)(c_e(t)+c_g(t))+\nonumber \\
f1(t)(c_g(t)+c_e(t))+|t|(c_e(t)-c_g(t))=0
\end{eqnarray}
and substracting we have
\begin{eqnarray}
i \hbar \frac{ d}{dt}(c_e(t)-c_e(t))=(E_p)(c_e(t)-c_g(t))+ \nonumber \\
f1(t)(c_e(t)-c_e(t))+|t|(c_e(t)+c_g(t))=0
\end{eqnarray}
%\onecolumn
The last two equations can be rewritten to the operator format. We obtain

\begin{eqnarray}
[ i \hbar \frac{ d}{dt} - E_p - f1(t) ](c_e(t)+c_g(t))= \nonumber \\
 \hat{O}(c_e(t)+c_g(t))=+|t|(c_e(t)-c_g(t))=0
\end{eqnarray}
and
\begin{eqnarray}
[i \hbar \frac{ d}{dt} -E_p - f1(t)] (c_e(t)-c_g(t))=  \nonumber \\
\hat{O}(c_e(t)-c_g(t))=+|t|(c_e(t)+c_g(t))=0,
\end{eqnarray}
where $\hat{O}= [i \hbar \frac{ d}{dt} -E_p - f1(t)] $.
It is interesting to observe that equations have propagator format.
Let us define the object
\begin{equation}
G(1,2,t=time)=(c_e(t)+c_g(t))(c_e(t)-c_g(t))^{*}.
\end{equation}
%\onecolumn
We apply the operator $[ i \hbar \frac{ d}{dt} - E_p - f_1(t) ]$ to the $G(1,2)$. We obtain
\begin{eqnarray}
\hat{O} G(1,2,t=time)=[i \hbar \frac{ d}{dt} -E_p - f_1(t)]G(1,2,t=time)= \nonumber \\
=[i \hbar \frac{ d}{dt} -E_p - f_1(t)][(c_e(t)+c_g(t))(c_e(t)-c_g(t))^{*}] = \nonumber \\
=[(i \hbar \frac{ d}{dt} -E_p - f_1(t))(c_e(t)+c_g(t))](c_e(t)-c_g(t))^{*} + \nonumber \\
(c_e(t)+c_g(t))[(i \hbar \frac{ d}{dt} -E_p - f_1(t))(c_e(t)-c_g(t))^{*}] = \nonumber \\
=|t|(c_e(t)-c_g(t))(c_e(t)-c_g(t))^{*}+|t|(c_e(t)+c_g(t))(c_e(t)-c_g(t))^{*} \nonumber \\
=|t|.
%(c_e(t)+c_e(t))[(i \hbar \frac{ d}{dt} -E_p - f1(t))(c_e(t)-c_e(t))^{*)]=
\end{eqnarray}
\onecolumn
Final propagator can be written as
\begin{equation}
G(1,2,t=time)=\frac{1}{i \hbar \frac{ d}{dt} -E_p - f1(t)}|t|.
\end{equation}
We can extend our definition of propagator to the form
\begin{equation}
G(x1,x2,t=time)=\frac{1}{i \hbar \frac{ d}{dt} - (-\frac{\hbar^2}{2m}\frac{d^2}{dx^2}+V(x,t)) - f_1(t)}|t|.
\end{equation}
%\onecolumn
It is interesting to observe that set of equations
\begin{eqnarray}
[ i \hbar \frac{ d}{dt} - E_p - f_1(t) ](c_e(t)+c_g(t))= \nonumber \\
\hat{O}(c_e(t)+c_g(t))=+|t|(c_e(t)-c_g(t))=0
\end{eqnarray}
and
\begin{eqnarray}
[i \hbar \frac{ d}{dt} -E_p - f_1(t)] (c_e(t)-c_g(t))= \nonumber \\
\hat{O}(c_e(t)-c_g(t))=+|t|(c_e(t)+c_g(t))=0,
\end{eqnarray}
can be solved by applying additional operator $\hat{O}$ to each of sides.
We obtain using new introduced functions $u_1(t)=c_e(t)+c_g(t)$ and $u_2(t)=c_e(t)-c_g(t)$:
\begin{eqnarray}
[ i \hbar \frac{ d}{dt} - E_p - f_1(t) ]^2(c_e(t)+c_g(t))= \nonumber \\
 +|t|[ i \hbar \frac{ d}{dt} - E_p - f_1(t) ](c_e(t)-c_g(t))= \nonumber \\
|t|^2(c_e(t)+c_g(t))=|t|^2u_1(t).
\end{eqnarray}
In similar fashion we obtain
\begin{eqnarray}
[ i \hbar \frac{ d}{dt} - E_p - f_1(t) ]^2(c_e(t)-c_g(t))= \nonumber \\
+|t|[ i \hbar \frac{ d}{dt} - E_p - f_1(t) ](c_e(t)+c_g(t))=\nonumber \\
|t|^2(c_e(t)-c_g(t))=|t|^2u_2(t).
\end{eqnarray}
We need to evaluate the operator
\begin{eqnarray*}
[ i \hbar \frac{ d}{dt} - (E_p + f_1(t)) ]^2= 
\end{eqnarray*}
\begin{eqnarray}
[-\hbar^2 \frac{d^2}{dt^2}+(E_p+f_1(t))^2-2(E_p+f1(t))i \hbar \frac{ d}{dt} 
-i\hbar (\frac{d}{dt}f1(t))]
\end{eqnarray}
We therefore end up in
\begin{eqnarray}
[-\hbar^2 \frac{d^2}{dt^2}+(E_p+f_1(t))^2-2(E_p+f1(t))i \hbar \frac{ d}{dt}-i\hbar (\frac{d}{dt}f_1(t))]u_1(t) \nonumber \\
=|t|^2u_1(t)
\end{eqnarray}
and
\begin{eqnarray}
[-\hbar^2 \frac{d^2}{dt^2}+(E_p+f_1(t))^2-2(E_p+f1(t))i \hbar \frac{ d}{dt}-i\hbar (\frac{d}{dt}f_1(t))]u_2(t) \nonumber  \\
=|t|^2u_2(t).
\end{eqnarray}
If $f_1=\sin(\gamma t + \phi)$ both functions $u_1(t)$ and $u_2(t)$ have the analytic solutions.
We also notice that $|u_1(t)|^2+|u_2(t)|^2=2$ at any time instant t.
%\onecolumn
\section{Describing the decoherence effects by tight binding model}
\subsection{Describing energy flow between 2 interacting qubits}

The Hamiltonian for 2 electrostatically  interacting qubits A and B (or in general any interacting physical systems A and B) is of the form
\begin{eqnarray}
\hat{H}=(E(t)_{1a}\ket{E_{1a}}_t\bra{E_{1a}}_t+E(t)_{2a} \ket{E_{2a}}\bra{E_{2a}}) \times \hat{I}_{qA}+ %\nonumber \\
\hat{I}_{qB}\times (E(t)_{1b}\ket{E_{1b}}\bra{E_{1b}}+E(t)_{2b}\ket{E_{2b}}\bra{E_{2b}})+\nonumber \\
+(f_{1a}(t))\ket{E_{1a}}_t\bra{E_{1a}}+f_{2a}^{*}(t)\ket{E_{2a}}_t\bra{E_{1a}})\times \hat{I}_{qB}
+\hat{I}_{qA} \times (f_{1b}(t))\ket{E_{1b}}_t\bra{E_{1b}}+f_{2b}^{*}(t)\ket{E_{2b}}_t\bra{E_{1b}})+\nonumber \\
+(g(E_{1a},E_{2b},t)\ket{E_{1a},E_{2b}}\bra{E2a,E_{1b}}+g(E_{2a},E_{1b},t)^{*}\ket{E_{2a},E_{1b}}\bra{E_{1a},E_{2b}})+\nonumber \\
+(r(E_{1a},E_{2b},t)\ket{E_{1a},E_{1b}}\bra{E_{2a},E_{2b}}+r(E_{2a},E_{1b},t)^{*}\ket{E_{2a},E_{2b}}\bra{E_{1a},E_{1b}}).
\end{eqnarray}
Such structure of Hamiltonian is postulated and is fundamentally justified. At first we assume that there is no electron-electron interaction between two qubits. In such case in tight-binding model we have localized $E_{p1A(B)}, E_{p2A(B)}$ and delocalized energy $t_{12A(B)}, t_{21A(B)}$ associated with qubit A and with qubit B. 
Now we need to interlink this Hamiltonian structure with turning on Coulomb interaction. Coulomb interaction is responsible for energy exchange between qubit A and B. It is propagated by the photons exchange that have discrete values. However the electrostatic energy has the limited value so it shall bring certain renormalization to the initial qubit eigenstates. 
It is easier and methodologically justified to start from Coulomb interaction that has the following form for electron A and electron B at nodes $(x_k,x'_l)$ so we have $(1, 1')\rightarrow \ket{x_1}\ket{x_{1'}}\frac{q^2}{d(1,1')}\bra{x_1}\bra{x_{1'}}$, $(2, 2')\rightarrow \ket{x_2}\ket{x_{2'}}\frac{q^2}{d(2,2')}\bra{x_2}\bra{x_{2'}}$, $(1,2') \rightarrow \ket{x_1}\ket{x_{2'}}\frac{q^2}{d(1,2')}\bra{x_1}\bra{x_{2'}}$, $(2,1') \rightarrow \ket{x_2}\ket{x_{1'}}\frac{q^2}{d(2,1')}\bra{x_2}\bra{x_{1'}}$. 
Therefore total Coulomb Hamiltonian assoctated with qubit A and B electrostatic interaction is expressed by the Hamiltonian 
\begin{equation}
\hat{H}_{Coulomb}(1,2,1',2')=\ket{x_1}\ket{x_{1'}}\frac{q^2}{d(1,1')}\bra{x_1}\bra{x_{1'}}+\ket{x_1}\ket{x_{2'}}\frac{q^2}{d(1,2')}\bra{x_1}\bra{x_{2'}}+\ket{x_2}\ket{x_{1'}}\frac{q^2}{d(2,1')}\bra{x_2}\bra{x_{1'}}+\ket{x_2}\ket{x_{2'}}\frac{q^2}{d(2,2')}\bra{x_2}\bra{x_{2'}}.
\end{equation}
This Hamiltonian is responsible for generating entanglement between qubit A and B since Hilbert space of 2 non-interacting qubits is its tensor product of Hilbert space of qubit A and qubit B. It is instructive to notice that we can control the entanglement between qubits A and B in electrostatic way by having time-dependent control on coefficients $a_{A}(t),b_{A}(t),c_{A}(t),d_{A}(t)$ and $a_{B}(t),b_{B}(t),c_{B}(t),d_{B}(t)$ (8 complex value parameters) that are function of 6 voltages applied to qubit A and B (3 voltages for each qubit). 
We start from spectral decomposition of operator $\ket{x_1}\ket{x_{1'}}\frac{q^2}{d(1,1')}\bra{x_1}\bra{x_{1'}}$ into eigenergy represenation of qubits A and B and we obtain  , %$$  and we obtain    
\begin{eqnarray}
\left(\ket{x_1}\ket{x_{1'}}\frac{q^2}{d(1,1')}\bra{x_1}\bra{x_{1'}}\right)_{a_{A}(t),c_{A}(t),a_{B}(t),c_{B}(t)}=\nonumber \\
%\end{eqnarray}
%\begin{eqnarray}
(|E_{1a}><E_{1a}|+|E_{2a}><E_{2a}|)(|E_{1b}><E_{1b}|+|E_{2b}><E_{2b}|)(|x_1>|x_{1'}>\frac{q^2}{d(1,1')}<x_1|<x_{1'}|) \times \nonumber \\
\times (|E_{1a}><E_{1a}| 
+|E_{2a}><E_{2a}|)(|E_{1b}><E_{1b}|+|E_{2b}><E_{2b}|)=\nonumber \\
=\frac{q^2}{d(1,1')}(|E_{1a}><E_{1a}||x_1>+|E_{2a}><E_{2a}||x_1>)(|E_{1b}><E_{1b}||x_{1'}>+ 
|E_{2b}><E_{2b}||x_{1'}>)\times \nonumber \\
\times  (<x_1||E_{1a}><E_{1a}|+<x_1||E_{2a}><E_{2a}|)(<x_{1'}||E_{1b}><E_{1b}|+<x_{1'}||E_{2b}><E_{2b}|)= \nonumber \\
=\frac{q^2}{d(1,1')}(|E_{1a}>a_{A}(t)^{*}+|E_{2a}>c_{A}(t)^{*})(|E_{1b}>a_{B}(t)^{*}+ 
|E_{2b}>c_{B}(t)^{*})\times \nonumber \\
\times  (a_{A}(t)<E_{1a}|+c_{A}(t)<E_{2a}|)(a_{B}(t)<E_{1b}|+c_{B}(t)<E_{2b}|)=\nonumber \\
=\frac{q^2}{d(1,1')}(|\ket{E_{1a},E_{1b}}a_{B}(t)^{*}a_{A}(t)^{*}+\ket{E_{1a},E_{2b}}a_{A}(t)^{*}c_{B}(t)^{*}+\ket{E_{2a},E_{1b}}c_{A}(t)^{*}a_{B}(t)^{*}+\ket{E_{2a},E_{2b}}c_{A}(t)^{*}c_{B}(t)^{*})\times \nonumber \\ 
(\bra{E_{1a},E_{1b}}a_{B}(t)a_{A}(t)+\bra{E_{1a},E_{2b}}a_{A}(t)c_{B}(t)+\bra{E_{2a},E_{1b}}c_{A}(t)a_{B}(t)+\bra{E_{2a},E_{2b}}c_{A}(t)c_{B}(t))=\nonumber \\
=\Bigg[\frac{q^2}{d(1,1')} (\ket{E_{1a},E_{1b}}\bra{E_{1a},E_{1b}}|a_{B}(t)|^2|a_{A}(t)|^{2}+\ket{E_{1a},E_{2b}}\bra{E_{1a},E_{2b}}|a_{B}(t)|^2|c_{A}(t)|^{2}+ \nonumber \\ +\ket{E_{2a},E_{1b}}\bra{E_{2a},E_{1b}}|c_{B}(t)|^2|a_{A}(t)|^{2}+\ket{E_{2a},E_{2b}}\bra{E_{2a},E_{2b}}|c_{B}(t)|^2|c_{A}(t)|^{2}) \Bigg]_{r1}+ \nonumber \\
\Bigg[\frac{q^2}{d(1,1')} (\ket{E_{1a},E_{1b}}\bra{E_{1a},E_{2b}}|a_{A}(t)|^2c_{B}(t)a_{B}^{*}(t)+\ket{E_{1a},E_{2b}}\bra{E_{1a},E_{1b}}|a_{A}(t)|^2c_{B}^{*}(t)a_{B}(t)+ \nonumber \\ +\ket{E_{2a},E_{1b}}\bra{E_{2a},E_{2b}}|c_{A}(t)|^2c_{B}(t)a_{B}^{*}(t)+\ket{E_{2a},E_{2b}}\bra{E_{2a},E_{1b}}|c_{A}(t)|^2c_{B}^{*}(t)a_{B}(t)) \Bigg]_{r2}+ \nonumber \\
\Bigg[\frac{q^2}{d(1,1')} (\ket{E_{1a},E_{1b}}\bra{E_{2a},E_{1b}}c_{A}(t)a_{b}(t)a_{A}^{*}(t)a_{B}^{*}(t)+\ket{E_{2a},E_{1b}}\bra{E_{1a},E_{1b}}c_{A}^{*}(t)a_{b}^{*}(t)a_{A}(t)a_{B}(t)+ \nonumber \\ +\ket{E_{2a},E_{2b}}\bra{E_{1a},E_{2b}}a_{A}(t)c_{B}(t)c_{A}^{*}(t)c_{B}^{*}(t)+\ket{E_{1a},E_{2b}}\bra{E_{2a},E_{2b}}a_{A}^{*}(t)c_{B}^{*}(t)c_{A}(t)c_{B}(t))\Bigg]_{r3}+ \nonumber \\
\Bigg[\frac{q^2}{d(1,1')} (\ket{E_{1a},E_{1b}}\bra{E_{2a},E_{2b}}c_{A}(t)c_{B}(t)a_{A}^{*}(t)a_{B}^{*}(t)+\ket{E_{2a},E_{2b}}\bra{E_{1a},E_{1b}}c_{A}^{*}(t)c_{B}^{*}(t)a_{A}(t)a_{B}(t)+ \nonumber \\ +\ket{E_{2a},E_{1b}}\bra{E_{1a},E_{2b}}a_{A}(t)c_{B}(t)^{*}c_{A}(t)^{*}a_{B}(t)^{*}+\ket{E_{1a},E_{2b}}\bra{E_{2a},E_{1b}}a_{A}^{*}(t)c_{B}(t)c_{A}(t)a_{B}(t))\Bigg]_{r4}
 %|E_{2a}>c_{A}(t)^{*})(|E_{1b}>a_{B}
%(t)^{*}+ 
%|E_{2b}>c_{B}(t)^{*})
\end{eqnarray}
We identify 4 types of renormalization coming to the Hamiltonian of non-interacting qubits A and B from interacting term $\ket{x_1}\ket{x_{1'}}\frac{q^2}{d(1,1')}\bra{x_1}\bra{x_{1'}}$. Renormalization 1 denoted by r1 is describing the change of total energy 
in the non-interacting qubit due to the apperance of Coulomb interaction. Renormalization 2 describes the process in which qubit A energy is unchanged and qubit B populates or depopulates energy levels $E_{1b}$ and $E_{2b}$ so qubit B is either heated up or cooled down.
Renormalization r3 is describing the same as renormalization 2 but in the case when qubit A is heated up or cooled down while energy of qubit B is unchanged. Finally renormalization 4 describes the process when both qubit A and B are heated up or cooled down by 
mutual exchange of energy due to Coulomb interaction. What is more processes r2-r4 are describing decoherence of qubit A and B due to existence of Coulomb interaction in analytical way due to internal system dynamics. Here we have omitted the interaction of external
world with our qubit systems. 
Furhtermore we can think about quantum state hybridization that is taking place in the presence of very strong Coulomb interaction. It be omitted in this work and it is the subject of future works. 
Similarly as it was done before we have spectral decomposition of operator $\ket{x_2}\ket{x_{2'}}\frac{q^2}{d(2,2')}\bra{x_2}\bra{x_{2'}}$ and we obtain 
\begin{eqnarray}
\left(\ket{x_2}\ket{x_{2'}}\frac{q^2}{d(2,2')}\bra{x_2}\bra{x_{2'}}\right)_{b_{A}(t),d_{A}(t),b_{B}(t),d_{B}(t)}=\nonumber \\
%\end{eqnarray}
%\begin{eqnarray}
(|E_{1a}><E_{1a}|+|E_{2a}><E_{2a}|)(|E_{1b}><E_{1b}|+|E_{2b}><E_{2b}|)(|x_2>|x_{2'}>\frac{q^2}{d(2,2')}<x_2|<x_{2'}|) \times \nonumber \\
\times (|E_{1a}><E_{1a}| 
+|E_{2a}><E_{2a}|)(|E_{1b}><E_{1b}|+|E_{2b}><E_{2b}|)=\nonumber \\
=\frac{q^2}{d(2,2')}(|E_{1a}><E_{1a}||x_2>+|E_{2a}><E_{2a}||x_2>)(|E_{1b}><E_{1b}||x_{2'}>+ 
|E_{2b}><E_{2b}||x_{2'}>)\times \nonumber \\
\times  (<x_2||E_{1a}><E_{1a}|+<x_2||E_{2a}><E_{2a}|)(<x_{2'}||E_{1b}><E_{1b}|+<x_{2'}||E_{2b}><E_{2b}|)= \nonumber \\
=\frac{q^2}{d(2,2')}(|E_{1a}>b_{A}(t)^{*}+|E_{2a}>d_{A}(t)^{*})(|E_{1b}>b_{B}(t)^{*}+ 
|E_{2b}>d_{B}(t)^{*})\times \nonumber \\
\times  (b_{A}(t)<E_{1a}|+d_{A}(t)<E_{2a}|)(b_{B}(t)<E_{1b}|+d_{B}(t)<E_{2b}|)=\nonumber \\
=\frac{q^2}{d(2,2')}(|\ket{E_{1a},E_{1b}}b_{B}(t)^{*}b_{A}(t)^{*}+\ket{E_{1a},E_{2b}}b_{A}(t)^{*}d_{B}(t)^{*}+\ket{E_{2a},E_{1b}}d_{A}(t)^{*}b_{B}(t)^{*}+\ket{E_{2a},E_{2b}}d_{A}(t)^{*}d_{B}(t)^{*})\times \nonumber \\ 
(\bra{E_{1a},E_{1b}}b_{B}(t)b_{A}(t)+\bra{E_{1a},E_{2b}}b_{A}(t)d_{B}(t)+\bra{E_{2a},E_{1b}}d_{A}(t)b_{B}(t)+\bra{E_{2a},E_{2b}}d_{A}(t)d_{B}(t))=\nonumber \\
=\Bigg[\frac{q^2}{d(2,2')} (\ket{E_{1a},E_{1b}}\bra{E_{1a},E_{1b}}|b_{B}(t)|^2|b_{A}(t)|^{2}+\ket{E_{1a},E_{2b}}\bra{E_{1a},E_{2b}}|b_{B}(t)|^2|d_{A}(t)|^{2}+ \nonumber \\ +\ket{E_{2a},E_{1b}}\bra{E_{2a},E_{1b}}|d_{B}(t)|^2|b_{A}(t)|^{2}+\ket{E_{2a},E_{2b}}\bra{E_{2a},E_{2b}}|b_{B}(t)|^2|d_{A}(t)|^{2}) \Bigg]_{r1}+ \nonumber \\
\Bigg[\frac{q^2}{d(2,2')} (\ket{E_{1a},E_{1b}}\bra{E_{1a},E_{2b}}|b_{A}(t)|^2d_{B}(t)b_{B}^{*}(t)+\ket{E_{1a},E_{2b}}\bra{E_{1a},E_{1b}}|b_{A}(t)|^2d_{B}^{*}(t)b_{B}(t)+ \nonumber \\ +\ket{E_{2a},E_{1b}}\bra{E_{2a},E_{2b}}|d_{A}(t)|^2d_{B}(t)b_{B}^{*}(t)+\ket{E_{2a},E_{2b}}\bra{E_{2a},E_{1b}}|d_{A}(t)|^2d_{B}^{*}(t)b_{B}(t)) \Bigg]_{r2}+ \nonumber \\
\Bigg[\frac{q^2}{d(2,2')} (\ket{E_{1a},E_{1b}}\bra{E_{2a},E_{1b}}d_{A}(t)b_{b}(t)b_{A}^{*}(t)b_{B}^{*}(t)+\ket{E_{2a},E_{1b}}\bra{E_{1a},E_{1b}}d_{A}^{*}(t)b_{b}^{*}(t)b_{A}(t)b_{B}(t)+ \nonumber \\ +\ket{E_{2a},E_{2b}}\bra{E_{1a},E_{2b}}b_{A}(t)d_{B}(t)d_{A}^{*}(t)d_{B}^{*}(t)+\ket{E_{1a},E_{2b}}\bra{E_{2a},E_{2b}}b_{A}^{*}(t)d_{B}^{*}(t)d_{A}(t)d_{B}(t))\Bigg]_{r3}+ \nonumber \\
\Bigg[\frac{q^2}{d(2,2')} (\ket{E_{1a},E_{1b}}\bra{E_{2a},E_{2b}}d_{A}(t)d_{B}(t)b_{A}^{*}(t)b_{B}^{*}(t)+\ket{E_{2a},E_{2b}}\bra{E_{1a},E_{1b}}d_{A}^{*}(t)d_{B}^{*}(t)b_{A}(t)b_{B}(t)+ \nonumber \\ +\ket{E_{2a},E_{1b}}\bra{E_{1a},E_{2b}}b_{A}(t)d_{B}(t)^{*}d_{A}(t)^{*}b_{B}(t)^{*}+\ket{E_{1a},E_{2b}}\bra{E_{2a},E_{1b}}b_{A}^{*}(t)d_{B}(t)d_{A}(t)b_{B}(t))\Bigg]_{r4}
 %|E_{2a}>c_{A}(t)^{*})(|E_{1b}>a_{B}
%(t)^{*}+ 
%|E_{2b}>c_{B}(t)^{*})
\end{eqnarray}

Again we identify 4 types of renormalization coming to the Hamiltonian of non-interacting qubits A and B from interacting term $\ket{x_2}\ket{x_{2'}}\frac{q^2}{d(2,2')}\bra{x_2}\bra{x_{2'}}$. They have the same interpretation as it was in case of operator $\ket{x_1}\ket{x_{1'}}\frac{q^2}{d(1,1')}\bra{x_1}\bra{x_{1'}}$. 

It is important to notice resonant states coming from changes of voltages at each separated qubits as in infite distance forms the matrix 
\begin{eqnarray}
H_{0+resonant}(t)=
\begin{pmatrix}
E_{1A}(t)+E_{1B}(t) & E_{12,B}(t) &  E_{12,A}(t) & 0 \\
E_{12,B}(t)^{*} &  E_{1A}(t)+E_{2B}(t) & 0 &  E_{12,A}(t) \\
E_{12,A}(t)^{*} & 0 & E_{2A}(t)+E_{1B}(t) &  E_{12,B}(t)  \\
0 &  E_{12,A}(t)^{*} & E_{12,B}(t)^{*} & E_{2A}(t)+E_{2B}(t)  \\
\end{pmatrix}
\end{eqnarray}

Now we are making spectral decomposition of operator $\ket{x_1}\ket{x_{2'}}\frac{q^2}{d(1,2')}\bra{x_1}\bra{x_{2'}}$ and we obtain 
\begin{eqnarray}
\left(\ket{x_1}\ket{x_{2'}}\frac{q^2}{d(1,2')}\bra{x_1}\bra{x_{2'}}\right)_{a_{A}(t),c_{A}(t),b_{B}(t),d_{B}(t)}=\nonumber \\
%\end{eqnarray}
%\begin{eqnarray}
(|E_{1a}><E_{1a}|+|E_{2a}><E_{2a}|)(|E_{1b}><E_{1b}|+|E_{2b}><E_{2b}|)(|x_1>|x_{2'}>\frac{q^2}{d(1,2')}<x_1|<x_{2'}|) \times \nonumber \\
\times (|E_{1a}><E_{1a}| 
+|E_{2a}><E_{2a}|)(|E_{1b}><E_{1b}|+|E_{2b}><E_{2b}|)=\nonumber \\
=\frac{q^2}{d(1,2')}(|E_{1a}><E_{1a}||x_1>+|E_{2a}><E_{2a}||x_1>)(|E_{1b}><E_{1b}||x_{2'}>+ 
|E_{2b}><E_{2b}||x_{2'}>)\times \nonumber \\
\times  (<x_1||E_{1a}><E_{1a}|+<x_1||E_{2a}><E_{2a}|)(<x_{2'}||E_{1b}><E_{1b}|+<x_{2'}||E_{2b}><E_{2b}|)= \nonumber \\
=\frac{q^2}{d(1,2')}(|E_{1a}>a_{A}(t)^{*}+|E_{2a}>c_{A}(t)^{*})(|E_{1b}>b_{B}(t)^{*}+ 
|E_{2b}>d_{B}(t)^{*})\times \nonumber \\
\times  (a_{A}(t)<E_{1a}|+c_{A}(t)<E_{2a}|)(b_{B}(t)<E_{1b}|+d_{B}(t)<E_{2b}|)=\nonumber \\
=\frac{q^2}{d(1,2')}(|\ket{E_{1a},E_{1b}}a_{B}(t)^{*}b_{A}(t)^{*}+\ket{E_{1a},E_{2b}}a_{A}(t)^{*}d_{B}(t)^{*}+\ket{E_{2a},E_{1b}}c_{A}(t)^{*}b_{B}(t)^{*}+\ket{E_{2a},E_{2b}}c_{A}(t)^{*}d_{B}(t)^{*})\times \nonumber \\ 
(\bra{E_{1a},E_{1b}}a_{B}(t)b_{A}(t)+\bra{E_{1a},E_{2b}}a_{A}(t)d_{B}(t)+\bra{E_{2a},E_{1b}}c_{A}(t)b_{B}(t)+\bra{E_{2a},E_{2b}}c_{A}(t)d_{B}(t))=\nonumber \\
=\Bigg[\frac{q^2}{d(1,2')} (\ket{E_{1a},E_{1b}}\bra{E_{1a},E_{1b}}|a_{A}(t)|^2|b_{B}(t)|^{2}+\ket{E_{1a},E_{2b}}\bra{E_{1a},E_{2b}}|a_{A}(t)|^2|d_{B}(t)|^{2}+ \nonumber \\ +\ket{E_{2a},E_{1b}}\bra{E_{2a},E_{1b}}|c_{A}(t)|^2|b_{B}(t)|^{2}+\ket{E_{2a},E_{2b}}\bra{E_{2a},E_{2b}}|c_{A}(t)|^2|d_{B}(t)|^{2}) \Bigg]_{r1}+ \nonumber \\
\Bigg[\frac{q^2}{d(1,2')} (\ket{E_{1a},E_{1b}}\bra{E_{1a},E_{2b}}|a_{A}(t)|^2b_{B}^{*}(t)d_{B}(t)+\ket{E_{1a},E_{2b}}\bra{E_{1a},E_{1b}}|a_{A}(t)|^2d_{B}^{*}(t)b_{B}(t)+ \nonumber \\ +\ket{E_{2a},E_{1b}}\bra{E_{2a},E_{2b}}|c_{A}|^2b_{B}^{*}(t)d_{B}(t)+\ket{E_{2a},E_{2b}}\bra{E_{2a},E_{1b}}|c_{A}(t)|^2c_{B}^{*}(t)b_{B}(t)) \Bigg]_{r2}+ \nonumber \\
\Bigg[\frac{q^2}{d(1,2')} (\ket{E_{1a},E_{1b}}\bra{E_{2a},E_{1b}}a_{A}^{*}(t)c_{A}(t)|b_{B}(t)|^2+\ket{E_{2a},E_{1b}}\bra{E_{1a},E_{1b}}a_{A}(t)c_{A}^{*}(t)|b_{B}(t)|^2+ \nonumber \\ +\ket{E_{2a},E_{2b}}\bra{E_{1a},E_{2b}}a_{A}(t)c_{A}^{*}(t)|d_{B}(t)|^{2}+\ket{E_{1a},E_{2b}}\bra{E_{2a},E_{2b}}a_{A}^{*}(t)c_{A}(t)|d_{B}(t)|^{2})\Bigg]_{r3}+ \nonumber \\
\Bigg[\frac{q^2}{d(1,2')} (\ket{E_{1a},E_{1b}}\bra{E_{2a},E_{2b}}a_{A}^{*}(t)b_{B}^{*}(t)c_{A}(t)d_{B}(t)+\ket{E_{2a},E_{2b}}\bra{E_{1a},E_{1b}}a_{A}(t)b_{B}(t)c_{A}^{*}(t)d_{B}^{*}(t)+ \nonumber \\ +\ket{E_{2a},E_{1b}}\bra{E_{1a},E_{2b}}a_{A}(t)d_{B}(t)c_{A}^{*}(t)b_{B}^{*}(t)+\ket{E_{1a},E_{2b}}\bra{E_{2a},E_{1b}}a_{A}^{*}(t)d_{B}^{*}(t)c_{A}(t)b_{B}(t))\Bigg]_{r4}
 %|E_{2a}>c_{A}(t)^{*})(|E_{1b}>a_{B}
%(t)^{*}+ 
%|E_{2b}>c_{B}(t)^{*})
\end{eqnarray}
and 
\begin{eqnarray}
\left(\ket{x_2}\ket{x_{1'}}\frac{q^2}{d(2,1')}\bra{x_2}\bra{x_{1'}}\right)_{d_{A}(t),a_{B}(t),d_{B}(t),a_{B}(t)}=\nonumber \\
%\end{eqnarray}
%\begin{eqnarray}
(|E_{1a}><E_{1a}|+|E_{2a}><E_{2a}|)(|E_{1b}><E_{1b}|+|E_{2b}><E_{2b}|)(|x_1>|x_{1'}>\frac{q^2}{d(2,1')}<x_1|<x_{1'}|) \times \nonumber \\
\times (|E_{1a}><E_{1a}| 
+|E_{2a}><E_{2a}|)(|E_{1b}><E_{1b}|+|E_{2b}><E_{2b}|)=\nonumber \\
=\frac{q^2}{d(2,1')}(|E_{1a}><E_{1a}||x_2>+|E_{2a}><E_{2a}||x_2>)(|E_{1b}><E_{1b}||x_{1'}>+ 
|E_{2b}><E_{2b}||x_{1'}>)\times \nonumber \\
\times  (<x_2||E_{1a}><E_{1a}|+<x_2||E_{2a}><E_{2a}|)(<x_{1'}||E_{1b}><E_{1b}|+<x_{1'}||E_{2b}><E_{2b}|)= \nonumber \\
=\frac{q^2}{d(2,1')}(|E_{1a}>b_{A}(t)^{*}+|E_{2a}>d_{A}(t)^{*})(|E_{1b}>a_{B}(t)^{*}+ 
|E_{2b}>c_{B}(t)^{*})\times \nonumber \\
\times  (b_{A}(t)<E_{1a}|+d_{A}(t)<E_{2a}|)(a_{B}(t)<E_{1b}|+c_{B}(t)<E_{2b}|)=\nonumber \\
=\frac{q^2}{d(2,1')}(\ket{E_{1a},E_{1b}}b_{A}^{*}(t)a_{B}^{*}(t)+\ket{E_{1a},E_{2b}}b_{A}^{*}c_{B}^{*}(t)+\ket{E_{2a},E_{1b}}d_{A}^{*}(t)a_{B}^{*}(t)+\ket{E_{2a},E_{2b}}d_{A}^{*}(t)c_{B}^{*}(t))\times \nonumber \\ 
(\bra{E_{1a},E_{1b}}b_{A}(t)a_{B}(t)+\bra{E_{1a},E_{2b}}b_{A}c_{B}(t)+\bra{E_{2a},E_{1b}}d_{A}(t)a_{B}(t)+\bra{E_{2a},E_{2b}}d_{A}(t)c_{B}(t))=\nonumber \\
=\Bigg[\frac{q^2}{d(2,1')} (\ket{E_{1a},E_{1b}}\bra{E_{1a},E_{1b}}|b_{A}(t)|^2|a_{B}(t)|^{2}+\ket{E_{1a},E_{2b}}\bra{E_{1a},E_{2b}}|b_{A}(t)|^2|c_{B}(t)|^{2}+ \nonumber \\ +\ket{E_{2a},E_{1b}}\bra{E_{2a},E_{1b}}|d_{A}(t)|^2|a_{B}(t)|^{2}+\ket{E_{2a},E_{2b}}\bra{E_{2a},E_{2b}}|d_{A}(t)|^2|c_{B}(t)|^{2}) \Bigg]_{r1}+ \nonumber \\
\Bigg[\frac{q^2}{d(2,1')} (\ket{E_{1a},E_{1b}}\bra{E_{1a},E_{2b}}|b_{A}(t)|^2a_{B}^{*}(t)c_{B}(t)+\ket{E_{1a},E_{2b}}\bra{E_{1a},E_{1b}}|b_{A}(t)|^2a_{B}(t)c_{B}^{*}(t)+ \nonumber \\ +\ket{E_{2a},E_{1b}}\bra{E_{2a},E_{2b}}|d_{A}(t)|^2c_{B}(t)a^{*}_{B}(t)+\ket{E_{2a},E_{2b}}\bra{E_{2a},E_{1b}}|d_{A}(t)|^2c_{B}^{*}(t)a_{B}(t)) \Bigg]_{r2}+ \nonumber \\
\Bigg[\frac{q^2}{d(2,1')} (\ket{E_{1a},E_{1b}}\bra{E_{2a},E_{1b}}b_{A}^{*}(t)b_{B}^{*}(t)d_{A}(t)a_{B}(t)+\ket{E_{2a},E_{1b}}\bra{E_{1a},E_{1b}}b_{A}(t)b_{B}(t)d_{A}^{*}(t)a_{B}^{*}(t)+ \nonumber \\ +\ket{E_{2a},E_{2b}}\bra{E_{1a},E_{2b}}b_{A}(t)c_{B}(t)d_{A}^{*}(t)c_{B}^{*}(t)+\ket{E_{1a},E_{2b}}\bra{E_{2a},E_{2b}}b_{A}^{*}(t)c_{B}^{*}(t)d_{A}(t)c_{B}(t))\Bigg]_{r3}+ \nonumber \\
\Bigg[\frac{q^2}{d(2,1')} (\ket{E_{1a},E_{1b}}\bra{E_{2a},E_{2b}}b_{A}^{*}(t)a_{B}^{*}(t)d_{A}(t)c_{B}(t)+\ket{E_{2a},E_{2b}}\bra{E_{1a},E_{1b}}b_{A}(t)a_{B}(t)d_{A}^{*}(t)c_{B}^{*}(t)+ \nonumber \\ +\ket{E_{2a},E_{1b}}\bra{E_{1a},E_{2b}}b_{A}(t)c_{B}(t)d_{A}^{*}(t)a_{B}^{*}(t)+\ket{E_{1a},E_{2b}}\bra{E_{2a},E_{1b}}b_{A}^{*}(t)c_{B}^{*}(t)d_{A}(t)a_{B}(t))\Bigg]_{r4}
 %|E_{2a}>c_{A}(t)^{*})(|E_{1b}>a_{B}
%(t)^{*}+ 
%|E_{2b}>c_{B}(t)^{*})
\end{eqnarray} 
We can easily identify 4 renormalized eigenenergies to be of the form
\begin{eqnarray}
 E_{1A}+E_{1B} \rightarrow E_{1A}+E_{1B}+\frac{q^2}{d(2,1')}|b_{A}(t)|^2|a_{B}(t)|^{2} +\frac{q^2}{d(1,2')}|a_{A}(t)|^2|b_{B}(t)|^{2}+\frac{q^2}{d(2,2')}|b_{B}(t)|^2|b_{A}(t)|^{2}+\frac{q^2}{d(1,1')}|a_{B}(t)|^2|a_{A}(t)|^{2}, \nonumber \\ 
E_{1A}+E_{2B} \rightarrow E_{1A}+E_{2B}+ \frac{q^2}{d(2,1')}|b_{A}|^2|c_{B}|^2 +\frac{q^2}{d(1,2')}|a_{A}|^2|d_{B}|^2+ \frac{q^2}{d(2,2')}|d_{A}|^2|b_{B}|^2 + \frac{q^2}{d(1,1')}|c_{A}|^2|a_{B}|^2, \nonumber \\
E_{2A}+E_{1B} \rightarrow E_{2A}+E_{1B}+ \frac{q^2}{d(2,1')}|d_{A}|^2 |a_{B}|^2+\frac{q^2}{d(2,1')}|a_{A}|^2|b_{B}|^2+ \frac{q^2}{d(2,2')}|b_{A}|^2|d_B|^2+ \frac{q^2}{d(1,1')}|a_A|^2|c_B|^2, \nonumber \\
 E_{2A}+E_{2B} \rightarrow E_{2A}+E_{2B}+\frac{q^2}{d(2,1')}|d_{A}(t)|^2|c_{B}(t)|^{2} +\frac{q^2}{d(1,2')}|c_{A}|^2|d_{B}|^2 +\frac{q^2}{d(2,2')}|d_{A}|^2|b_{B}|^2 +\frac{q^2}{d(1,1')}|c_{A}(t)|^2|c_{B}(t)|^2,
\end{eqnarray}
under assumption that decoherence effects described by terms spanned by $\ket{E_{k,l}}\bra{E_{s,w}}$ are small as in comparison with $ E_{1A}+E_{1B}, E_{1A}+E_{2B}, E_{2A}+E_{1B},  E_{2A}+E_{2B}$ and under assumption that $\delta_{k,s}\delta_{l,w}=0$. 
Now we are going to identify decoherence matrix arrising from natural non-dissipative equations of motion.

We are have the following decoherence matrix  describing heating of qubit B and keeping unchanged qubit A as
%\begin{eqnarray}
%H_{\ket{E_i,E_j}\bra{E_k,E_s}}=
%\begin{pmatrix}
%%\begin{eqnarray}
%%H_{(E_{1a} \rightarrow E_{1a}, E_{1b} \rightarrow E_{2b})}=\frac{q^2}{d(2,2')}(|b_A|^2d_Bb_B^{*}+d_A|b_B|^2b_A^{*}+d_Ad_Bb_{A}^{*}b_B^{*}) %%%+ \nonumber \\ 
%%+\frac{q^2}{d(1,1')}(|a_A|^2c_Ba_B^{*}+|a_B|^2a_A^{*}+a_{A}^{*}a_{B}^{*}c_A c_B )+ \nonumber \\
%%\frac{q^2}{d(2,1')}( |b_A|^2 a_B^{*}c_B + b_A^{*}b_B^{*}d_Aa_B+b_A^{*}a_B^{*}d_Ac_B)+\frac{q^2}{d(1',2')}(|a_A|^2b_B^{*}d_B+|b_B|^2a_A^{*}c_A+a_A^{*}b_B^{*}c_Ad_B)=H_{(E_{1a} \rightarrow E_{1a}, E_{2b} \rightarrow E_{1b})}^{*}.
%%\end{eqnarray}
%%and the following decoherence matrix describing cooling of qubit A and keeping unchanged qubit B as
\begin{eqnarray}
H_{(E_{1a} \rightarrow E_{1a}, E_{2b} \rightarrow E_{1b})}=\frac{q^2}{d(1,2')}(|a_A|^2b_B^{*}d_B)+\frac{q^2}{d(2,1')}(|b_A|^2a_B^{*}c_B)+  %%%+ \nonumber \\ 
\frac{q^2}{d(1,1')}(|a_A|^2c_Ba_B^{*})+\frac{q^2}{d(2,2')}(|b_A|^2d_Bb_B^{*}), \nonumber \\
H_{(E_{2a} \rightarrow E_{2a}, E_{2b} \rightarrow E_{1b})}=\frac{q^2}{d(1,2')}(|c_A|^2b_B^{*}d_B)+\frac{q^2}{d(2,1')}(|d_A|^2c_Ba_B^{*})+  %%%+ \nonumber \\ 
\frac{q^2}{d(1,1')}(|c_A|^2c_Ba_B^{*})+\frac{q^2}{d(2,2')}(|d_A|^2d_Bb_B^{*}), \nonumber \\
H_{(E_{2a} \rightarrow E_{1a}, E_{1b} \rightarrow E_{1b})}=\frac{q^2}{d(1,2')}(a_A^{*}c_A|b_B|^2)+\frac{q^2}{d(2,1')}(b_A^{*}b_{B}^{*}d_Aa_B)+  %%%+ \nonumber \\ 
\frac{q^2}{d(1,1')}(c_Aa_A^{*}|a_B|^2)+\frac{q^2}{d(2,2')}(d_Ab_A^{*}|b_B|^2), \nonumber \\
H_{(E_{2a} \rightarrow E_{1a}, E_{2b} \rightarrow E_{2b})}=\frac{q^2}{d(1,2')}(d_Ab_A^{*}|c_B|^2)+\frac{q^2}{d(2,1')}(d_Ab_A^{*}|c_B|^2)+  %%%+ \nonumber \\ 
\frac{q^2}{d(1,1')}(|c_B|^2c_Aa_A^{*})+\frac{q^2}{d(2,2')}(b_A^{*}d_A|d_B|^2), \nonumber \\
H_{(E_{1a} \rightarrow E_{2a}, E_{1b} \rightarrow E_{2b})}=\frac{q^2}{d(1,2')}(a_Ab_Bc_A^{*}d_B^{*})+\frac{q^2}{d(2,1')}(b_Aa_Bd_A^{*}c_B^{*})+  %%%+ \nonumber \\ 
\frac{q^2}{d(1,1')}(c_A^{*}c_B^{*}a_Aa_B)+\frac{q^2}{d(2,2')}(d_A^{*}d_B^{*}b_Ab_B), \nonumber \\
H_{(E_{2a} \rightarrow E_{1a}, E_{1b} \rightarrow E_{2b})}=\frac{q^2}{d(1,2')}(a_A^{*}d_B^{*}c_Ab_B)+\frac{q^2}{d(2,1')}(b_A^{*}c_{B}^{*}d_Aa_B)  %%%+ \nonumber \\ 
+\frac{q^2}{d(1,1')}(a_A^{*}c_Bc_Aa_B)+\frac{q^2}{d(2,2')}(b_A^{*}d_Bd_Ab_B), \nonumber \\
%%\frac{q^2}{d(2,1')}( )+\frac{q^2}{d(1',2')}()=H_{(E_{1a} \rightarrow E_{1a}, E_{2b} \rightarrow E_{1b})}^{*}.
\end{eqnarray}
and we have $H_{(E_{1a} \rightarrow E_{1a}, E_{2b} \rightarrow E_{1b})}=H_{(E_{1a} \rightarrow E_{1a}, E_{1b} \rightarrow E_{2b})}^{*}$, $H_{(E_{2a} \rightarrow E_{2a}, E_{2b} \rightarrow E_{1b})}=H_{(E_{2a} \rightarrow E_{2a}, E_{1b} \rightarrow E_{2b})}^{*}$, $H_{(E_{2a} \rightarrow E_{1a}, E_{1b} \rightarrow E_{1b})}=H_{(E_{1a} \rightarrow E_{2a}, E_{1b} \rightarrow E_{1b})}^{*}$, $H_{(E_{2a} \rightarrow E_{1a}, E_{2b} \rightarrow E_{2b})}=H_{(E_{1a} \rightarrow E_{2a}, E_{2b} \rightarrow E_{2b})}^{*}$, $H_{(E_{1a} \rightarrow E_{2a}, E_{1b} \rightarrow E_{2b})}=H_{(E_{2a} \rightarrow E_{1a}, E_{2b} \rightarrow E_{1b})}^{*}$ . The decoherence terms can be expressed by the matrix 

\begin{eqnarray}
H_{Coulomb, \ket{E_i,E_j}\bra{E_k,E_s}}=
\begin{pmatrix}
(E_{1a}+E_{1b})_{r1} & H_{(E_{1a} \rightarrow E_{1a}, E_{2b} \rightarrow E_{1b})} & H_{(E_{2a} \rightarrow E_{1a}, E_{2b} \rightarrow E_{1b})} & H_{(E_{2a} \rightarrow E_{1a}, E_{2b} \rightarrow E_{1b})} \\
H_{(E_{1a} \rightarrow E_{1a}, E_{1b} \rightarrow E_{2b})} & (E_{1a}+E_{2b})_{r1} &  H_{(E_{2a} \rightarrow E_{1a}, E_{1b}\rightarrow E_{2b}} & H_{(E_{2a} \rightarrow E_{1a}, E_{2b}\rightarrow E_{2b}} \\
 H_{(E_{1a} \rightarrow E_{2a}, E_{1b} \rightarrow E_{2b})} &  H_{(E_{1a} \rightarrow E_{2a}, E_{2b}\rightarrow E_{1b}} & (E_{2a}+E_{1b})_{r1} & H_{(E_{2a} \rightarrow E_{2a}, E_{2b} \rightarrow E_{1b})} \\
H_{(E_{1a} \rightarrow E_{2a}, E_{1b} \rightarrow E_{2b})} &  H_{(E_{1a} \rightarrow E_{2a}, E_{2b}\rightarrow E_{2b}}  & H_{(E_{2a} \rightarrow E_{2a}, E_{1b} \rightarrow E_{2b})} & (E_{2a}+E_{2b})_{r1} \\
\end{pmatrix},
\end{eqnarray}
where 
\begin{eqnarray}
(E_{1A}+E_{1B})_{r1} =+\frac{q^2}{d(2,1')}|b_{A}(t)|^2|a_{B}(t)|^{2} +\frac{q^2}{d(1,2')}|a_{A}(t)|^2|b_{B}(t)|^{2}+\frac{q^2}{d(2,2')}|b_{B}(t)|^2|b_{A}(t)|^{2}+\frac{q^2}{d(1,1')}|a_{B}(t)|^2|a_{A}(t)|^{2}, \nonumber \\ 
(E_{1A}+E_{2B})_{r1} =+ \frac{q^2}{d(2,1')}|b_{A}(t)|^2|c_{B}(t)|^2 +\frac{q^2}{d(1,2')}|a_{A}(t)|^2|d_{B}(t)|^2+ \frac{q^2}{d(2,2')}|d_{A}(t)|^2|b_{B}(t)|^2 + \frac{q^2}{d(1,1')}|c_{A(t)}|^2|a_{B}(t)|^2, \nonumber \\
(E_{2A}+E_{1B})_{r1} =+ \frac{q^2}{d(2,1')}|d_{A}(t)|^2 |a_{B}(t)|^2+\frac{q^2}{d(2,1')}|a_{A}(t)|^2|b_{B}(t)|^2+ \frac{q^2}{d(2,2')}|b_{A}(t)|^2|d_B(t)|^2+ \frac{q^2}{d(1,1')}|a_A(t)|^2|c_B(t)|^2, \nonumber \\
(E_{2A}+E_{2B})_{r1} =+\frac{q^2}{d(2,1')}|d_{A}(t)|^2|c_{B}(t)|^{2} +\frac{q^2}{d(1,2')}|c_{A}(t)|^2|d_{B}(t)|^2 +\frac{q^2}{d(2,2')}|d_{A}(t)|^2|b_{B}(t)|^2 +\frac{q^2}{d(1,1')}|c_{A}(t)|^2|c_{B}(t)|^2.
\end{eqnarray}
Similar reasoning can be conducted for 3 and more interacting qubits and thus we have hint how to describe the decoherence processes in the system with N-N1 inputs (qubits) and N1 outputs (qubits) having 2 energy levels (or more). 
Decoherence effects are due to qubit-qubit interaction and they arise from Coulomb interaction and they are inevitable. 
Finally we arrive to the Hamiltonian of the system of the following form

\begin{eqnarray}
H_{0+resonant}(t)+H_{Coulomb, \ket{E_i,E_j}\bra{E_k,E_s}}(t)=
\begin{pmatrix}
E_{1A}(t)+E_{1B}(t) & E_{12,B}(t) &  E_{12,A}(t) & 0 \\
E_{12,B}(t)^{*} &  E_{1A}(t)+E_{2B}(t) & 0 &  E_{12,A}(t) \\
E_{12,A}(t)^{*} & 0 & E_{2A}(t)+E_{1B}(t) &  E_{12,B}(t)  \\
0 &  E_{12,A}(t)^{*} & E_{12,B}(t)^{*} & E_{2A}(t)+E_{2B}(t)  \\
\end{pmatrix}
+ \nonumber \\
\begin{pmatrix}
(E_{1a}+E_{1b})_{r1} & H_{(E_{1a} \rightarrow E_{1a}, E_{2b} \rightarrow E_{1b})} & H_{(E_{2a} \rightarrow E_{1a}, E_{2b} \rightarrow E_{1b})} & H_{(E_{2a} \rightarrow E_{1a}, E_{2b} \rightarrow E_{1b})} \\
H_{(E_{1a} \rightarrow E_{1a}, E_{1b} \rightarrow E_{2b})} & (E_{1a}+E_{2b})_{r1} &  H_{(E_{2a} \rightarrow E_{1a}, E_{1b}\rightarrow E_{2b}} & H_{(E_{2a} \rightarrow E_{1a}, E_{2b}\rightarrow E_{2b}} \\
 H_{(E_{1a} \rightarrow E_{2a}, E_{1b} \rightarrow E_{2b})} &  H_{(E_{1a} \rightarrow E_{2a}, E_{2b}\rightarrow E_{1b}} & (E_{2a}+E_{1b})_{r1} & H_{(E_{2a} \rightarrow E_{2a}, E_{2b} \rightarrow E_{1b})} \\
H_{(E_{1a} \rightarrow E_{2a}, E_{1b} \rightarrow E_{2b})} &  H_{(E_{1a} \rightarrow E_{2a}, E_{2b}\rightarrow E_{2b}}  & H_{(E_{2a} \rightarrow E_{2a}, E_{1b} \rightarrow E_{2b})} & (E_{2a}+E_{2b})_{r1} \\
\end{pmatrix}.
\end{eqnarray}
In the presented Hamiltonian charge-charge interaction by electric field is accounted for all possible cases under assumption of the fact that system of 2 qubits preseverves 4 energetic levels. However it is possible to account the existence of magnetic field in qubit-qubit interaction as specified in \cite{PomorskiInternet}.  

We recognize that Coulomb interaction can be treated as the perturbation to the initial quantum state. In such case we can write the evolution of the quantum system as 
\begin{equation}
\ket{\psi(t)}= e^{\frac{1}{\hbar i}\int_{t_0}^{t}dt' H_{Coulomb}(t')}e^{\frac{1}{\hbar i}\int_{t_0}^{t}dt' H_{0+resonant}(t')}\ket{\psi(t_0)}. %%%e^{\frac{1}{\hbar i}\int_{t_0}^{t}dt'\ket{E_i,E_j}\bra{E_k,E_s}}(t')}\ket{\psi}\ket{\psi}(t_0).
\end{equation}
Particular simple analytical form can be obtained if $|a_A(t)|^2=|b_A(t)|^2=|c_A(t)|^2=|d_A(t)|^2$ 
and $|a_B(t)|^2=|b_B(t)|^2=|c_B(t)|^2=|d_B(t)|^2=|a_{A}(t)|^2$ 
that implies $E_{p1,A}(t)=E_{p2,A}(t)=E_{p1,B}(t)=E_{p2,B}(t)=E_{p}(t) \in R$ and $|t_{s12,A}(t)|=|t_{s12,B}(t)|=|t_{s12}(t)|$
so $a_{(A)B}=\frac{1}{\sqrt{2}}=-b_{(A)B}$ 
and $c_{(A)B}=d_{(A)B}=\frac{1}{\sqrt{2}}$. Having such simplifications we immediately recognize that all 4 renormalized eigenenergies brings the same renormalized value to each among 4 eigenenergies in the way as 

\begin{eqnarray}
(E_{1A}+E_{1B})_{r1} %=+\frac{1}{4}(\frac{q^2}{d(2,1')} +\frac{q^2}{d(1,2')}+\frac{q^2}{d(2,2')}+\frac{q^2}{d(1,1')}) %=, \nonumber \\ 
=(E_{1A}+E_{2B})_{r1} %+\frac{1}{4} (\frac{q^2}{d(2,1')}|b_{A}(t)|^2|c_{B}(t)|^2 +\frac{q^2}{d(1,2')}|a_{A}(t)|^2|d_{B}(t)|^2+ \frac{q^2}{d(2,2')}|d_{A}(t)|^2|b_{B}(t)|^2 + \frac{q^2}{d(1,1')}|c_{A(t)}|^2|a_{B}(t)|^2), \nonumber \\
=(E_{2A}+E_{1B})_{r1} %+\frac{1}{4} (\frac{q^2}{d(2,1')}|d_{A}(t)|^2 |a_{B}(t)|^2+\frac{q^2}{d(2,1')}|a_{A}(t)|^2|b_{B}(t)|^2+ \frac{q^2}{d(2,2')}|b_{A}(t)|^2|d_B(t)|^2+ \frac{q^2}{d(1,1')}|a_A(t)|^2|c_B(t)|^2), \nonumber \\
=(E_{2A}+E_{2B})_{r1}= +\frac{1}{4}(\frac{q^2}{d(2,1')} +\frac{q^2}{d(1,2')}+\frac{q^2}{d(2,2')}+\frac{q^2}{d(1,1')})= \nonumber \\
=(E_{A-B})_{r1}.
%+\frac{1}{4}(\frac{q^2}{d(2,1')}|d_{A}(t)|^2|c_{B}(t)|^{2} +\frac{q^2}{d(1,2')}|c_{A}(t)|^2|d_{B}(t)|^2 +\frac{q^2}{d(2,2')}|d_{A}(t)|^2|b_{B}(t)|^2 +\frac{q^2}{d(1,1')}|c_{A}(t)|^2|c_{B}(t)|^2).
\end{eqnarray}
We also have 
\begin{eqnarray}
H_{(E_{1a} \rightarrow E_{1a}, E_{2b} \rightarrow E_{1b})}=\frac{1}{4}(\frac{q^2}{d(1,2')}(-1)+\frac{q^2}{d(2,1')}(+1)+  %%%+ \nonumber \\ 
\frac{q^2}{d(1,1')}(+1)+\frac{q^2}{d(2,2')}(-1))=H(Q_1), \nonumber \\
H_{(E_{2a} \rightarrow E_{2a}, E_{2b} \rightarrow E_{1b})}=\frac{1}{4}(\frac{q^2}{d(1,2')}(-1)+\frac{q^2}{d(2,1')}(+1)+  %%%+ \nonumber \\ 
\frac{q^2}{d(1,1')}(+1)+\frac{q^2}{d(2,2')}(-1))=H(Q_1), \nonumber \\
H_{(E_{2a} \rightarrow E_{1a}, E_{1b} \rightarrow E_{1b})}=\frac{1}{4}(\frac{q^2}{d(1,2')}(+1)+\frac{q^2}{d(2,1')}(+1)+  %%%+ \nonumber \\ 
\frac{q^2}{d(1,1')}(+1)+\frac{q^2}{d(2,2')}(-1))=H(Q_2), \nonumber \\
H_{(E_{2a} \rightarrow E_{1a}, E_{2b} \rightarrow E_{2b})}=\frac{1}{4}(\frac{q^2}{d(1,2')}(-1)+\frac{q^2}{d(2,1')}(-1)+  %%%+ \nonumber \\ 
\frac{q^2}{d(1,1')}(+1)+\frac{q^2}{d(2,2')}(-1))=H(Q_3), \nonumber \\
H_{(E_{1a} \rightarrow E_{2a}, E_{1b} \rightarrow E_{2b})}=\frac{1}{4}(\frac{q^2}{d(1,2')}(-1)+\frac{q^2}{d(2,1')}(-1)+  %%%+ \nonumber \\ 
\frac{q^2}{d(1,1')}(+1)+\frac{q^2}{d(2,2')}(+1))=H(Q_4), \nonumber \\
H_{(E_{2a} \rightarrow E_{1a}, E_{1b} \rightarrow E_{2b})}=\frac{1}{4}(\frac{q^2}{d(1,2')}(-1)+\frac{q^2}{d(2,1')}(-1)  %%%+ \nonumber \\ 
+\frac{q^2}{d(1,1')}(+1)+\frac{q^2}{d(2,2')}(+1))=H(Q_4), \nonumber \\
%%\frac{q^2}{d(2,1')}( )+\frac{q^2}{d(1',2')}()=H_{(E_{1a} \rightarrow E_{1a}, E_{2b} \rightarrow E_{1b})}^{*}.
\end{eqnarray}
and this brings 
\begin{eqnarray}
H_{Coulomb, \ket{E_i,E_j}\bra{E_k,E_s}}=
\begin{pmatrix}
(E_{A-B})_{r1} & H_{(E_{1a} \rightarrow E_{1a}, E_{2b} \rightarrow E_{1b})} & H_{(E_{2a} \rightarrow E_{1a}, E_{2b} \rightarrow E_{1b})} & H_{(E_{2a} \rightarrow E_{1a}, E_{2b} \rightarrow E_{1b})} \\
H_{(E_{1a} \rightarrow E_{1a}, E_{1b} \rightarrow E_{2b})} & (E_{A-B})_{r1}  &  H_{(E_{2a} \rightarrow E_{1a}, E_{1b}\rightarrow E_{2b}} & H_{(E_{2a} \rightarrow E_{1a}, E_{2b}\rightarrow E_{2b}} \\
 H_{(E_{1a} \rightarrow E_{2a}, E_{1b} \rightarrow E_{2b})} &  H_{(E_{1a} \rightarrow E_{2a}, E_{2b}\rightarrow E_{1b}} & (E_{A-B})_{r1}  & H_{(E_{2a} \rightarrow E_{2a}, E_{2b} \rightarrow E_{1b})} \\
H_{(E_{1a} \rightarrow E_{2a}, E_{1b} \rightarrow E_{2b})} &  H_{(E_{1a} \rightarrow E_{2a}, E_{2b}\rightarrow E_{2b}}  & H_{(E_{2a} \rightarrow E_{2a}, E_{1b} \rightarrow E_{2b})} & (E_{A-B})_{r1} \\
\end{pmatrix}= \nonumber \\ =
\begin{pmatrix}
0 & H_{(E_{2a} \rightarrow E_{1a}, E_{2b} \rightarrow E_{1b})} & H_{(E_{2a} \rightarrow E_{1a}, E_{2b} \rightarrow E_{1b})} &  H_{(E_{2a} \rightarrow E_{1a}, E_{2b} \rightarrow E_{1b})}   \\
H_{(E_{1a} \rightarrow E_{1a}, E_{1b} \rightarrow E_{2b})} & 0  &  H_{(E_{2a} \rightarrow E_{1a}, E_{1b}\rightarrow E_{2b}} & H_{(E_{2a} \rightarrow E_{1a}, E_{2b}\rightarrow E_{2b}} \\
 H_{(E_{1a} \rightarrow E_{2a}, E_{1b} \rightarrow E_{2b})} &  H_{(E_{1a} \rightarrow E_{2a}, E_{2b}\rightarrow E_{1b}} & 0  & H_{(E_{2a} \rightarrow E_{2a}, E_{2b} \rightarrow E_{1b})} \\
H_{(E_{1a} \rightarrow E_{2a}, E_{1b} \rightarrow E_{2b})} &  H_{(E_{1a} \rightarrow E_{2a}, E_{2b}\rightarrow E_{2b}}  & H_{(E_{2a} \rightarrow E_{2a}, E_{1b} \rightarrow E_{2b})} & 0  \\
\end{pmatrix}+
(E_{A-B})_{r1} 
\begin{pmatrix}
1 & 0 & 0 & 0 \\
0 & 1 & 0 & 0 \\
0 & 0 & 1 & 0 \\
0 & 0 & 0 & 1 \\
\end{pmatrix}.
\end{eqnarray}

Finally the quantum state is the subject to the evolution with time of the following form

\begin{eqnarray}
\ket{\psi}(t)=\exp \left( \frac{1}{\hbar i}\int_{t0}^{t}dt'
\begin{pmatrix}
0 & H_{(E_{2a} \rightarrow E_{1a}, E_{2b} \rightarrow E_{1b})} & H_{(E_{2a} \rightarrow E_{1a}, E_{2b} \rightarrow E_{1b})} &  H_{(E_{2a} \rightarrow E_{1a}, E_{2b} \rightarrow E_{1b})}   \\
H_{(E_{1a} \rightarrow E_{1a}, E_{1b} \rightarrow E_{2b})} & 0  &  H_{(E_{2a} \rightarrow E_{1a}, E_{1b}\rightarrow E_{2b}} & H_{(E_{2a} \rightarrow E_{1a}, E_{2b}\rightarrow E_{2b}} \\
 H_{(E_{1a} \rightarrow E_{2a}, E_{1b} \rightarrow E_{2b})} &  H_{(E_{1a} \rightarrow E_{2a}, E_{2b}\rightarrow E_{1b}} & 0  & H_{(E_{2a} \rightarrow E_{2a}, E_{2b} \rightarrow E_{1b})} \\
H_{(E_{1a} \rightarrow E_{2a}, E_{1b} \rightarrow E_{2b})} &  H_{(E_{1a} \rightarrow E_{2a}, E_{2b}\rightarrow E_{2b}}  & H_{(E_{2a} \rightarrow E_{2a}, E_{1b} \rightarrow E_{2b})} & 0  \\
\end{pmatrix} \right) \times \nonumber \\
\times 
\begin{pmatrix}
e^{\frac{1}{\hbar i}((E_{A-B})_{r1} +E_{1A}+E_{1B})(t-t_0)} & 0 & 0 & 0 \\
0 & e^{\frac{1}{\hbar i}((E_{A-B})_{r1}+E_{1A}+E_{2B})(t-t_0)} & 0 & 0 \\
0 & 0 & e^{\frac{1}{\hbar i}((E_{A-B})_{r1}+E_{2A}+E_{1B})(t-t_0)} & 0 \\
0 & 0 & 0 & e^{\frac{1}{\hbar i}((E_{A-B})_{r1}+E_{2A}+E_{2B})(t-t_0)} \\
\end{pmatrix}
\begin{pmatrix}
\gamma_1(t_0) \\
\gamma_2(t_0) \\
\gamma_3(t_0) \\
\gamma_4(t_0)
\end{pmatrix},
\end{eqnarray}
where normalization condition takes place $|\gamma_1(t_0)|^2+|\gamma_2(t_0)|^2+|\gamma_3(t_0)|^2+|\gamma_4(t_0)|^2=1$ and $\gamma$ coeffients determine initial state of quantum system. 
It is therefore quite straighforward to obtain density matrix of the quantum system with time that has the structure
\begin{eqnarray}
\rho(t)=\ket{\psi}(t)\bra{\psi}(t)=\exp \left( \frac{1}{\hbar i}\int_{t0}^{t}dt'
\begin{pmatrix}
0 & H_{(E_{2a} \rightarrow E_{1a}, E_{2b} \rightarrow E_{1b})} & H_{(E_{2a} \rightarrow E_{1a}, E_{2b} \rightarrow E_{1b})} &  H_{(E_{2a} \rightarrow E_{1a}, E_{2b} \rightarrow E_{1b})}   \\
H_{(E_{1a} \rightarrow E_{1a}, E_{1b} \rightarrow E_{2b})} & 0  &  H_{(E_{2a} \rightarrow E_{1a}, E_{1b}\rightarrow E_{2b}} & H_{(E_{2a} \rightarrow E_{1a}, E_{2b}\rightarrow E_{2b}} \\
 H_{(E_{1a} \rightarrow E_{2a}, E_{1b} \rightarrow E_{2b})} &  H_{(E_{1a} \rightarrow E_{2a}, E_{2b}\rightarrow E_{1b}} & 0  & H_{(E_{2a} \rightarrow E_{2a}, E_{2b} \rightarrow E_{1b})} \\
H_{(E_{1a} \rightarrow E_{2a}, E_{1b} \rightarrow E_{2b})} &  H_{(E_{1a} \rightarrow E_{2a}, E_{2b}\rightarrow E_{2b}}  & H_{(E_{2a} \rightarrow E_{2a}, E_{1b} \rightarrow E_{2b})} & 0  \\
\end{pmatrix} \right) \times \nonumber \\
\times 
\begin{pmatrix}
e^{\frac{1}{\hbar i}((E_{A-B})_{r1} +E_{1A}+E_{1B})(t-t_0)} & 0 & 0 & 0 \\
0 & e^{\frac{1}{\hbar i}((E_{A-B})_{r1}+E_{1A}+E_{2B})(t-t_0)} & 0 & 0 \\
0 & 0 & e^{\frac{1}{\hbar i}((E_{A-B})_{r1}+E_{2A}+E_{1B})(t-t_0)} & 0 \\
0 & 0 & 0 & e^{\frac{1}{\hbar i}((E_{A-B})_{r1}+E_{2A}+E_{2B})(t-t_0)} \\
\end{pmatrix}
\begin{pmatrix}
\gamma_1(t_0) \\
\gamma_2(t_0) \\
\gamma_3(t_0) \\
\gamma_4(t_0)
\end{pmatrix} \times \nonumber \\
\begin{pmatrix}
\gamma_1(t_0)^{*} & \gamma_2(t_0)^{*} & \gamma_3(t_0)^{*} & \gamma_4(t_0)^{*}
\end{pmatrix} \times \nonumber \\
\begin{pmatrix}
e^{-\frac{1}{\hbar i}((E_{A-B})_{r1} +E_{1A}+E_{1B})(t-t_0)} & 0 & 0 & 0 \\
0 & e^{-\frac{1}{\hbar i}((E_{A-B})_{r1}+E_{1A}+E_{2B})(t-t_0)} & 0 & 0 \\
0 & 0 & e^{-\frac{1}{\hbar i}((E_{A-B})_{r1}+E_{2A}+E_{1B})(t-t_0)} & 0 \\
0 & 0 & 0 & e^{-\frac{1}{\hbar i}((E_{A-B})_{r1}+E_{2A}+E_{2B})(t-t_0)} \\
\end{pmatrix} \times \nonumber \\
\times \exp \left( -\frac{1}{\hbar i}\int_{t0}^{t}dt'
\begin{pmatrix}
0 & H_{(E_{2a} \rightarrow E_{1a}, E_{2b} \rightarrow E_{1b})} & H_{(E_{2a} \rightarrow E_{1a}, E_{2b} \rightarrow E_{1b})} &  H_{(E_{2a} \rightarrow E_{1a}, E_{2b} \rightarrow E_{1b})}   \\
H_{(E_{1a} \rightarrow E_{1a}, E_{1b} \rightarrow E_{2b})} & 0  &  H_{(E_{2a} \rightarrow E_{1a}, E_{1b}\rightarrow E_{2b}} & H_{(E_{2a} \rightarrow E_{1a}, E_{2b}\rightarrow E_{2b}} \\
 H_{(E_{1a} \rightarrow E_{2a}, E_{1b} \rightarrow E_{2b})} &  H_{(E_{1a} \rightarrow E_{2a}, E_{2b}\rightarrow E_{1b}} & 0  & H_{(E_{2a} \rightarrow E_{2a}, E_{2b} \rightarrow E_{1b})} \\
H_{(E_{1a} \rightarrow E_{2a}, E_{1b} \rightarrow E_{2b})} &  H_{(E_{1a} \rightarrow E_{2a}, E_{2b}\rightarrow E_{2b}}  & H_{(E_{2a} \rightarrow E_{2a}, E_{1b} \rightarrow E_{2b})} & 0  \\
\end{pmatrix} \right)= \nonumber \\
\begin{pmatrix}
1 & \frac{1}{\hbar i}\int_{t0}^{t}dt'H_{(E_{2a} \rightarrow E_{1a}, E_{2b} \rightarrow E_{1b})} & \frac{1}{\hbar i}\int_{t0}^{t}dt'H_{(E_{2a} \rightarrow E_{1a}, E_{2b} \rightarrow E_{1b})} & \frac{1}{\hbar i}\int_{t0}^{t}dt' H_{(E_{2a} \rightarrow E_{1a}, E_{2b} \rightarrow E_{1b})}   \\
\frac{1}{\hbar i}\int_{t0}^{t}dt'H_{(E_{1a} \rightarrow E_{2a}, E_{1b} \rightarrow E_{2b})} & 1  &  H_{(E_{2a} \rightarrow E_{1a}, E_{1b}\rightarrow E_{2b}} & H_{(E_{2a} \rightarrow E_{1a}, E_{2b}\rightarrow E_{2b}} \\
 \frac{1}{\hbar i}\int_{t0}^{t}dt'H_{(E_{1a} \rightarrow E_{2a}, E_{1b} \rightarrow E_{2b})} & \frac{1}{\hbar i}\int_{t0}^{t}dt' H_{(E_{1a} \rightarrow E_{2a}, E_{2b}\rightarrow E_{1b}} & 1  & \frac{1}{\hbar i}\int_{t0}^{t}dt'H_{(E_{2a} \rightarrow E_{2a}, E_{2b} \rightarrow E_{1b})} \\
\frac{1}{\hbar i}\int_{t0}^{t}dt'H_{(E_{1a} \rightarrow E_{2a}, E_{1b} \rightarrow E_{2b})} &  \frac{1}{\hbar i}\int_{t0}^{t}dt'H_{(E_{1a} \rightarrow E_{2a}, E_{2b}\rightarrow E_{2b}}  & \frac{1}{\hbar i}\int_{t0}^{t}dt'H_{(E_{2a} \rightarrow E_{2a}, E_{1b} \rightarrow E_{2b})} & 1  \\
\end{pmatrix}  \times \nonumber \\
\times 
\begin{pmatrix}
e^{\frac{1}{\hbar i}((E_{A-B})_{r1} +E_{1A}+E_{1B})(t-t_0)} & 0 & 0 & 0 \\
0 & e^{\frac{1}{\hbar i}((E_{A-B})_{r1}+E_{1A}+E_{2B})(t-t_0)} & 0 & 0 \\
0 & 0 & e^{\frac{1}{\hbar i}((E_{A-B})_{r1}+E_{2A}+E_{1B})(t-t_0)} & 0 \\
0 & 0 & 0 & e^{\frac{1}{\hbar i}((E_{A-B})_{r1}+E_{2A}+E_{2B})(t-t_0)} \\
\end{pmatrix} \times \nonumber \\
\begin{pmatrix}
|\gamma_1(t_0)|^2 & \gamma_1(t_0)\gamma_2(t_0)^{*}  & \gamma_1(t_0)\gamma_3(t_0)^{*} & \gamma_1(t_0)\gamma_4(t_0)^{*} \\
\gamma_2(t_0)\gamma_1(t_0)^{*} & |\gamma_2(t_0)|^2  & \gamma_2(t_0)\gamma_3(t_0)^{*} & \gamma_2(t_0)\gamma_4(t_0)^{*} \\
\gamma_3(t_0) & \gamma_3(t_0)\gamma_2(t_0)^{*}  & |\gamma_3(t_0)|^2  &  \gamma_3(t_0)\gamma_4(t_0)^{*} \\
\gamma_4(t_0)\gamma_1(t_0)^{*} & \gamma_4(t_0)\gamma_2(t_0)^{*}  & \gamma_4(t_0)\gamma_3(t_0)^{*} & |\gamma_4(t_0)|^2 
\end{pmatrix} \times \nonumber \\
\begin{pmatrix}
e^{-\frac{1}{\hbar i}((E_{A-B})_{r1} +E_{1A}+E_{1B})(t-t_0)} & 0 & 0 & 0 \\
0 & e^{-\frac{1}{\hbar i}((E_{A-B})_{r1}+E_{1A}+E_{2B})(t-t_0)} & 0 & 0 \\
0 & 0 & e^{-\frac{1}{\hbar i}((E_{A-B})_{r1}+E_{2A}+E_{1B})(t-t_0)} & 0 \\
0 & 0 & 0 & e^{-\frac{1}{\hbar i}((E_{A-B})_{r1}+E_{2A}+E_{2B})(t-t_0)} \\
\end{pmatrix} \times \nonumber \\
\times
\begin{pmatrix}
1 & -\frac{1}{\hbar i}\int_{t0}^{t}dt' H_{(E_{2a} \rightarrow E_{1a}, E_{2b} \rightarrow E_{1b})} & -\frac{1}{\hbar i}\int_{t0}^{t}dt' H_{(E_{2a} \rightarrow E_{1a}, E_{2b} \rightarrow E_{1b})} &  -\frac{1}{\hbar i}\int_{t0}^{t}dt' H_{(E_{2a} \rightarrow E_{1a}, E_{2b} \rightarrow E_{1b})}   \\
 -\frac{1}{\hbar i}\int_{t0}^{t}dt'H_{(E_{1a} \rightarrow E_{2a}, E_{1b} \rightarrow E_{2b})} & 1  &   -\frac{1}{\hbar i}\int_{t0}^{t}dt'H_{(E_{2a} \rightarrow E_{1a}, E_{1b}\rightarrow E_{2b}} &  -\frac{1}{\hbar i}\int_{t0}^{t}dt'H_{(E_{2a} \rightarrow E_{1a}, E_{2b}\rightarrow E_{2b}} \\
 -\frac{1}{\hbar i}\int_{t0}^{t}dt' H_{(E_{1a} \rightarrow E_{2a}, E_{1b} \rightarrow E_{2b})} &  -\frac{1}{\hbar i}\int_{t0}^{t}dt' H_{(E_{1a} \rightarrow E_{2a}, E_{2b}\rightarrow E_{1b}} & 1  &  -\frac{1}{\hbar i}\int_{t0}^{t}dt' H_{(E_{2a} \rightarrow E_{2a}, E_{2b} \rightarrow E_{1b})} \\
 -\frac{1}{\hbar i}\int_{t0}^{t}dt'H_{(E_{1a} \rightarrow E_{2a}, E_{1b} \rightarrow E_{2b})} &   -\frac{1}{\hbar i}\int_{t0}^{t}dt'H_{(E_{1a} \rightarrow E_{2a}, E_{2b}\rightarrow E_{2b}}  &  -\frac{1}{\hbar i}\int_{t0}^{t}dt'H_{(E_{2a} \rightarrow E_{2a},  -\frac{1}{\hbar i}\int_{t0}^{t}dt'E_{1b} \rightarrow E_{2b})} & 1  \\
\end{pmatrix}  = \nonumber \\
\exp \left( \frac{1}{\hbar i}\int_{t0}^{t}dt'
\begin{pmatrix}
0 & H_{(E_{2a} \rightarrow E_{1a}, E_{2b} \rightarrow E_{1b})} & H_{(E_{2a} \rightarrow E_{1a}, E_{2b} \rightarrow E_{1b})} &  H_{(E_{2a} \rightarrow E_{1a}, E_{2b} \rightarrow E_{1b})}   \\
H_{(E_{1a} \rightarrow E_{1a}, E_{1b} \rightarrow E_{2b})} & 0  &  H_{(E_{2a} \rightarrow E_{1a}, E_{1b}\rightarrow E_{2b}} & H_{(E_{2a} \rightarrow E_{1a}, E_{2b}\rightarrow E_{2b}} \\
 H_{(E_{1a} \rightarrow E_{2a}, E_{1b} \rightarrow E_{2b})} &  H_{(E_{1a} \rightarrow E_{2a}, E_{2b}\rightarrow E_{1b}} & 0  & H_{(E_{2a} \rightarrow E_{2a}, E_{2b} \rightarrow E_{1b})} \\
H_{(E_{1a} \rightarrow E_{2a}, E_{1b} \rightarrow E_{2b})} &  H_{(E_{1a} \rightarrow E_{2a}, E_{2b}\rightarrow E_{2b}}  & H_{(E_{2a} \rightarrow E_{2a}, E_{1b} \rightarrow E_{2b})} & 0  \\
\end{pmatrix} \right) \times \nonumber \\ 
\end{eqnarray}
\tiny
\begin{eqnarray*}
\times
\begin{pmatrix}
e^{-\frac{1}{\hbar i}((E_{A-B})_{r1} +E_{1A}+E_{1B})(t-t_0)}  |\gamma_1(t_0)|^2 & e^{-\frac{1}{\hbar i}((E_{A-B})_{r1} +E_{1A}+E_{1B})(t-t_0)}  \gamma_1(t_0)\gamma_2(t_0)^{*}  & e^{-\frac{1}{\hbar i}((E_{A-B})_{r1} +E_{1A}+E_{1B})(t-t_0)}  \gamma_1(t_0)\gamma_3(t_0)^{*} & e^{-\frac{1}{\hbar i}((E_{A-B})_{r1} +E_{1A}+E_{1B})(t-t_0)}  \gamma_1(t_0)\gamma_4(t_0)^{*} \\
 e^{-\frac{1}{\hbar i}((E_{A-B})_{r1}+E_{1A}+E_{2B})(t-t_0)} \gamma_2(t_0)\gamma_1(t_0)^{*} & e^{-\frac{1}{\hbar i}((E_{A-B})_{r1}+E_{1A}+E_{2B})(t-t_0)}|\gamma_2(t_0)|^2  & e^{-\frac{1}{\hbar i}((E_{A-B})_{r1}+E_{1A}+E_{2B})(t-t_0)} \gamma_2(t_0)\gamma_3(t_0)^{*} & e^{-\frac{1}{\hbar i}((E_{A-B})_{r1}+E_{1A}+E_{2B})(t-t_0)} \gamma_2(t_0)\gamma_4(t_0)^{*} \\
e^{-\frac{1}{\hbar i}((E_{A-B})_{r1}+E_{2A}+E_{1B})(t-t_0)} \gamma_3(t_0) & e^{-\frac{1}{\hbar i}((E_{A-B})_{r1}+E_{2A}+E_{1B})(t-t_0)} \gamma_3(t_0)\gamma_2(t_0)^{*}  & e^{-\frac{1}{\hbar i}((E_{A-B})_{r1}+E_{2A}+E_{1B})(t-t_0)} |\gamma_3(t_0)|^2  &  e^{-\frac{1}{\hbar i}((E_{A-B})_{r1}+E_{2A}+E_{1B})(t-t_0)} \gamma_3(t_0)\gamma_4(t_0)^{*} \\
e^{-\frac{1}{\hbar i}((E_{A-B})_{r1}+E_{2A}+E_{2B})(t-t_0)} \gamma_4(t_0)\gamma_1(t_0)^{*} & e^{-\frac{1}{\hbar i}((E_{A-B})_{r1}+E_{2A}+E_{2B})(t-t_0)} \gamma_4(t_0)\gamma_2(t_0)^{*}  &e^{-\frac{1}{\hbar i}((E_{A-B})_{r1}+E_{2A}+E_{2B})(t-t_0)}  \gamma_4(t_0)\gamma_3(t_0)^{*} & e^{-\frac{1}{\hbar i}((E_{A-B})_{r1}+E_{2A}+E_{2B})(t-t_0)}  |\gamma_4(t_0)|^2 
\end{pmatrix} \times \nonumber \\
\end{eqnarray*}
\normalsize
\begin{eqnarray*}
\times
\begin{pmatrix}
e^{-\frac{1}{\hbar i}((E_{A-B})_{r1} +E_{1A}+E_{1B})(t-t_0)} & 0 & 0 & 0 \\
0 & e^{-\frac{1}{\hbar i}((E_{A-B})_{r1}+E_{1A}+E_{2B})(t-t_0)} & 0 & 0 \\
0 & 0 & e^{-\frac{1}{\hbar i}((E_{A-B})_{r1}+E_{2A}+E_{1B})(t-t_0)} & 0 \\
0 & 0 & 0 & e^{-\frac{1}{\hbar i}((E_{A-B})_{r1}+E_{2A}+E_{2B})(t-t_0)} \\
\end{pmatrix} \times \nonumber \\
\times \exp \left(- \frac{1}{\hbar i}\int_{t0}^{t}dt'
\begin{pmatrix}
0 & H_{(E_{2a} \rightarrow E_{1a}, E_{2b} \rightarrow E_{1b})}^{*} & H_{(E_{2a} \rightarrow E_{1a}, E_{2b} \rightarrow E_{1b})}^{*} &  H_{(E_{2a} \rightarrow E_{1a}, E_{2b} \rightarrow E_{1b})}^{*}   \\
H_{(E_{1a} \rightarrow E_{1a}, E_{1b} \rightarrow E_{2b})}^{*} & 0  &  H_{(E_{2a} \rightarrow E_{1a}, E_{1b}\rightarrow E_{2b}}^{*} & H_{(E_{2a} \rightarrow E_{1a}, E_{2b}\rightarrow E_{2b}}^{*} \\
 H_{(E_{1a} \rightarrow E_{2a}, E_{1b} \rightarrow E_{2b})}^{*} &  H_{(E_{1a} \rightarrow E_{2a}, E_{2b}\rightarrow E_{1b}}^{*} & 0  & H_{(E_{2a} \rightarrow E_{2a}, E_{2b} \rightarrow E_{1b})}^{*} \\
H_{(E_{1a} \rightarrow E_{2a}, E_{1b} \rightarrow E_{2b})}^{*} &  H_{(E_{1a} \rightarrow E_{2a}, E_{2b}\rightarrow E_{2b}}^{*}  & H_{(E_{2a} \rightarrow E_{2a}, E_{1b} \rightarrow E_{2b})}^{*} & 0  \\
\end{pmatrix} \right)=  \nonumber  \\
= 
\begin{pmatrix}
1 & e^{\frac{1}{\hbar i}\int_{t0}^{t}dt' H_{(E_{1a} \rightarrow E_{1a}, E_{2b} \rightarrow E_{1b})}} & e^{\frac{1}{\hbar i}\int_{t0}^{t}dt'H_{(E_{2a} \rightarrow E_{1a}, E_{1b} \rightarrow E_{1b})}} &  e^{\frac{1}{\hbar i}\int_{t0}^{t}dt'H_{(E_{2a} \rightarrow E_{1a}, E_{2b} \rightarrow E_{1b})}}   \\
e^{\frac{1}{\hbar i}\int_{t0}^{t}dt'H_{(E_{1a} \rightarrow E_{1a}, E_{1b} \rightarrow E_{2b})}} & 1  &  e^{\frac{1}{\hbar i}\int_{t0}^{t}dt'H_{(E_{2a} \rightarrow E_{1a}, E_{1b}\rightarrow E_{2b})}} & e^{\frac{1}{\hbar i}\int_{t0}^{t}dt'H_{(E_{2a} \rightarrow E_{1a}, E_{2b}\rightarrow E_{2b}}} \\
e^{\frac{1}{\hbar i}\int_{t0}^{t}dt'H_{(E_{1a} \rightarrow E_{2a}, E_{1b} \rightarrow E_{1b})}} &  e^{\frac{1}{\hbar i}\int_{t0}^{t}dt'H_{(E_{1a} \rightarrow E_{2a}, E_{2b}\rightarrow E_{1b})}} & 1  & e^{\frac{1}{\hbar i}\int_{t0}^{t}dt'H_{(E_{2a} \rightarrow E_{2a}, E_{2b} \rightarrow E_{1b})}} \\
e^{\frac{1}{\hbar i}\int_{t0}^{t}dt'H_{(E_{1a} \rightarrow E_{2a}, E_{1b} \rightarrow E_{2b})}} &  e^{\frac{1}{\hbar i}\int_{t0}^{t}dt'H_{(E_{1a} \rightarrow E_{2a}, E_{2b}\rightarrow E_{2b}}}  & e^{\frac{1}{\hbar i}\int_{t0}^{t}dt'H_{(E_{2a} \rightarrow E_{2a}, E_{1b} \rightarrow E_{2b})}} & 1  \\
\end{pmatrix}  \times \nonumber \\
\end{eqnarray*}
\tiny
\begin{eqnarray*}
\times
\begin{pmatrix}
 |\gamma_1(t_0)|^2 & e^{\frac{1}{\hbar i}(E_{1B}-E_{2B})(t-t_0)}  \gamma_1(t_0)\gamma_2(t_0)^{*}  & e^{\frac{1}{\hbar i}(E_{1A}-E_{2A})(t-t_0)}  \gamma_1(t_0)\gamma_3(t_0)^{*} & e^{\frac{1}{\hbar i}((E_{1A}-E_{2A})+(E_{1B}-E_{2B}))(t-t_0)}  \gamma_1(t_0)\gamma_4(t_0)^{*} \\
 e^{\frac{1}{\hbar i}(E_{2B}-E_{1B})(t-t_0)} \gamma_2(t_0)\gamma_1(t_0)^{*} & |\gamma_2(t_0)|^2  & e^{\frac{1}{\hbar i}((E_{1A}-E_{2A})+(E_{2B}-E_{1B}))(t-t_0)} \gamma_2(t_0)\gamma_3(t_0)^{*} & e^{\frac{1}{\hbar i}(E_{1A}-E_{2A})(t-t_0)} \gamma_2(t_0)\gamma_4(t_0)^{*} \\
e^{\frac{1}{\hbar i}(E_{2A}-E_{1A})(t-t_0)} \gamma_3(t_0)\gamma_1^{*}(t_0) & e^{\frac{1}{\hbar i}((E_{2A}-E_{1A})-(E_{2B}-E_{1B}))(t-t_0)} \gamma_3(t_0)\gamma_2(t_0)^{*}  &  |\gamma_3(t_0)|^2  &  e^{\frac{1}{\hbar i}(E_{1B}-E_{2B})(t-t_0)} \gamma_3(t_0)\gamma_4(t_0)^{*} \\
e^{\frac{1}{\hbar i}((E_{2A}-E_{1A})+(E_{2B}-E_{1B}))(t-t_0)} \gamma_4(t_0)\gamma_1(t_0)^{*} & e^{\frac{1}{\hbar i}(E_{2A}-E_{1A})(t-t_0)} \gamma_4(t_0)\gamma_2(t_0)^{*}  &e^{\frac{1}{\hbar i}(E_{2B}-E_{1B})(t-t_0)}  \gamma_4(t_0)\gamma_3(t_0)^{*} &  |\gamma_4(t_0)|^2 
\end{pmatrix} \times \nonumber \\
\end{eqnarray*}
\normalsize
\begin{eqnarray*}
\times
\begin{pmatrix}
1 & e^{\frac{-1}{\hbar i}\int_{t0}^{t}dt'H_{(E_{1a} \rightarrow E_{1a}, E_{2b} \rightarrow E_{1b})}^{*}} & e^{\frac{-1}{\hbar i}\int_{t0}^{t}dt'H_{(E_{2a} \rightarrow E_{1a}, E_{1b} \rightarrow E_{1b})}^{*}} &  e^{\frac{-1}{\hbar i}\int_{t0}^{t}dt'H_{(E_{2a} \rightarrow E_{1a}, E_{2b} \rightarrow E_{1b})}^{*}}   \\
e^{\frac{-1}{\hbar i}\int_{t0}^{t}dt'H_{(E_{1a} \rightarrow E_{1a}, E_{1b} \rightarrow E_{2b})}^{*}} & 1  &  e^{\frac{-1}{\hbar i}\int_{t0}^{t}dtH_{(E_{2a} \rightarrow E_{1a}, E_{1b}\rightarrow E_{2b}}^{*}} & e^{\frac{-1}{\hbar i}\int_{t0}^{t}dt'H_{(E_{2a} \rightarrow E_{1a}, E_{2b} \rightarrow E_{2b})}^{*}} \\
e^{\frac{-1}{\hbar i}\int_{t0}^{t}dt'H_{(E_{1a} \rightarrow E_{2a}, E_{1b} \rightarrow E_{1b})}^{*}}  & e^{\frac{-1}{\hbar i}\int_{t0}^{t}dt' H_{(E_{1a} \rightarrow E_{2a}, E_{2b}\rightarrow E_{1b}}^{*}} & 1  & e^{\frac{-1}{\hbar i}\int_{t0}^{t}dt'H_{(E_{2a} \rightarrow E_{2a}, E_{2b} \rightarrow E_{1b})}^{*}} \\
e^{\frac{-1}{\hbar i}\int_{t0}^{t}dt'H_{(E_{1a} \rightarrow E_{2a}, E_{1b} \rightarrow E_{2b})}^{*}} &  e^{\frac{-1}{\hbar i}\int_{t0}^{t}dt'H_{(E_{1a} \rightarrow E_{2a}, E_{2b} \rightarrow E_{2b})}^{*}}  & e^{\frac{-1}{\hbar i}\int_{t0}^{t}dt'H_{(E_{2a} \rightarrow E_{2a}, E_{1b} \rightarrow E_{2b})}^{*}} & 1  \\
\end{pmatrix} = \nonumber \\
\begin{pmatrix}
1 & e^{i \Theta_{12}(t)} & e^{i \Theta_{13}(t)} & e^{i \Theta_{14}(t)} \\
 e^{i \Theta_{12}(t)^{*}} &  1 &  e^{i \Theta_{23}(t)} & e^{i \Theta_{24}(t)} \\
 e^{i \Theta_{13}(t)^{*}} &  e^{i \Theta_{23}^{*}(t)}   & 1  & e^{i \Theta_{34}(t)} \\
 e^{i \Theta_{14}(t)^{*}} &  e^{i \Theta_{24}^{*}(t)}   & e^{i \Theta_{34}^{*}(t)}   & 1 \\
\end{pmatrix} 
\begin{pmatrix}
e^{i \alpha_1(t)} & 0 & 0 & 0 \\
0 & e^{i \alpha_2(t)} & 0 & 0 \\
0 & 0 & e^{i \alpha_3(t)} & 0 \\
0 & 0 & 0 & e^{i \alpha_4(t)}
\end{pmatrix} 
\begin{pmatrix}
\rho_{11}(t_0) &  \rho_{12}(t_0) & \rho_{13}(t_0) &  \rho_{14}(t_0) \\
\rho_{21}(t_0) &  \rho_{22}(t_0) & \rho_{23}(t_0) &  \rho_{24}(t_0) \\
\rho_{31}(t_0) &  \rho_{32}(t_0) & \rho_{33}(t_0) &  \rho_{34}(t_0) \\
\rho_{41}(t_0) &  \rho_{42}(t_0) & \rho_{43}(t_0) &  \rho_{44}(t_0)
\end{pmatrix} \times \nonumber \\
\times
\begin{pmatrix}
e^{-i \alpha_1(t)} & 0 & 0 & 0 \\
0 & e^{-i \alpha_2(t)} & 0 & 0 \\
0 & 0 & e^{-i \alpha_3(t)} & 0 \\
0 & 0 & 0 & e^{-i \alpha_4(t)}
\end{pmatrix}
\begin{pmatrix}
1 & e^{-i \Theta_{12}(t)} & e^{-i \Theta_{13}(t)} & e^{-i \Theta_{14}(t)} \\
 e^{-i \Theta_{12}(t)^{*}} &  1 &  e^{-i \Theta_{23}(t)} & e^{-i \Theta_{24}(t)} \\
 e^{-i \Theta_{13}(t)^{*}} &  e^{-i \Theta_{23}^{*}(t)}   & 1  & e^{-i \Theta_{34}(t)} \\
 e^{-i \Theta_{14}(t)^{*}} &  e^{-i \Theta_{24}^{*}(t)}   & e^{-i \Theta_{34}^{*}(t)}   & 1 \\
\end{pmatrix}
%\end{eqnarray}
%\begin{eqnarray}
=\begin{pmatrix}
\rho_{11}(t) &  \rho_{12}(t) & \rho_{13}(t) &  \rho_{14}(t) \\
\rho_{21}(t) &  \rho_{22}(t) & \rho_{23}(t) &  \rho_{24}(t) \\
\rho_{31}(t) &  \rho_{32}(t) & \rho_{33}(t) &  \rho_{34}(t) \\
\rho_{41}(t) &  \rho_{42}(t) & \rho_{43}(t) &  \rho_{44}(t)
\end{pmatrix}
=\rho(t)_{A-B}, \nonumber \\
\end{eqnarray*}
\begin{eqnarray}
\rho(t)_{A}= 
\begin{pmatrix}
\rho_{11}+\rho_{22} & \rho_{13}+\rho_{24} \\
\rho_{31}+\rho_{42} & \rho_{33}+\rho_{44}
\end{pmatrix}. 
\rho(t)_{B}= 
\begin{pmatrix}
\rho_{11}+\rho_{33}  & \rho_{12}+\rho_{34}  \\
\rho_{21}+\rho_{43} & \rho_{22}+\rho_{44}
\end{pmatrix}. 
%%Tr\begin{pmatrix} \rho_{11} & \rho_{12} \\  \rho_{21} & \rho_{22} \end{pmatrix} & Tr\begin{pmatrix} \rho_{13} & \rho_{14} \\ \rho_{23} & \rho_{24} \end{pmatrix}  \\
%%Tr\begin{pmatrix} \rho_{31} & \rho_{32} \\ \rho_{41} & \rho_{42} \end{pmatrix} & Tr \begin{pmatrix} \rho_{33} & \rho_{34} \\ \rho_{43} & \rho_{44} \end{pmatrix}
%%\end{pmatrix}. 
\end{eqnarray}
From the above considerations we recognize  that evolution of the quantum state is equivalent to rotation by 4 real values angles $\alpha_1(t)$, $\alpha_2(t)$, $\alpha_3(t)$ and $\alpha_4(t)$ that correspond to 4 eigenergies of the system and by 6 complex valued angles 
$\Theta_{12}(t)$, $\Theta_{13}(t)$ , $\Theta_{14}(t)$, $\Theta_{23}(t)$, $\Theta_{24}(t)$, $\Theta_{34}(t)$ that correspons to transition between eigenenergies of non-interacting system. 
\normalsize
\section{Describing the decoherence effects by Schroedinger formalism}

\section{Introduction to spectral representation of Coulomb energy in Schroedinger formalism}

We have two weakly electrostatically interacting particles that are confined in separate wells by potentials $V_{p1}(x_1)$ and $V_{p2}(x_2)$. We neglect spin presence.
In order to avoid certain infinities we set small and non-zero d ($|d|<<1$). Then sightly mishaped Coulomb operator is given

\begin{equation}
    \label{simple_equation}
V_{Coulomb} = \frac{e^2}{\sqrt{|x_1-x_2|^2+d^2}}
\end{equation}
In formalized way we have
\begin{equation}
    \label{simple_equation}
\hat{V}_{Coulomb} = \int dx_1 \int dx_2 \frac{e^2}{\sqrt{|x_1-x_2|^2+d^2}}|x_1,x_2><x_1,x_2|
\end{equation}
.
The given quantum state can be written as having N energetic levels for each of perturbatively interacting particle p1 and p2.
\begin{eqnarray}
|\psi>=q_{0,0}|E_0,p_1>|E_0,p_2>+q_{1,0}|E_1,p_1>|E,p_2>+..+q_{N,0}|E_N,p1>|E_0,p2>+.. \nonumber \\
+q_{0,1}|E_0,p_1>|E_1,p_2>+q_{1,1}|E_1,p_1>|E,p_2>+..+q_{N,1}|E_N,p_1>|E_1,p_2>+..  \nonumber \\
+q_{0,2}|E_0,p_1>|E_2,p_2>+q_{1,2}|E_1,p_1>|E_2,p_2>+..+q_{N,2}|E_N,p1>|E_2,p2>+..  \nonumber \\
.. \nonumber \\
+q_{0,N}|E0,p1>|E_N,p2>+q_{1,n}|E1,p1>|E_N,p2>+..+q_{N,N}|E_N,p1>|E_N,p2> .
\label{state}
\end{eqnarray}

Normalization of quantum state requires $|q_{0,0}(t)|^2+|q_{0,1}(t)|^2+..+|q_{0,n}(t)|^2+|q_{1,0}(t)|^2+|q_{1,1}(t)|^2+..+|q_{1,n}(t)|^2+...+|q_{n,0}(t)|^2+|q_{n,1}(t)|^2+..+|q_{n,n}(t)|^2=1$.
We work in Schroedinger picture with preassumption that particle-particle interaction will change only values of $q_{k,l}(t)$ coefficients while all energetic level values remain the same. Therefore we have
\begin{eqnarray}
   \frac{d}{dt}|\psi>= \sum_{k,l} [\frac{d}{dt}(q_{k,l}(t)) |E_k,p1>|E_l,p2> ]= \sum_{k,l} [\frac{d}{dt}(q_{k,l}(t)) |E_{k,p1},E_{l,p2}>]
\end{eqnarray}
Single particle Hamiltonians acting on particle p1 and p2 are denoted as $H_{0,p1}=H_{k,p1}+V_{p1}$ and $H_{0,p2}=H_{k,p2}+V_{p2}$. They give
\begin{eqnarray}
   H_{0,p1}|\psi>= \sum_{k,l} [H_{0,p1}(q_{k,l}(t)) |E_k,p1>|E_l,p2> ]= \sum_l \sum_k (q_{k,l}(t)) E_{k,p1}|E_k,p1>|E_l,p2> %= \sum_{k,l} [\frac{d}{dt}(q_{k,l}(t)) |E_{k,p1},E_{l,p2}>]
\end{eqnarray}
and consequently
\begin{eqnarray}
   H_{0,p2}|\psi>= \sum_{k,l} [H_{0,p2}(q_{k,l}(t)) |E_k,p1>|E_l,p2> ]= \sum_k \sum_l (q_{k,l}(t)) E_{l,p2}|E_k,p1>|E_l,p2>.
    %= \sum_{k,l} [\frac{d}{dt}(q_{k,l}(t)) |E_{k,p1},E_{l,p2}>]
\end{eqnarray}

In our notation p1 and p2 stands for first and second particle.
We are using identities \\ $(\sum_n |E_n>_{p_1}<E_n|_{p_1} )=1$, \\ $(\sum_n |E_n>_{p_2}<E_n|_{p_2} )=1$, $\sum_n \sum_m |E_n(p_1),E_m (p_2) > <E_n(p_1), E_m(p_2)| =1$ and $\int dx'_1 \int dx'_2 |x'_1,x'_2><x'_1,x'_2|=1$.

We are making spectral decomposition of Coulomb operator as
\begin{eqnarray}
\label{main}
\hat{V}(x_1,x_2)=V(x_1,x_2) (\sum_n |E_n>_{p_1}<E_n|_{p_1} )(\sum_m |E_m>_{p_2}<E_m|_{p_2} ) = \nonumber \\
=V(x_1,x_2) (\sum_n \sum_m |E_n(p_1),E_m (p_2) > <E_n(p_1), E_m(p_2)| ) = \nonumber \\
=\sum_{i,j}g_{i,j}\psi_i(x1(p1))\psi_j(x2(p2))(\sum_n \sum_m |E_n(p_1),E_m (p_2) > <E_n(p_1), E_m(p_2)| )  = \nonumber \\
=\sum_{i,j}(\sum_n \sum_m g_{i,j}\psi_i(x1(p1))\psi_j(x2(p2))|E_n(p_1),E_m (p_2) > <E_n(p_1), E_m(p_2)| \int dx'_1 \int dx'_2 |x'_1,x'_2><x'_1,x'_2|) \nonumber \\
=\sum_{i,j}(\sum_n \sum_m g_{i,j}\psi_i(x1(p1))\psi_j(x2(p2)) \int dx'_1 \int dx'_2 \psi_m^{*}(x'_1(p1))\psi_n^{*}(x'_2(p2))|E_n(p_1),E_m (p_2) ><x'_1,x'_2| ) \nonumber \\
=\hat{V}(x_1,x_2)=\sum_{i,j,n,m} g_{i,j} \psi_i(x_1)_{p1}\psi_j(x_2)_{p2} \int_{-\infty}^{+\infty} dx'_1 \psi_m^{*}(x'_1)_{p1} \int_{-\infty}^{+\infty} dx'_2 \psi_n^{*}(x'_2)_{p2}|E_n(p_1),E_m (p_2) ><x'_1,x'_2|
\end{eqnarray}
Here indices i,j,n,m are running from 0 to N. Here operators $\int_{-\infty}^{+\infty} dx'_1 \psi_m^{*}(x'_1)_{p1}$ [.] and $\int_{-\infty}^{+\infty} dx'_1 \psi_n^{*}(x'_2)_{p2}$ [.]  play role of annihilation operators of particle p1 with energetic index m (energetic index components) and particle p2 with energetic index n (n-th energy eigenvalue) . At the same time operators $\psi_i(x_1)_{p1}$ and $\psi_j(x_2)_{p2}$ play role of creation operators of particle p1 with i-th energy component and particle p2 with j-th energy component. We work only in first quantization picture although some similarities to second quantization procedure can be recognized. Now we are determining spectral coefficients present in last equation and denoted by $g_{i,j}$.
Essentially we can write
\begin{equation}\label{spectral_decomposition}
V_{Coulomb} = \frac{e^2}{\sqrt{|x_1-x_2|^2+d^2}}=\sum_{i,j} g_{i,j} \psi_i(x_1)_{p1}\psi_j(x_2)_{p2}
\end{equation}
Now we use orthonormality features of $\psi_i(x_1)_{p1}$ and $\psi_j(x_2)_{p2}$ functions and we obtain
\begin{equation}\label{spectral_decomposition}
\int_{-\infty}^{+\infty} dx_1  \int_{-\infty}^{+\infty} dx_2 \frac{e^2}{\sqrt{|x_1-x_2|^2+d^2}} \psi_i^{*}(x_1)_{p1}\psi_j^{*}(x_2)_{p2} =g_{i,j}.
\end{equation}
Obviously having knowledge of confiment potentials $V_{p1}$ and $V_{p2}$  we have the knowledge of all eigenstates of p1 and p2 that is $\psi_i(x_1)_{p1}$ and $\psi_j(x_2)_{p2}$ and thus we have full knowledge on $g_{i,j}$ spectral coefficients in accordance to the last formula.
For 2 two level system of 2 weakly interacting particles we have i=(0,1) and j=(0,1) so we have 4 spectral coefficients $g_{i,j}$.
However we shall trace appearance of higher excited states in the process of particle-particle interaction. We can limit our considerations to N+1-th energetic levels for each interacting particle. Strict and most accurate results are with assumption $N+1 \rightarrow +\infty$. Practically we make certain cutoff for certain value of N+1 when for big values of i and j we observe that $g_{i,j}$ becomes really small.

Now we inspect the action of Coulomb operator in spectral representation as given by equation \ref{main} on the quantum state $|\psi>$. We have

\begin{eqnarray}
\hat{V}(x_1,x_2)|\psi>= \nonumber \\
=(\sum_{i,j,n,m} g_{i,j} \psi_i(x_1)_{p_1}\psi_j(x_2)_{p_2} \int_{-\infty}^{+\infty} dx'_1 \psi_n^{*}(x'_1)_{p_1} \int_{-\infty}^{+\infty} dx'_2 \psi_m^{*}(x'_2)_{p_2}|E_n(p_1),E_m (p_2) ><x'_1,x'_2|) \nonumber \\
(\int \int dx''_1 dx''_2 \psi(x''_1,x''_2,t) |x''_1,x''_2>)= \nonumber \\
=(\sum_{i,j,n,m} g_{i,j} \psi_i(x_1)_{p_1}\psi_j(x_2)_{p_2} \int_{-\infty}^{+\infty} dx'_1 \psi_n^{*}(x'_1)_{p_1} \int_{-\infty}^{+\infty} dx'_2 \psi_m^{*}(x'_2)_{p_2}|E_n(p_1),E_m (p_2) ><x'_1,x'_2|) \nonumber \\
(\int_{-\infty}^{+\infty} \int_{-\infty}^{+\infty} dx''_1 dx''_2 (\sum_{s,e}q_{s,e}(t)\psi_s(x''_1)_{p_1}\psi_e(x''_2)_{p_2}) |x''_1,x''_2>)= \nonumber \\
(\sum_{i,j,n,m} g_{i,j} \psi_i(x_1)_{p_1}\psi_j(x_2)_{p_2} q_{n,m}(t)|E_n,E_m >)=i \hbar \frac{d}{dt}|\psi>-H_{0,p_1}|\psi>-H_{0,p_2}|\psi> = \nonumber \\
=i\hbar \sum_{n,m} [ \frac{d}{dt}(q_{n,m}(t)) |E_{n},E_{m}>]- \sum_{n,m} ( q_{n,m}(t) E_{n,p_1})|E_n,E_m> \nonumber \\ -\sum_{n,m} (q_{n,m}(t) E_{m,p_2})|E_n,E_m> = (\sum_{i,j,n,m} g_{i,j} \psi_i(x_1)_{p_1}\psi_j(x_2)_{p_2} q_{n,m}(t)|E_n,E_m >).
\end{eqnarray}

The last two lines are equivalent to the following
\begin{eqnarray}
\int \int|x_1,x_2><x_1,x_2|dx_1dx_2 (i\hbar \sum_{n,m} [ \frac{d}{dt}(q_{n,m}(t)) |E_{n},E_{m}>]- \sum_{n,m} ( q_{n,m}(t) E_{n,p1})|E_n,E_m> ) \nonumber \\ = \int \int|x_1,x_2><x_1,x_2|dx_1dx_2|(\sum_{n,m} (q_{n,m}(t) E_{m,p2})|E_n,E_m> + (\sum_{i,j,n,m} g_{i,j} \psi_i(x_1)_{p1}\psi_j(x_2)_{p2} q_{n,m}(t)|E_n,E_m >)).
\end{eqnarray}
and can be written as
\begin{eqnarray}
\int \int|x_1,x_2> dx_1dx_2 (i\hbar \sum_{n,m} [ \frac{d}{dt}(q_{n,m}(t)) \psi(x_1)_{E_{n},p1}\psi(x_2)_{E_{m},p2} ] \nonumber \\ - \sum_{n,m} ( q_{n,m}(t) E_{n,p1}) ) \nonumber \\ = \int \int|x_1,x_2> dx_1dx_2\sum_{n,m}\psi(x_1)_{E_{n},p1}\psi(x_2)_{E_{m},p2} ((q_{n,m}(t) E_{m,p2})\psi(x_1)_{E_{N},p1}\psi(x_2)_{E_{m},p2} + \nonumber \\  (\sum_{i,j,n,m} g_{i,j} \psi_i(x_1)_{p1}\psi_j(x_2)_{p2} q_{n,m}(t))).
\end{eqnarray}
and further simplified as
\begin{eqnarray}
\int \int|x_1,x_2> dx_1dx_2 \sum_{n,m}\psi(x_1)_{E_{n},p1}\psi(x_2)_{E_{m},p2} \nonumber \\ (i\hbar [ \frac{d}{dt}(q_{n,m}(t))  ] - ( q_{n,m}(t) E_{n,p1}) ) \nonumber \\ = \int \int|x_1,x_2> dx_1dx_2\sum_{n,m}\psi(x_1)_{E_{N},p1}\psi(x_2)_{E_{m},p2} [(q_{n,m}(t) E_{m,p2}) + \nonumber \\  (\sum_{i,j} g_{i,j} \psi_i(x_1)_{p1}\psi_j(x_2)_{p2} q_{n,m}(t))].
\end{eqnarray}
Now we can write the $(N+1)^2$ equations since n and m run from 0 to N:
\begin{eqnarray}
i\hbar \frac{d}{dt}(q_{n,m}(t)) - q_{n,m}(t) E_{n,p1}- (q_{n,m}(t) E_{m,p2}) = (\sum_{i,j} g_{i,j}  q_{n,m}(t)\psi_i(x_1)_{p1}\psi_j(x_2)_{p2}).
\end{eqnarray}
We recognize that $q_{n,m}(t)$ coefficient is also position dependent. Basically having given $q_{n,m}(t)$ at given time instant we can evolve them to the next time step $t+\delta t$ and they will be position dependent. It is interesting result that is absent in lack of Coulomb interaction.

This procedure is straighforwad for the case of Np weakly interacting particles (as by electrostatic potential) confined by local potentials. However this procedure is more accurate than integro-differential equations that are not giving the full picture of quantum situation.  In particular we can trace development and emergence of entanglement between interacting particles. Since tight-binding model can be regarded as simplified version of Schroedinger equation it is desirable to describe the decoherence effects with usage of Schroedinger formalism as it is done in the next section.  
%%\subsection{
\section{Conclusion}

By imposing an occupancy of energetic state on one position-based qubit entangled to a radiation coming from a quantum coherent resonant cavity, we are enforcing the other qubit to change its state accordingly. It can be the base for the quantum communication and quantum internet. The generalization of the reasoning for $N$ qubits coupled to the resonant cavity as via a superconducting waveguide (that has a high quality factor) is quite straightforward. In most considerations, we need to go beyond the rotation phase approximation.
The concept of quantum internet was shown in this work. The more detailed picture requires taking into account various effects as decoherence processes that drive the quantum position-base qubit out of its coherence as well as decoherence processes that destroy the coherence of qEC (quantum Electromagnetic Cavity). It is quite important to underline that in order to bring the interaction of qEC with the position-based qubit, we need to place the position-based qubit either in the interior of qEC or in its proximity.
In the first case, bringing the position based qubit into the interior of qEC we are changing the resonant modes of the qEC and we are thus naturally bringing additional decoherence to the qEC. In the second case, in order to force the interaction between qEC and position-based qubit, we need to make a hole in the qEC wall. There is a non-zero electromagnetic radiation emitted outside from that hole, which brings the internal decoherence to the qEC. The larger the hole the stronger the interaction between the position-based qubit and qEC.
Therefore, the presented mathematical results shall be treated as a preliminary work on implementing a quantum communication with the position-based qubits.
In the conducted work, the simplistic approach is attempted as we are using a tight-binding model for the description of position-based qubits or a simplistic model for the matter-radiation interaction. This methodology shall be extended to take into account the Schr\"odinger description
of the position-based qubits as a more refined Quantum Electrodynamical Models and thus it is the subject of future work. The presented results open perspectives for implementing quantum Internet-of-Things devices. However, it shall be stressed that the conducted considerations are implementable when semiconductor qubits are quantum coherent and when the electromagnetic cavity maintains quantum coherence as well as when there is quantum interaction between position-based qubits and quantum electromagnetic cavity. It is achievable at very low temperatures as in range of 10\,mK. Quite obviously, we can extend the presented results to a quantum waveguide interacting with the position-based qubits, since the waveguide is a special case of the electromagnetic resonator.
We can use the obtained approach to study time crystals in CMOS structures \cite{Wilczek}, \cite{Sacha}.
%In this section we welcome you to include a summary of the end results of your research.

%\begin{acknowledgments}
%We wish to acknowledge the support of the author community in using
%REV\TeX{}, offering suggestions and encouragement, testing new versions,
%% The activity was suppored by the grant by Science Foundation Ireland under Grant 14/RP/I2921. %Sections 1-4 were presented to UCD group and to EQUAL 1 company on 16 February 2019.
%%I would like to thank to professor Jakub Rembielinski from University of Lodz for teaching me quantum mechanics as expressed in terms of projector operators. Special thanks are also given to Adam Bednorz from Univerity of Warsaw and to professor Andrew Mitchell from University College Dublin for lengthy discussions on tight-binding model. The assistance in picture preparation were done by Erik Staszewski from University College Dublin.
%\end{acknowledgments}
%\nocite{*}
%\bibliography{aipsamp}% Produces the bibliography via BibTeX.

\begin{thebibliography}{00}

\bibitem{Bashir19}
I. Bashir, M. Asker, C. Cetintepe, D. Leipold, A. Esmailiyan, H. Wang,
T. Siriburanon, P. Giounanlis, E. Blokhina, K. Pomorski, and R.~B.~Staszewski,
``A mixed-signal control core for a fully integrated semiconductor quantum computer system-on-chip,''
\emph{Proc. of IEEE European Solid-State Circuits Conf. (ESSCIRC)},
sec.~A2L-C4, pp.~125--128, 24~Sept.~2019.

\bibitem{Fujisawa} T. Fujisawa, T. Hayashi, HD Cheong, YH Jeong, and Y. Hirayama.
Rotation and phase-shift operations for a charge qubit in a double
quantum dot. Physica E: Low-dimensional Systems and Nanostructures,
21(2-4):10461052, 2004.

\bibitem{Petta} K. D. Petersson, J. R. Petta, H. Lu, and A. C. Gossard. Quantum
coherence in a one-electron semiconductor charge qubit. Phys. Rev. Lett.,
105:246804, 2010.

\bibitem{Dirk}
D. Leipold, Controlled Rabi Oscillations as foundation for entangled quantum aperture logic, Seminar
at UC Berkley Quantum Labs, 25th July 2018.

\bibitem{Panos} P.Giounanlis, E.Blokhina, K.Pomorski, D.R.Leipold, R.B.Staszewski, Modeling of Semiconductor Electrostatic Qubits Realized Through Coupled Quantum Dots,
10.1109/ACCESS.2019.2909489,IEEE Access, 2019

\bibitem{Pomorski_spie} Krzysztof Pomorski, Panagiotis Giounanlis, Elena Blokhina, Dirk Leipold, Pawel Peczkowski, Robert Bogdan Staszewski, From two types of electrostatic position-dependent semiconductor qubits to quantum universal gates and hybrid semiconductor-superconducting quantum computer, Proc. SPIE 11054, Superconductivity and Particle Accelerators 2018, 110540M, 2019.

\bibitem{Spalek}
Jozef Spalek, Wstep do fizyki materii skondensowanej, PWN, 2015.

\bibitem{Jaynes}
 E. T. Jaynes, and F. W. Cummings, Proc. IEEE51, 89(1963).

\bibitem{EntanglementMR}
Dimitris G. Angelakis, Stefano Mancini, Sougato Bose,
Steady state entanglement between hybrid light-matter qubits, arXiv:0711.1830, 2008.

\bibitem{Pomorski_compel} K.Pomorski, H.Akaike, A.Fujimaki, and K.Rusek. Relaxation method
in description of ram memory cell in rsfq computer, COMPEL, 38(1):395414, 2019.
\bibitem{Choi}
M.S.Choi, J.Yi, M.Y.Choi, J.Choi, and S.I.Lee. Quantum phase
transitions in josephson-junction chains. Phys. Rev. B, 57:R716R719, 1998.
\bibitem{QPT}
S.Sachdev. Quantum phase transitions. Cambridge Univ. Press, 2011.
\bibitem{Xu} H. Q. Xu. Method of calculations for electron transport in multiterminal quantum systems based on real-space lattice models. Phys. Rev. B, 66:165305.
\bibitem{Belzig}
D. Maile, S. Andergassen, and W. Belzig. Quantum phase transition
with dissipative frustration. Phys. Rev. B, 97, 2018.
\bibitem{PSSB2012}
K.Pomorski, P.Prokopow, Possible existence of field-induced Josephson junctions, Vol.249, No. 9, Physica Status Solidi B, 2012
\bibitem{Statistics}
C. Wetterich, Quantum mechanics from classical statistics,
Arxiv:0906.4919
\bibitem{Statistics1}
$http://math.ucr.edu/home/baez/quantropy.pdf$
\bibitem{Nbody}
Krzysztof Pomorski, Robert Bogdan Staszewski, Analytical Solutions for N-Electron Interacting System Confined in Graph of Coupled Electrostatic Semiconductor and Superconducting Quantum Dots in Tight-Binding Model with Focus on Quantum Information Processing, 22 October 2019, $https://arxiv.org/abs/1907.03180$.
\bibitem{Wilczek}
Frank Wilczek, Quantum Time Crystals, Phys. Rev. Lett. 109, 2012
\bibitem{Sacha}
Krzysztof Sacha, Jakub Zakrzewski,Time crystals: a review, Reports on Progress in Physics, Volume 81, Number 1, 2017
\bibitem{Birula}
I. Bialynicki-Birula, On the wavefunction of the photon,
Vol. 86, Acta Physical Polonica A, No. 1-2, Proceedings of the International Conference "Quantum Optics III", Szczyrk, Poland, 1993
\bibitem{2SEL}
K.Pomorski, P.Giounanlis, E. Blokhina, D.Leipold, R.B. Staszewski, "Analytic view on coupled single-electron lines", 
Semiconductor Science and Technology, Vol.34, Nr.12, 2019
\bibitem{PomorskiInternet}
K.Pomorski, R.B.Staszewski, Towards quantum internet and non-local communication in position based qubits, Arxiv:1911.02094, 2019
%%Imran Bashir, Mike Asker, Cagri Cetintepey, Dirk Leipold, Ali Esmailiyany, Hongying Wangy,
%%Teerachot Siriburanony, Panagiotis Giounanlisy, Elena Blokhinay, Krzysztof Pomorskiy and R. Bogdan Staszewski , A Mixed-Signal Control Core for a Fully Integrated Semiconductor Quantum Computer System-on-Chip, ESSCIRC 2019 - IEEE 45th European Solid State Circuits Conference (ESSCIRC), 2019
\end{thebibliography}
\section*{Acknowledgment}
 The activity was suppored by the grant by Science Foundation Ireland under Grant 14/RP/I2921. %Sections 1-4 were presented to UCD group and to EQUAL 1 company on 16 February 2019.
I would like to thank to professor Robert Bogdan Staszewski for invitation to this project and to professor Jakub Rembielinski from University of Lodz for teaching me quantum mechanics as expressed in terms of projector operators. Special thanks are also given to Adam Bednorz from Univerity of Warsaw and to professor Andrew Mitchell from University College Dublin for lengthy discussions on tight-binding model. The assistance in picture preparation were done by Erik Staszewski from University College Dublin. Section IIa, III and V was presented to professor Robert Bogdan Staszewski on 16 th February 2019 and to EQUAL 1 company at the same time.

\vspace{12pt}

\end{document}